\documentclass[twocolumn,linenumbers]{aastex701}

\usepackage{amsmath}
\usepackage{amssymb}
\usepackage{enumitem}
\usepackage[ruled,vlined,linesnumbered]{algorithm2e}
\usepackage{tikz}
\usetikzlibrary{arrows.meta,positioning,shapes.geometric}
\usepackage{pifont}
\setlist{nosep}

\definecolor{checkgreen}{RGB}{0,130,0}
\newcommand{\yes}{\textcolor{checkgreen}{\ding{51}}}
\newcommand{\no}{\textcolor{red}{\ding{55}}}

\newcommand{\RICH}{\textsc{rich}}
\newcommand{\STORM}{\textsc{storm}}
\newcommand{\MadVoro}{\textsc{MadVoro}}
\newcommand{\IMC}{\textsc{IMC}}
\newcommand{\DDMC}{\textsc{DDMC}}
\newcommand{\Ddop}{D}
\newcommand{\Ddiff}{\mathcal{D}}
\newcommand{\cV}{C_V}

\numlinesfalse\nolinenumbers  % Uncomment this line to disable line numbers

\shorttitle{Multigroup IMC on Moving Voronoi Meshes}
\shortauthors{Mizrachi et al.}

\begin{document}

\title{Multigroup Implicit Monte Carlo Radiation Hydrodynamics on Moving Voronoi Meshes with STORM and RICH}

\correspondingauthor{Maor Mizrachi}
\author{Maor Mizrachi}
\affiliation{School of Computer Science and Engineering, The Hebrew University, 9190401 Jerusalem, Israel}
\email[show]{maormiz@cs.huji.ac.il}

\author{Barak Raveh}
\affiliation{School of Computer Science and Engineering, The Hebrew University, 9190401 Jerusalem, Israel}
\email{barak.raveh@cs.huji.ac.il}

\author{Elad Steinberg}
\affiliation{Racah Institute of Physics, The Hebrew University, 9190401 Jerusalem, Israel}
\email{elad.steinberg@mail.huji.ac.il}

\begin{abstract}
We present and validate an Implicit Monte Carlo (\IMC) radiation-transport solver in the
open-source, mesh-agnostic \STORM{} library, coupled to the moving-mesh hydrodynamics code \RICH{}
on three-dimensional Voronoi tessellations.  The solver is multigroup, or continuous in frequency,
throughout; it transforms photon packets exactly between the comoving and laboratory frames, books
every exchange with the material as a laboratory four-momentum so that the \IMC{} path conserves
energy to roundoff, formulates the Fleck factor in the fluid's proper time, and accelerates optically thick
cells with multigroup Discrete Diffusion Monte Carlo (\DDMC{}) behind an emissivity-preserving
interface.  We tested the solver on a broad suite of benchmark problems, and it performed well
on all of them.  The solver is built on the MPI-parallel Monte Carlo framework of \STORM{}, which
scales to $13{,}440$ cores.
\end{abstract}


%%%%%%%%%%%%%%%%%%%%%%%%%%%%%%%%%%%%%%%%%%%%%%%%%%

\section{Introduction}\label{sec:introduction}

Time-dependent radiation transport plays a central role in a broad range of astrophysical and
laboratory high-energy-density phenomena.  It sets the thermal and ionization structure of stellar
atmospheres \citep{Mihalas1978,HubenyMihalas2014} and can drive eruptions from the
radiation-dominated envelopes of massive stars \citep{Jiang2015,Jiang2018}.  Radiation pressure,
photon trapping, and radiative cooling regulate accretion disks and launch outflows from
super-Eddington flows \citep{ShakuraSunyaev1973,Ohsuga2005,Jiang2014,Jiang2019,Sadowski2015,DavisTchekhovskoy2020},
and limit accretion onto forming massive stars and mediate stellar feedback
\citep{Krumholz2009,Kuiper2010,Rosen2016,KrumholzThompson2012,Davis2014}.  Explosive transients
are shaped by radiation transport from shock breakout
\citep{NakarSari2010,Sapir2011,WaxmanKatz2017,LevinsonNakar2020} through the diffusion-dominated
light curves of supernovae
\citep{Arnett1982,Kasen2006,KromerSim2009,KasenWoosley2009,KasenBildsten2010,ChevalierIrwin2011,Smith2014};
the emission of tidal disruption events is produced by shocks in, and reprocessed through,
optically thick and highly asymmetric debris
\citep{Rees1988,Piran2015,Roth2016,MetzgerStone2016,Dai2018,LuBonnerot2020,Gezari2021,Thomsen2022,SteinbergStone2024,Price2024};
kilonovae are governed by photon diffusion through $r$-process ejecta with extreme line opacities
\citep{LiPaczynski1998,Metzger2010,Kasen2013,BarnesKasen2013,TanakaHotokezaka2013,Kasen2017,Metzger2019};
and radiative shocks and recombination fronts shape classical and luminous red novae
\citep{Metzger2014,Ivanova2013,MatsumotoMetzger2022}.  In the laboratory, photon transport through
heterogeneous, high-opacity materials controls inertial-confinement-fusion implosions
\citep{Lindl1995,Lindl2004,Hurricane2014,AbuShawareb2022} and the supersonic radiation (Marshak)
waves of high-energy-density experiments \citep{Remington2006,Drake2006,Moore2015}.  In all of these
settings the radiation is coupled to the material through absorption, emission, and scattering,
with opacities spanning many orders of magnitude in space and photon energy
\citep{Pomraning1973,Mihalas1984,Castor2004}.

The mathematical description of this coupling is the time-dependent Boltzmann transport equation
for the specific intensity, a seven-dimensional quantity depending on position, direction,
frequency, and time.  Approximate treatments---flux-limited diffusion (FLD), moment methods
($M_1$, the first-moment closure, and variable Eddington tensor), and the $P_N$ spherical-harmonics expansion of order $N$---reduce the
dimensionality by integrating over angle, but each introduces a closure that can fail in a
specific regime.  FLD cannot reproduce shadows or beaming in optically thin regions, while
low-order moment closures may produce unphysical radiation fronts near material interfaces.
Discrete-ordinates ($S_N$, with $N$ denoting the angular quadrature order) methods retain directional dependence, but suffer from ray effects in
multidimensional problems and require substantial memory at high angular resolution
\citep{Lathrop1968,LewisMiller1984}.

Monte Carlo (MC) methods provide a natural framework for the full transport equation
\citep{Noebauer2019}.  By sampling photon-packet trajectories stochastically, MC avoids angular
discretization and accommodates complicated three-dimensional geometries, frequency-dependent
opacities, and anisotropic scattering.  The method is free of ray effects, treats boundaries of
arbitrary shape without special meshing, and converges at a rate independent of dimensionality.
Once the material state is known, packets can be propagated independently, which maps efficiently
onto distributed-memory architectures.

Na\"ive explicit MC time-stepping is nonetheless impractical when matter and radiation are tightly
coupled.  In optically thick regions the photon mean free path is much smaller than the cell size,
and the material--radiation equilibration time is much shorter than the hydrodynamic time step.
An explicit scheme must resolve that fast timescale, producing prohibitively small steps and an
enormous number of absorption--re-emission events per packet.

The Implicit Monte Carlo (\IMC{}) method of \citet{FleckCummings1971} removes this stiffness by
linearizing the Planck emission source over the time step and converting part of the true
absorption into effective isotropic scattering, distinct from the elastic scattering of the
medium.  The resulting scheme is unconditionally stable in
the usual stiff sense and approaches the correct diffusion limit in optically thick media.  Its
properties, extensions, and limitations---including possible violations of the maximum
principle---have been reviewed by \citet{Wollaber2016}.

Even with that linearization, optically thick cells remain expensive: each packet may undergo many
effective-scattering events before leaving a cell.  Discrete Diffusion Monte Carlo (\DDMC{})
replaces the random walk inside a thick cell by a diffusion jump to a face
\citep{Densmore2007,Densmore2012,Wollaeger2013}.

Many astrophysical flows---stellar winds \citep{CastorAbbottKlein1975,Owocki1988,Stevens1992},
supernova ejecta \citep{Kifonidis2003,Wongwathanarat2015}, accretion streams
\citep{GuillochonRamirezRuiz2013,Shiokawa2015,Bonnerot2016,SteinbergStone2024}, and merger
remnants \citep{Hotokezaka2013,Radice2018,Ohlmann2016,Schneider2019}---have large bulk
velocities and evolving geometry that static grids capture poorly.
Moving-mesh hydrodynamics advects computational cells with the fluid, combining Lagrangian
geometric flexibility with Eulerian robustness
\citep{Springel2010,DuffellMacFadyen2011,Yalinewich2015}.  Coupling
thermal radiation to such a mesh lets Doppler and aberration effects be treated in the comoving
frame of each cell.  Monte Carlo packets themselves suffer no advection error, but quasi-Lagrangian
motion reduces the numerical diffusion of shocks and material interfaces, and the radiation
solution improves through the sharper material structure it is coupled to
(Section~\ref{sec:mach2}).

The open-source code \RICH{} \citep{Yalinewich2015} already provides this hydrodynamics
infrastructure on a moving Voronoi mesh, together with grey FLD \citep{SteinbergStone2024} and
multigroup FLD \citep{Giron2026}.  Those diffusion solvers are efficient in the optically thick
limit, but they cannot represent shadows, beaming, or streaming through narrow optically thin
channels.  A complementary Monte Carlo treatment of thermal radiative transfer is therefore
needed inside the same hydrodynamics framework.  Closely related Monte Carlo radiation
hydrodynamics on a moving Voronoi mesh has been developed in \textsc{Arepo-mcrt}
\citep{Smith2020}, which is built into the \textsc{Arepo} code and combines implicit Monte Carlo
thermal emission, discrete-diffusion acceleration of unresolved scattering, path-based estimators,
and face-centered momentum coupling.  Our code is novel in its multigroup, or continuous-in-frequency,
handling: the transport and its \DDMC{} acceleration are both multigroup, as described in
Section~\ref{sec:multigroup} and Section~\ref{sec:ddmc}.  We demonstrate this on the Densmore
step-opacity problem with $32$ groups (Section~\ref{sec:densmore}), the Olson two-dimensional
multimaterial problem with continuous frequency (Section~\ref{sec:olson}), and the McClarren--Gentile
moving slab with $124$ groups (Section~\ref{sec:moving_slab}).  Our code is also open source,
whereas, to our knowledge, \textsc{Arepo-mcrt} has not been publicly released.

The principal contribution of this work is not a new derivation of \IMC{}, but the
implementation and validation of a Monte Carlo radiation--hydrodynamics capability with four
distinguishing features.  First, the transport kernel lives in a standalone, mesh-agnostic
library and runs unchanged on Cartesian grids and on three-dimensional unstructured Voronoi
tessellations, with \RICH{} supplying the moving-mesh hydrodynamics as one client.  Second, the
transport is multigroup, or continuous in frequency, throughout, and its \DDMC{} acceleration is
multigroup as well, with partially diffusive frequency bands, an emissivity-preserving
\IMC{}--\DDMC{} interface \citep{Densmore2006}, and moving-interface corrections
\citep{Wollaeger2013} on unstructured cells.  Third, photon packets are Lorentz-transformed
exactly between the comoving and laboratory frames, every exchange with the material is booked as a
laboratory-frame four-momentum so that the \IMC{} path conserves energy to roundoff, and the Fleck
factor is
formulated in the fluid's proper time.  Fourth, the solver is validated on a broad suite that
includes three-dimensional, heterogeneous problems---the crooked pipe, a binary stochastic medium,
and the spherical RMTV shock on fixed and moving meshes---as well as an axisymmetric cylindrical
hohlraum run as a full three-dimensional calculation.

We implement the radiation physics in \STORM{} (Scalable Transport via One-sided Remote Memory)
\citep{Mizrachi_2026}, an open-source Monte
Carlo library whose transport kernel does not depend on a particular mesh.  \STORM{} accesses the
grid only through a small set of geometric queries (Section~\ref{sec:architecture}); the same
\IMC{} solver can therefore run on Cartesian grids or unstructured tessellations.  The Voronoi constructor used by \RICH{} is
\MadVoro{} \citep{Mizrachi2025MadVoro}, which is also available to \STORM{} as a standalone
tessellation backend.  \RICH{} supplies the operator-split coupling of that transport to
three-dimensional moving-mesh hydrodynamics: it advances the Euler equations, reconstructs the
Voronoi mesh with \MadVoro{}, and then calls \STORM{} to propagate photon packets on the current
tessellation.  Packet migration across MPI ranks, RDMA communication, termination detection, and
load balancing are described in the companion \STORM{} paper and are
deliberately decoupled from the physics described here.

This paper is structured as follows.  Section~\ref{sec:physical_model} presents the
multigroup transport equation, including the comoving-to-laboratory frame transformations that
moving media require, and its \IMC{} discretization.  Section~\ref{sec:implementation} describes
the Monte Carlo implementation: the software architecture, the radiation substep, the frequency
groups, thermal emission, energy deposition, and the per-packet event loop on a Voronoi cell.
Section~\ref{sec:radhydro} describes the operator-split radiation--hydrodynamics coupling.
Section~\ref{sec:ddmc} describes \DDMC{} acceleration.
Section~\ref{sec:pop_control} discusses population control.  Section~\ref{sec:results} validates
the solver against a suite of benchmarks, and Section~\ref{sec:conclusion} summarizes our
conclusions.

%%%%%%%%%%%%%%%%%%%%%%%%%%%%%%%%%%%%%%%%%%%%%%%%%%
\section{Physical Model and IMC Formulation}\label{sec:physical_model}

\subsection{The Radiation-Transport Equation}\label{sec:transport_eq}

We consider the time-dependent, frequency-dependent radiation-transport equation for the specific
intensity $I(\mathbf{x}, \hat{\mathbf{n}}, \nu, t)$ of photons propagating through a material
medium:
\begin{equation}\label{eq:transport}
    \frac{1}{c}\frac{\partial I}{\partial t}
    + \hat{\mathbf{n}} \cdot \nabla I
    = \eta - \kappa_a I - \kappa_s I
    + \frac{\kappa_s}{4\pi} \int I(\hat{\mathbf{n}}')\, d\Omega',
\end{equation}
where $\mathbf{x}$ is position, $t$ is time, $d\Omega$ is an element of solid angle, and the primed direction
$\hat{\mathbf{n}}'$ is integrated over the full sphere.  Here $c$ is the speed of light, $\hat{\mathbf{n}}$ is the propagation direction, $\nu$ is the
photon frequency, $\kappa_a(\nu, \rho, T)$ and $\kappa_s(\rho, T)$ are the macroscopic absorption
and scattering opacities, $\rho$ is material mass density, $T$ is material
temperature and $\eta(\nu, \rho, T)$ is the thermal emission coefficient.  The
scattering carried by $\kappa_s$ is assumed to be \emph{elastic} (coherent) and isotropic: the scattering
integral of Eq.~\eqref{eq:transport} redistributes photons in angle at fixed frequency, so a
scattered photon retains its comoving energy.  We assume that there are no physical inelastic processes.  The only
scattering that changes a photon's frequency is the effective scattering introduced by the \IMC{}
linearization in Section~\ref{sec:imc}, which redistributes photons in frequency as well as in angle.
Under local thermodynamic equilibrium (LTE), Kirchhoff's law gives
$\eta(\nu, \rho, T) = \kappa_a(\nu, \rho, T)\, B_\nu(T)$, where $B_\nu(T)$ is the
Planck specific intensity per unit frequency.

Equation~\eqref{eq:transport} is the radiative transfer equation written in the local material
rest frame, with its coefficients held fixed during transport.  Emission and absorption are naturally described in that local frame, whereas packet positions are advanced in the
laboratory frame of the mesh.  For a photon of laboratory direction $\hat{\mathbf{n}}$ in a cell of
velocity $\mathbf{v}$ the two frames are related by the Doppler factor
\begin{equation}\label{eq:doppler}
    \Ddop = \gamma\left(1 - \frac{\hat{\mathbf{n}}\cdot\mathbf{v}}{c}\right),\qquad
    \gamma = \bigl(1-|\mathbf{v}|^2/c^2\bigr)^{-1/2},
\end{equation}
Here $\gamma$ is the Lorentz factor and $v=|\mathbf{v}|$ is the material speed; the comoving
and laboratory frequencies satisfy $\nu_{\mathrm{co}} = \Ddop\, \nu_{\mathrm{lab}}$.  Opacities are therefore
evaluated at the Doppler-shifted frequency, and every laboratory-frame interaction rate carries a
factor $\Ddop$.  The transformations are applied as full Lorentz boosts of packet frequency, energy
weight, and direction, and they enter the solver at the three points
where a packet meets the material: emission (Section~\ref{sec:emission}), continuous absorption
(Section~\ref{sec:energy_dep}), and scattering (Section~\ref{sec:event_loop}).  In a static medium
$\gamma = \Ddop = 1$ and every boost below is the identity.  Whatever the frame in which an
interaction is evaluated, the energy and momentum it hands to the material are booked in the
laboratory frame, and the material internal energy is derived from the pair by conservation
(Eq.~\eqref{eq:material_exchange} in Section~\ref{sec:rad_momentum}).

The radiation field couples to the material through
\begin{equation}\label{eq:material_energy}
    \frac{\partial U_m}{\partial t}
    = \int_0^\infty \kappa_a(\nu, \rho, T)
      \left[\oint I\, d\Omega - 4\pi B_\nu(T)\right] d\nu,
\end{equation}
where $U_m(\rho, T)$ is the material internal energy per unit volume and
$\cV = \partial U_m/\partial T$ is the volumetric heat capacity (energy per unit volume per unit
temperature).

Rather than reduce Eq.~\eqref{eq:material_energy} to a grey, Planck-mean equivalent
\citep{Pomraning1973,Mihalas1984}, the frequency range is divided into $G$ groups
$[\nu_g, \nu_{g+1}]$, $g = 0,\ldots,G-1$ (Section~\ref{sec:multigroup}), and the expressions that
follow are written group by group, with the grey formulation as the special case $G=1$.  Let $\kappa_{a,g}(\rho,T)$ and $\kappa_{s,g}(\rho,T)$ be the group absorption and
elastic scattering opacities, and define the group radiation energy density and the normalized
Planck weight of group~$g$,
\begin{equation}\label{eq:group_defs}
    \begin{aligned}
    E_g &= \frac{1}{c}\oint \int_{\nu_g}^{\nu_{g+1}} I\, d\nu\, d\Omega, \\
    b_g(T) &= \frac{4\pi}{c\, a_R T^4}\int_{\nu_g}^{\nu_{g+1}} B_\nu(T)\, d\nu,
    \end{aligned}
\end{equation}
with $a_R$ the radiation constant; the weights sum to unity when the group grid spans the Planck
spectrum (Section~\ref{sec:multigroup}).  Equation
\eqref{eq:material_energy} then reads
\begin{equation}\label{eq:mat_balance}
    \frac{\partial U_m}{\partial t}
    = c \sum_{g} \kappa_{a,g}\, \bigl[E_g - b_g(T)\, a_R T^4\bigr]
\end{equation}
Summed over groups, the emission term reduces to the frequency-integrated Planck emission
$c\, a_R T^4$ multiplied by the Planck-mean absorption opacity,
\begin{equation}\label{eq:kappa_planck}
    \begin{aligned}
    \kappa_P(\rho, T) &\equiv \sum_{g} \kappa_{a,g}\, b_g(T), \\
    c \sum_{g} \kappa_{a,g}\, b_g(T)\, a_R T^4 &= \kappa_P\, c\, a_R T^4
    \end{aligned}
\end{equation}
The absorption term $c\sum_g \kappa_{a,g} E_g$ admits no such reduction, since the in-cell
radiation field $E_g$ is in general not Planckian, and it retains its full group structure
throughout.  Hydrodynamic work and advection are applied in the operator-split hydro step
(Section~\ref{sec:radhydro}) and do not appear in Eq.~\eqref{eq:mat_balance}.

\subsection{Implicit Monte Carlo Discretization}\label{sec:imc}

The nonlinear coupling between Eqs.~\eqref{eq:transport} and~\eqref{eq:material_energy} makes explicit time discretization
impractical in optically thick media.  The \IMC{} method of \citet{FleckCummings1971} linearizes
the emission source over a single time step and converts part of the true absorption into
isotropic effective scattering.

\subsubsection{Derivation of the Fleck factor}

The opacities $\kappa_{a,g}(\rho,T)$, the Planck mean $\kappa_P$, and the spectral weights
$b_g(T)$ also depend on temperature, but they are frozen at their beginning-of-step values; the
nonlinearity that then remains in Eq.~\eqref{eq:mat_balance}, and makes it stiff, is the emission
term $a_R T^4$.  \citet{FleckCummings1971} remove it by expanding the equilibrium radiation energy
density
\begin{equation}\label{eq:ur_def}
    U_r \equiv a_R T^4
\end{equation}
in the material energy density $U_m$, to which the equation of state ties it.  With the derivative
denoted $\beta$ by \citet{FleckCummings1971},
\begin{equation}\label{eq:beta_def}
    \beta \equiv \frac{\partial U_r}{\partial U_m}
    = \frac{\partial \bigl(a_R T^4\bigr)/\partial T}{\partial U_m/\partial T}
    = \frac{4\, a_R\, T^3}{\cV}
\end{equation}
The denominator $\cV \equiv \partial U_m/\partial T$ is the volumetric heat capacity introduced after
Eq.~\eqref{eq:material_energy}: the energy per unit volume required to raise the material
temperature by one unit at fixed density.  With $\beta$ in hand,
the end-of-step emission is approximated to first order about the known beginning-of-step state
(superscripts $(n)$ and $(n+1)$ denote the beginning and end of time step $n$),
\begin{equation}\label{eq:t4_linearization}
    U_r^{(n+1)} \approx U_r^{(n)}
    + \beta^{(n)} \bigl(U_m^{(n+1)} - U_m^{(n)}\bigr)
\end{equation}
In this form the material physics enters only through the single derivative $\beta$, taken from
the equation of state, so that neither a constant heat capacity nor a temperature-linear internal
energy need be assumed.  For $U_m = \cV T$ with constant $\cV$, Eq.~\eqref{eq:t4_linearization}
reduces to the familiar expansion
$a_R T^4 \to a_R (T^{(n)})^4 + 4 a_R (T^{(n)})^3 (T-T^{(n)})$; the ionizing and strongly
temperature-dependent equations of state of Section~\ref{sec:results} lie outside that case.
\STORM{} evaluates $\beta$ from the equation-of-state derivative $\cV(\rho,T)$ supplied by
\RICH{} at the beginning-of-step state.

A second factor enters when the material moves.  Equation~\eqref{eq:mat_balance} is a statement in
the frame comoving with the cell: $U_m$ is the internal energy density of the material in its rest
frame, and the time derivative is taken with respect to the fluid's proper time $\tau$.  The Monte
Carlo sweep instead advances packets over a laboratory time step $\Delta t$ on the moving mesh,
during which a fluid element moving with Lorentz factor
$\gamma = \bigl(1-|\mathbf{v}|^2/c^2\bigr)^{-1/2}$ (Eq.~\ref{eq:doppler}) ages by only
$\Delta\tau = \Delta t/\gamma$.  With Newtonian hydrodynamics, in which cell volumes and masses are
frame independent, the internal energy of a cell therefore changes per unit laboratory time at
\begin{equation}\label{eq:proper_time}
    \frac{dU_m}{dt} = \frac{1}{\gamma}\,\frac{dU_m}{d\tau}
\end{equation}
Total derivatives are used here because the relation $dt = \gamma\,d\tau$ holds only along the
worldline of a fluid element.  Within the radiation substep the mesh is frozen and advection is
deferred to the hydrodynamic step (Section~\ref{sec:radhydro}), so this derivative following the cell
coincides with the partial derivative $\partial U_m/\partial t$ of Eq.~\eqref{eq:mat_balance}.
The same factor follows from the four-momentum bookkeeping.  The laboratory total energy exchanged
with a source that is isotropic in the comoving frame is $\gamma$ times its comoving value, but the
accompanying momentum does work $\mathbf{v}\cdot\Delta\mathbf{p}$ on the fluid, and the internal
energy receives only the comoving value divided by $\gamma$; this is exactly the
internal-energy change produced by the emission and absorption updates of
Sections~\ref{sec:emission} and~\ref{sec:energy_dep}.  Following \citet{Roth2022}, who derive the
relativistic Fleck factor in terms of the proper-time step, the factor $1/\gamma$ is retained
throughout the following; it equals unity in the static limit $|\mathbf{v}| \to 0$.

The discrete material update is assembled from the preceding results in four steps.  First,
Eq.~\eqref{eq:mat_balance} is discretized implicitly over the step
$\Delta t = t^{(n+1)}-t^{(n)}$, with the group radiation energy densities $E_g$ and the
temperature in the emission term evaluated at the end of the step (the former as the Monte Carlo
transport sweep will supply them).  Second, its right-hand side is multiplied by $1/\gamma$
according to Eq.~\eqref{eq:proper_time}, because the step $\Delta t$ is measured in laboratory
time.  Third, with the emission shape $b_g$ held at the beginning-of-step temperature, the emission
sum collapses onto the Planck mean by Eq.~\eqref{eq:kappa_planck},
$\sum_g \kappa_{a,g}\, b_g\, a_R T^4 = \kappa_P\, a_R T^4$.  Fourth, $a_R T^4$ is written as $U_r$
by Eq.~\eqref{eq:ur_def}.  Together these give
\begin{equation}\label{eq:imc_discrete}
    U_m^{(n+1)} - U_m^{(n)}
    = \frac{c\, \Delta t}{\gamma} \Bigl[\sum_g \kappa_{a,g}\, E_g^{(n+1)}
      - \kappa_P\, U_r^{(n+1)}\Bigr]
\end{equation}
The emission term involves only the Planck mean $\kappa_P$, whereas the absorption term keeps its
full group structure.  The stiff unknown is the end-of-step emission $U_r^{(n+1)}$, and it is for $U_r$, not
$U_m$, that the step is solved.  Multiplying Eq.~\eqref{eq:imc_discrete} by $\beta$ and replacing
$\beta\bigl(U_m^{(n+1)}-U_m^{(n)}\bigr)$ by $U_r^{(n+1)}-U_r^{(n)}$ according to
Eq.~\eqref{eq:t4_linearization} yields an equation that is linear in $U_r^{(n+1)}$ alone,
\begin{equation}\label{eq:ur_discrete}
    U_r^{(n+1)} - U_r^{(n)}
    = \frac{\beta\, c\, \Delta t}{\gamma} \Bigl[\sum_g \kappa_{a,g}\, E_g^{(n+1)}
      - \kappa_P\, U_r^{(n+1)}\Bigr]
\end{equation}
Equation~\eqref{eq:ur_discrete} is solved for $U_r^{(n+1)}$ in three elementary steps.  First,
the term $-(\beta c\Delta t/\gamma)\,\kappa_P U_r^{(n+1)}$ on the right-hand side, which carries
the unknown, is moved to the left-hand side, and the known $U_r^{(n)}$ is moved to the right:
\begin{equation*}
    \Bigl(1 + \frac{\beta\,\kappa_P\, c\,\Delta t}{\gamma}\Bigr) U_r^{(n+1)}
    = U_r^{(n)} + \frac{\beta\, c\,\Delta t}{\gamma} \sum_g \kappa_{a,g}\, E_g^{(n+1)}
\end{equation*}
Second, the dimensionless combination in the parentheses is named
\[
    \zeta \equiv \frac{\beta\, \kappa_P\, c\, \Delta t}{\gamma}
      = \frac{4\, a_R\, \bigl(T^{(n)}\bigr)^3\, \kappa_P\, c\, \Delta t}{\gamma\,\cV}
\]
It is the ratio of the time step to the time
$\gamma/(\beta\kappa_P c)$ in which the emission responds to a change in the absorbed energy, and
it therefore measures the stiffness of the matter--radiation coupling over the step.  Writing the
coefficient of the absorption sum as $\beta c\Delta t/\gamma = \zeta/\kappa_P$, the equation reads
\begin{equation*}
    (1+\zeta)\, U_r^{(n+1)}
    = U_r^{(n)} + \zeta\, \frac{1}{\kappa_P} \sum_g \kappa_{a,g}\, E_g^{(n+1)}
\end{equation*}
Third, the \emph{Fleck factor} is introduced,
\[
    f \equiv \frac{1}{1+\zeta},
\]
for which $\zeta/(1+\zeta) = 1-f$.  Dividing by $1+\zeta$ then gives the end-of-step emission in
closed form,
\begin{equation}\label{eq:ur_update}
    U_r^{(n+1)} = f\, U_r^{(n)}
      + (1-f)\, \frac{1}{\kappa_P} \sum_g \kappa_{a,g}\, E_g^{(n+1)}
\end{equation}
Both coefficients lie between 0 and 1 and sum to unity, so the implicit emission is a weighted mean
of two quantities.  The first, $U_r^{(n)} = a_R (T^{(n)})^4$, is the known beginning-of-step
emission.  The second, $\sum_g \kappa_{a,g} E_g^{(n+1)}/\kappa_P$, is the value of $U_r$ for which
the bracket of Eq.~\eqref{eq:imc_discrete} vanishes, i.e.\ the emission that exactly balances the
radiation absorbed during the step.  In an optically thin cell or for a short step
($\zeta \ll 1$, $f \to 1$) the emission stays at its old value, which is the explicit limit; in a
stiff cell ($\zeta \gg 1$, $f \to 0$) it relaxes to radiative equilibrium within the step.

Equation~\eqref{eq:ur_update} now determines the emission source of the transport equation.
Integrating the emissivity $\eta = \kappa_a B_\nu(T)$ of Eq.~\eqref{eq:transport} over group~$g$,
with the opacity taken as the group value $\kappa_{a,g}$, gives $\kappa_{a,g} B_g(T)$, where
\begin{equation}\label{eq:Bg_def}
    B_g(T) \equiv \int_{\nu_g}^{\nu_{g+1}} B_\nu(T)\, d\nu
    = \frac{c}{4\pi}\, b_g(T)\, a_R T^4 = \frac{c}{4\pi}\, b_g(T)\, U_r
\end{equation}
is the group-integrated Planck intensity, the second equality following from the definition of
$b_g$ in Eq.~\eqref{eq:group_defs}.  In the implicit step the emissivity is evaluated at the
end-of-step temperature.  The factor $b_g$ is frozen at $T^{(n)}$, like the opacities, so all of the
temperature dependence sits in $U_r^{(n+1)}$:
\begin{equation*}
    \kappa_{a,g}\, B_g\bigl(T^{(n+1)}\bigr)
    \approx \frac{c}{4\pi}\, \kappa_{a,g}\, b_g\bigl(T^{(n)}\bigr)\, U_r^{(n+1)}
\end{equation*}
Substituting Eq.~\eqref{eq:ur_update} for $U_r^{(n+1)}$ splits this source into two terms.  In the
first, $(c/4\pi)\,\kappa_{a,g}\, b_g\, f\, U_r^{(n)}$, Eq.~\eqref{eq:Bg_def} applied at $T^{(n)}$
gives $f\,\kappa_{a,g} B_g(T^{(n)})$.  In the second, the group energy densities are written in
terms of the group intensities $I_{g'} = \int_{\nu_{g'}}^{\nu_{g'+1}} I\, d\nu$ through
Eq.~\eqref{eq:group_defs}, $c\,E_{g'} = \oint I_{g'}(\hat{\mathbf{n}}')\, d\Omega'$, which cancels
the factor $c$.  The end-of-step group emissivity is therefore
\begin{equation}\label{eq:emission_split}
    \begin{aligned}
    \kappa_{a,g}\, B_g\bigl(T^{(n+1)}\bigr)
    \approx{}& f\, \kappa_{a,g}\, B_g\bigl(T^{(n)}\bigr) \\
    &+ \frac{\kappa_{a,g}\, b_g}{\kappa_P}\, \frac{1-f}{4\pi} \sum_{g'} \kappa_{a,g'}
      \oint I_{g'}(\hat{\mathbf{n}}')\, d\Omega'
    \end{aligned}
\end{equation}
The two terms have different characters.  The first is an ordinary thermal source at the known
temperature $T^{(n)}$, reduced by the factor $f$; it is independent of the radiation field and is
sampled as emitted packets at the start of the step.  The second contains no temperature at all:
it is proportional to the intensity itself.  Read from right to left, it removes radiation from
every group~$g'$ at the rate $(1-f)\,\kappa_{a,g'}$ (the integral $\oint I_{g'}\, d\Omega'$ collects
it from all directions), and returns that energy at once, isotropically (the factor $1/4\pi$), into
group~$g$ with the fraction $\kappa_{a,g} b_g/\kappa_P$.  These fractions sum to unity over $g$ by
Eq.~\eqref{eq:kappa_planck}, so no energy is created or lost.  An absorption followed by immediate
isotropic re-emission is a scattering event, so the second term acts as an \emph{effective
scattering} which, unlike the elastic scattering of Eq.~\eqref{eq:transport}, also redistributes
photons across groups.  Both this re-emission and the thermal emission of the first term follow the
opacity-weighted Planck spectrum, whose normalized group weights are
\begin{equation}\label{eq:emission_weights}
    p_g(T) \equiv \frac{\kappa_{a,g}\, b_g(T)}{\kappa_P(T)},
    \qquad \sum_g p_g(T) = 1,
\end{equation}
the normalization following from Eq.~\eqref{eq:kappa_planck}; the re-emission fraction in the
second term of Eq.~\eqref{eq:emission_split} is $p_g(T^{(n)})$.  Integrating
Eq.~\eqref{eq:transport} over group~$g$ and replacing its emissivity by the
split~\eqref{eq:emission_split} gives the transport equation solved during the Monte Carlo step,
for the group intensity $I_g = \int_{\nu_g}^{\nu_{g+1}} I\, d\nu$,
\begin{multline}\label{eq:imc_transport_split}
    \frac{1}{c}\frac{\partial I_g}{\partial t} + \hat{\mathbf{n}} \cdot \nabla I_g
    = f\, \kappa_{a,g}\, B_g\bigl(T^{(n)}\bigr)
      - \kappa_{a,g}\, I_g \\
      - \kappa_{s,g}\, I_g
      + \frac{\kappa_{s,g}}{4\pi} \oint I_g(\hat{\mathbf{n}}')\, d\Omega' \\
      + \frac{p_g(T^{(n)})}{4\pi}\, (1-f) \sum_{g'} \kappa_{a,g'}
        \oint I_{g'}(\hat{\mathbf{n}}')\, d\Omega'
\end{multline}
By Eq.~\eqref{eq:Bg_def}, the thermal source may equivalently be written
$f\,\kappa_P\, p_g\, a_R (T^{(n)})^4 c/(4\pi)$, and it sums over groups to
$f\,\kappa_P\, a_R (T^{(n)})^4 c/(4\pi)$, the Fleck-reduced emission rate implied by
Eq.~\eqref{eq:kappa_planck}.

Finally, substituting Eq.~\eqref{eq:ur_update} back into Eq.~\eqref{eq:imc_discrete}, the bracket
becomes
$f\bigl[\sum_g \kappa_{a,g} E_g^{(n+1)} - \kappa_P U_r^{(n)}\bigr]$, so that the material update is
\begin{equation}\label{eq:fleck}
    \begin{aligned}
    U_m^{(n+1)} - U_m^{(n)}
      &= f\, \frac{c\, \Delta t}{\gamma}
        \Bigl[\sum_g \kappa_{a,g}\, E_g^{(n+1)} \\
      &\qquad - \kappa_P\, a_R \bigl(T^{(n)}\bigr)^4\Bigr]
    \end{aligned}
\end{equation}
The Fleck factor therefore also sets the share of the net absorbed energy that the material
retains.  For $\zeta = 0$, Eq.~\eqref{eq:ur_update} keeps the
emission at its beginning-of-step value and the full absorbed energy heats the material; for
$\zeta > 0$ the fraction $1-f$ of the absorbed energy reappears in $U_r^{(n+1)}$ as the enhanced
emission that accompanies the temperature rise, already accounted for within the same step by the
linearization, and only the fraction $f$ remains in the material.

Equation~\eqref{eq:emission_split} supplies $f$ to the Monte Carlo transport
equation~\eqref{eq:imc_transport_split}.  A packet that is truly absorbed corresponds to the
fraction $f$ of $\kappa_{a,g}$ that changes the material energy; the complementary fraction $1-f$
represents absorption events whose energy is, in the linearized picture, re-emitted within the same
step, leaving the radiation field statistically unchanged apart from a new isotropic direction and
a frequency redrawn from the opacity-weighted Planck spectrum.  Such events act as scatterings that
also redistribute photons across groups, and $1-f$ is accordingly added to $\kappa_{s,g}$ to form
the effective scattering opacity of Eq.~\eqref{eq:eff_scat}.  Representing absorption and
re-emission this way allows the stiff coupling to be advanced without sub-cycling or Newton
iteration on the material state.

The Fleck factor is a single number per cell in the multigroup case as well, the linearization
being applied to the frequency-integrated balance~\eqref{eq:mat_balance}; the group structure
resides in the opacities that $f$ multiplies.  The time-centering parameter of
\citet{FleckCummings1971}, often also denoted $\alpha$, is taken to be unity (a fully implicit
linearization).  Physically, $f$ is the probability that a photon absorbed during the step is
truly absorbed, and $1-f$ is treated as effective isotropic scattering with a Planckian frequency
redistribution.

For a static medium $\gamma = 1$ and $f$ reduces to the classical
Fleck factor.  As \citet{Roth2022} discuss, an earlier flat-spacetime formulation used
$\gamma\Delta t$ in place of $\Delta t/\gamma$; the two differ at $O(v^2/c^2)$ and both give a
stable update.  In the
moving-media problems of
Sections~\ref{sec:mach2}, \ref{sec:mach45} and \ref{sec:moving_slab}, $\gamma$ is evaluated from
the cell velocity at every step (the RMTV problem of Section~\ref{sec:rmtv} runs its transport
without the moving-material corrections).

\subsubsection{The IMC transport equation}

The absorption term $-\kappa_{a,g} I_g$ of Eq.~\eqref{eq:imc_transport_split} can be divided in the
same proportions as the emission, $\kappa_{a,g} = f\,\kappa_{a,g} + (1-f)\,\kappa_{a,g}$.  The first
part is true absorption, which changes the material energy; the second removes exactly the
radiation that the effective-scattering term re-emits within the step.  Grouping the second part
with elastic scattering defines the effective opacities group by group,
\begin{align}
    \kappa_{a,g}^{\mathrm{eff}} &= f\, \kappa_{a,g}, \label{eq:eff_abs}\\
    \kappa_{s,g}^{\mathrm{eff}} &= (1 - f)\, \kappa_{a,g} + \kappa_{s,g}, \label{eq:eff_scat}
\end{align}
with the same cell Fleck factor $f$ in every group.  In terms of these, the transport
equation~\eqref{eq:imc_transport_split} reads
\begin{multline}\label{eq:imc_transport}
    \frac{1}{c}\frac{\partial I_g}{\partial t} + \hat{\mathbf{n}} \cdot \nabla I_g
    = f\, \kappa_{a,g}\, B_g\bigl(T^{(n)}\bigr)
      - \kappa_{a,g}^{\mathrm{eff}}\, I_g \\
      - \kappa_{s,g}^{\mathrm{eff}}\, I_g
      + \frac{\kappa_{s,g}}{4\pi} \oint I_g(\hat{\mathbf{n}}')\, d\Omega' \\
      + \frac{p_g(T^{(n)})}{4\pi}\, (1-f) \sum_{g'} \kappa_{a,g'}
        \oint I_{g'}(\hat{\mathbf{n}}')\, d\Omega',
\end{multline}
a standard transport equation with absorption opacity $\kappa_{a,g}^{\mathrm{eff}}$, the
Fleck-reduced thermal source, and two scattering processes.

The last two terms of Eq.~\eqref{eq:imc_transport} are the two halves of
$\kappa_{s,g}^{\mathrm{eff}}$: elastic scattering, which is in-group, and effective scattering,
which removes energy from every group $g'$ at the rate $(1-f)\kappa_{a,g'}$ and re-injects it into
group~$g$ with the weight $p_g(T^{(n)})$.  Effective scattering is thus a frequency-redistribution
operator rather than an in-group scattering operator, and the two terms merge into the grey form
$(\kappa_s^{\mathrm{eff}}/4\pi)\oint I\, d\Omega'$ only for $G=1$.  Since $\sum_g p_g = 1$, the
redistribution conserves energy exactly.

Packets that undergo effective scattering are re-emitted isotropically with their frequency
redrawn from $p_g(T^{(n)})$, the Monte Carlo realization of the last term of
Eq.~\eqref{eq:imc_transport}.  This formulation is unconditionally stable in the
stiff-coupling sense
\citep{FleckCummings1971,Wollaber2016}.  It does not automatically preserve the maximum principle.
Energy is conserved by construction within the radiation substep: every exchange with the material
is booked as a laboratory-frame energy and momentum, and the material internal energy is derived
from the two (Eq.~\eqref{eq:material_exchange}), so the sum of packet weights and material total
energy is preserved to roundoff for standard \IMC{} packets.  The operator split of
Section~\ref{sec:radhydro} remains first-order in time; we return to these caveats in
Section~\ref{sec:conclusion}.

%%%%%%%%%%%%%%%%%%%%%%%%%%%%%%%%%%%%%%%%%%%%%%%%%%
\section{Monte Carlo Implementation}\label{sec:implementation}

\subsection{Software Architecture}\label{sec:architecture}

The implementation is split across three layers, and only the middle layer is specific to Voronoi
hydrodynamics.

\begin{enumerate}
    \item \STORM{} owns the Monte Carlo engine: packet storage, the
        per-packet event loop, boundary-condition callbacks, population control, and---in
        parallel---rank-to-rank migration.  The physics module is a plugin.  The physics must decide, after
        each event, whether a packet remains in its cell, crosses a face, reaches census, or is
        removed.  It does not own communication.  The mesh interface is deliberately thin.  A grid
        can host the \IMC{} kernel if it satisfies the following requirements:
        \begin{itemize}
            \item its cells are convex polyhedra;
            \item it provides the volume of each cell;
            \item it provides the faces of each cell, with their areas, centroids, normals,
                vertices, and neighbor indices;
            \item it answers point-location queries: whether a point lies in a given cell, and
                which cell contains it;
            \item it provides the bounding box of the domain;
            \item under MPI, it provides the ghost-cell and neighboring-rank metadata.
        \end{itemize}
        \STORM{} orients each face normal inward itself from the face
        centroid and generating point, so the grid need not supply a consistent orientation.  Standalone \STORM{} calculations in
        this paper therefore run equally on Cartesian meshes and on Voronoi tessellations.
    \item \MadVoro{} constructs the three-dimensional Voronoi diagram
        from a set of mesh-generating points, in serial or in distributed memory.  It is the
        tessellation engine inside \RICH{}, and the same library is available to \STORM{} as a
        grid backend.
    \item \RICH{} owns the hydrodynamics: a second-order Godunov update of
        the Euler equations on the moving Voronoi mesh, mesh motion, and the operator-split
        coupling to radiation.  After the hydrodynamic step and the subsequent \MadVoro{}
        reconstruction, \RICH{} invokes \STORM{} to transport photon packets on the current
        tessellation and deposits the resulting energy and momentum into the conserved
        hydrodynamic variables.  Grey and multigroup FLD remain available as alternative radiation
        options in the same code \citep{SteinbergStone2024,Giron2026}.
\end{enumerate}

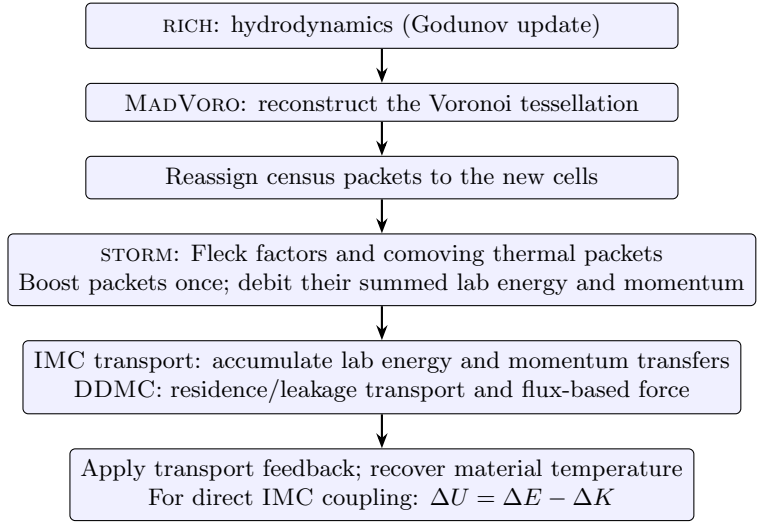
\begin{figure*}
\centering
\begin{tikzpicture}[
  font=\small,
  node distance=0.45cm,
  box/.style={draw,rounded corners=2pt,align=center,inner sep=4pt,fill=blue!6,minimum width=0.92\columnwidth},
  arr/.style={-{Stealth[length=2mm]},thick}
]
\node[box] (hydro) {\RICH{}: hydrodynamics (Godunov update)};
\node[box,below=of hydro] (mesh) {\MadVoro{}: reconstruct the Voronoi tessellation};
\node[box,below=of mesh] (reassign) {Reassign census packets to the new cells};
\node[box,below=of reassign] (source) {\STORM{}: Fleck factors and comoving thermal packets\\
Boost packets once; debit their summed lab energy and momentum};
\node[box,below=of source] (storm) {\IMC{} transport: accumulate lab energy and momentum transfers\\
\DDMC{}: residence/leakage transport and flux-based force};
\node[box,below=of storm] (update) {Apply transport feedback; recover material temperature\\
For direct \IMC{} coupling: $\Delta U=\Delta E-\Delta K$};
\draw[arr] (hydro) -- (mesh);
\draw[arr] (mesh) -- (reassign);
\draw[arr] (reassign) -- (source);
\draw[arr] (source) -- (storm);
\draw[arr] (storm) -- (update);
\end{tikzpicture}
\caption{Operator-split radiation--hydrodynamics cycle.  \RICH{} advances the fluid and moves the
mesh-generating points; \MadVoro{} rebuilds the tessellation; \STORM{} then transports photon
packets on that frozen mesh.  Thermal emission is debited using the actual emitted packets;
the subsequent transport tally supplies the remaining feedback.  Here $U$ and $K$ denote cell
internal and kinetic energies, and the direct \IMC{} update uses Eq.~\eqref{eq:material_exchange}.
The \DDMC{} force uses the diffusion closure of Section~\ref{sec:ddmc_hydro}.
The same \STORM{} physics kernel can be driven without hydrodynamics on any supported grid.}
\label{fig:coupling}
\end{figure*}

Figure~\ref{fig:coupling} summarizes one coupled cycle.  During the Monte Carlo substep the mesh
is held fixed; mesh motion enters the radiation solve only between substeps, through census-packet
reassignment (Section~\ref{sec:radhydro}).

\subsection{The Radiation Substep}\label{sec:substep}

Algorithm~\ref{alg:imc_step} combines the Fleck-factor linearization of Section~\ref{sec:imc}
with the emission, energy-deposition, and transport steps detailed in the remainder of this section
into the single \IMC{} radiation substep that \STORM{} executes once per cycle.  The coupling to
the fluid is shown alongside them: the census reassignment that follows mesh motion, the Lorentz
factor in the Fleck factor and in the emission bookkeeping, the Doppler factor and the frame
boosts during transport, and the momentum returned to the fluid.  Those pieces are derived in
Sections~\ref{sec:transport_eq}, \ref{sec:imc}, \ref{sec:emission}--\ref{sec:event_loop},
and~\ref{sec:radhydro}; the algorithm is given first so that the substep can be read as a whole
before its parts are described.  In the algorithm, $i$ labels cells,
$V_i$ and $m_i=\rho_i V_i$ are their volumes and masses, $N_i$ is the number of thermal packets
emitted by cell $i$, and $w$ is a packet's laboratory-frame energy weight.  The symbols
$\Delta E_i$ and $\Delta\mathbf{p}_i$ denote energy and momentum transferred to the cell, and the event times
$t_{\mathrm{face}}$, $t_{\mathrm{scat}}$, and $t_{\mathrm{left}}$ are the times to a face crossing,
a scatter, and census, respectively; $\mathrm{Unif}$ denotes a uniform distribution.  Subscripts
``before/after'' and ``old/new'' refer to the two sides of a scattering event.  Cell velocities used for packet boosts are held at their substep values; the velocity in the
kinetic-energy correction is computed from the current material momentum at each update.

\begin{algorithm*}[t]
\DontPrintSemicolon
\SetKwInOut{Input}{Input}
\SetKwInOut{Output}{Output}
\Input{material state
    $\{T_i^{(n)}, \rho_i, \mathbf{v}_i, \kappa_{a,g,i}, \kappa_{s,g,i}, \kappa_{P,i}, C_{V,i}\}$ on
    the tessellation rebuilt by the preceding hydrodynamic step, time step $\Delta t$, surviving
    census packets from the previous step}
\Output{updated material state $\{T_i^{(n+1)}\}$, fluid energy and momentum increments, census
    packets for the next step}
\BlankLine
\tcc{Census reassignment after mesh motion (Section~\ref{sec:radhydro})}
\ForEach{surviving census packet}{
    keep the packet in its previous cell if it still lies inside it; otherwise relocate it on the
        new tessellation (a containing-cell query in serial; under MPI a persistent cell-ID remap,
        a walk across faces, direct exchange to the owning rank, and an octree search for the
        remainder)\;
}
\BlankLine
\tcc{Per-cell setup (Sections~\ref{sec:imc}, \ref{sec:emission})}
\ForEach{cell $i$}{
    $\gamma_i \leftarrow (1-|\mathbf{v}_i|^2/c^2)^{-1/2}$\;
    $f_i \leftarrow$ Fleck factor $1/(1+\zeta_i)$ of Section~\ref{sec:imc}, evaluated with $\gamma_i$\;
    $\kappa_{a,g,i}^{\mathrm{eff}} \leftarrow f_i\,\kappa_{a,g,i}$,\quad
        $\kappa_{s,g,i}^{\mathrm{eff}} \leftarrow (1-f_i)\,\kappa_{a,g,i} + \kappa_{s,g,i}$
        \quad$\forall g$ \tcp*{Eqs.~\eqref{eq:eff_abs}--\eqref{eq:eff_scat}}
    $E_{\mathrm{emit},i} \leftarrow f_i\, V_i\, a_R\, (T_i^{(n)})^4\, \kappa_{P,i}\, c\, \Delta t$
        \tcp*{Eq.~\eqref{eq:emission_energy}}
    $\Delta E_i \leftarrow 0$,\quad $\Delta\mathbf{p}_i \leftarrow \mathbf{0}$
        \tcp*{laboratory-frame exchange accumulators for transport}
    source $N_i \propto E_{\mathrm{emit},i}$ thermal packets: random position, isotropic
        \emph{comoving} direction, emission time $\sim\mathrm{Unif}[0,\Delta t]$, comoving group
        $g$ with probability $p_g(T_i^{(n)})$\;
    boost each new packet to the laboratory frame,
        $\nu \leftarrow \nu_{\mathrm{co}}/\Ddop$ and
        $w \leftarrow E_{\mathrm{emit},i}/(N_i\, \Ddop)$\;
    sum the actual new packets: $E_{\mathrm{src},i}\leftarrow\sum_{k=1}^{N_i}w_k$,
        $\mathbf{P}_{\mathrm{src},i}\leftarrow\sum_{k=1}^{N_i}w_k\hat{\mathbf{n}}_k/c$\;
    debit the cell once by Eq.~\eqref{eq:material_exchange} with
        $(\Delta E,\Delta\mathbf{p})=(-E_{\mathrm{src},i},-\mathbf{P}_{\mathrm{src},i})$\;
}
source boundary packets (Section~\ref{sec:boundary})\;
\BlankLine
\tcc{Transport (Sections~\ref{sec:energy_dep}, \ref{sec:event_loop})}
\ForEach{packet (new, boundary, or census), in parallel across ranks}{
    \While{packet is alive and step time remains}{
        $\Ddop \leftarrow \gamma_i\,(1-\hat{\mathbf{n}}\cdot\mathbf{v}_i/c)$; evaluate
            $\kappa_{a,g}^{\mathrm{eff}}, \kappa_{s,g}^{\mathrm{eff}}$ in the group of
            $\nu_{\mathrm{co}} = \Ddop\,\nu$ \tcp*{Eq.~\eqref{eq:doppler}}
        $t_{\mathrm{face}}, t_{\mathrm{scat}}, t_{\mathrm{left}} \leftarrow$ competing event times,
            $t_{\mathrm{scat}}$ at the rate $\kappa_{s,g}^{\mathrm{eff}}\, c\, \Ddop$\;
        $\delta t \leftarrow \min(t_{\mathrm{face}}, t_{\mathrm{scat}}, t_{\mathrm{left}})$\;
        attenuate $w$ over $\delta t$ and tally the laboratory energy it loses,
            $\Delta E_i \mathrel{+}= \Delta E$
            (Eqs.~\eqref{eq:energy_dep}--\eqref{eq:weight_update})\;
        $\Delta\mathbf{p}_i \mathrel{+}= \Delta E\, \hat{\mathbf{n}}/c$
            \tcp*{Eq.~\eqref{eq:mom_propagation}}
        \If{$w$ fell below the cutoff}{
            $\Delta E_i \mathrel{+}= w$\;
            $\Delta\mathbf{p}_i \mathrel{+}= w\,\hat{\mathbf{n}}/c$; terminate the packet and stop\;
        }
        advance position by $\hat{\mathbf{n}}\, c\, \delta t$\;
        \uIf{cell boundary reached}{move to neighbor cell, migrating across ranks if needed, or
            apply the domain boundary condition\;}
        \uElseIf{elastic or effective scatter}{
            boost to the comoving frame and sample a new isotropic direction, preserving
                the comoving packet energy; preserve $\nu_{\mathrm{co}}$ for an elastic scatter or redraw the group from
                $p_g(T_i^{(n)})$ for an effective scatter; boost back\;
            tally $\Delta E_i \mathrel{+}= w_{\mathrm{before}} - w_{\mathrm{after}}$ and
                $\Delta\mathbf{p}_i \mathrel{+}= (w_{\mathrm{before}}\hat{\mathbf{n}}_{\mathrm{old}}
                - w_{\mathrm{after}}\hat{\mathbf{n}}_{\mathrm{new}})/c$\;
        }
        \Else{store the packet as census for the next step and stop\;}
    }
}
\BlankLine
\tcc{Population control, then material and fluid update}
apply comb population control to the census packets (Section~\ref{sec:pop_control})\;
\ForEach{cell $i$}{
    apply the accumulated $(\Delta E_i, \Delta\mathbf{p}_i)$ by
        Eq.~\eqref{eq:material_exchange}:\quad
        momentum $\mathrel{+}= \Delta\mathbf{p}_i$,\quad
        internal energy $\mathrel{+}= \Delta E_i - \mathbf{v}_i\cdot\Delta\mathbf{p}_i
        - |\Delta\mathbf{p}_i|^2/2m_i$\;
    invert the equation of state for $T_i^{(n+1)}$;\quad
        total energy $\leftarrow$ internal $+$ kinetic\;
}
\caption{One \IMC{} radiation substep with its hydrodynamic coupling.  The mesh is held fixed for
the duration of the substep; the Euler equations are advanced and the tessellation rebuilt by
\RICH{} beforehand (Section~\ref{sec:radhydro}).  In a prescribed static-material calculation
with momentum feedback disabled, $\gamma_i = \Ddop = 1$ and the internal-energy update is
$\Delta E_i$.  Zero initial velocity alone does not disable recoil.  The accelerated
\DDMC{} branch and its separate force estimator are described in Section~\ref{sec:ddmc}.}\label{alg:imc_step}
\end{algorithm*}

\subsection{Frequency Groups}\label{sec:multigroup}

The transport equations of Section~\ref{sec:physical_model} are written group by group; this
subsection describes how that group structure is realized.  The photon-energy range is divided
into $G$ groups $[\varepsilon_g, \varepsilon_{g+1}]$ for $g = 0, \ldots, G-1$, typically spaced
logarithmically, where $\varepsilon_g=h\nu_g$ is a group boundary energy and $h$ is Planck's
constant.  The symbol $E_g$ continues to denote group radiation energy density.  Each
photon packet carries a scalar frequency $\nu$ (equivalently, energy $E = h\nu$) that determines
which group it belongs to.  The grey solver is the single-group limit of the same scheme, run with
a cached cell Planck opacity instead of a per-packet opacity lookup.

The absorption opacity seen by a packet is obtained from the problem's opacity model at the
packet's own comoving photon energy, $\kappa_a(\varepsilon_{\mathrm{co}};\rho,T)$.  For a tabulated
multigroup model, such as the $124$-group aluminum table of \citet{McClarrenGentile2021} used in
the moving-slab test (Section~\ref{sec:moving_slab}), this is
the constant group value $\kappa_{a,g}$; an analytic model may instead vary continuously within a
group.  The same interface supports \emph{continuous-frequency} transport, used for the Olson
problem of Section~\ref{sec:olson}: the photon-energy range then forms a single bookkeeping group, every
absorption is evaluated from the exact $\kappa_a(\varepsilon, T)$ at the packet energy, the
Planck-mean opacity of the Fleck factor and emission rate is the integral
$\kappa_P = \int \kappa_a B_\nu\, d\nu / \int B_\nu\, d\nu$, and thermal emission and
effective-scattering redistribution sample the continuous opacity-weighted Planck spectrum
$\kappa_a(\varepsilon,T) B_\varepsilon(T)$ from a tabulated cumulative distribution.  The
multigroup formulas of this section are recovered when $\kappa_a$ is constant within each group.
The normalized Planck weights
$b_g(T)$ of Eq.~\eqref{eq:group_defs} are the incomplete Planck integrals
\begin{equation}\label{eq:bg_tabulated}
    b_g(T) = \frac{15}{\pi^4}\int_{\varepsilon_g/(k_B T)}^{\varepsilon_{g+1}/(k_B T)} \frac{x^3}{e^x - 1}\, dx,
\end{equation}
where $k_B$ is Boltzmann's constant and $x=E/(k_B T)$ is the dimensionless photon energy.  They
are evaluated analytically with the method of \citet{Clark1987}, a Taylor expansion for $x < 2$
and a rapidly converging five-term exponential series above, rather than from a table, and they
sum to unity, $\sum_{g=0}^{G-1} b_g(T) = 1$, when the group grid spans the Planck spectrum at
temperature~$T$.
If the group grid does not span the Planck spectrum, $\kappa_P$ is normalized by $\sum_g b_g(T)$,
consistently with the normalized sampling table below.  A second cumulative table, the running sum of
$\kappa_{a,g} b_g(T)$, is cached per cell and inverted to sample the emission weights $p_g$ of
Eq.~\eqref{eq:emission_weights}; it serves both thermal emission and effective-scattering
redistribution.  The Planck mean is the only frequency-averaged absorption opacity appearing in the \IMC{} scheme;
the \DDMC{} diffusion coefficient of Section~\ref{sec:ddmc} uses a Rosseland-type mean instead.

When a packet undergoes effective scattering (absorption--re-emission), its new group is drawn
from $p_g(T)$ at the cell temperature and generally differs from the group it arrived in, which
realizes the redistribution term of Eq.~\eqref{eq:imc_transport}.  Elastic scattering events, by
contrast, preserve the packet's comoving frequency and are strictly in-group up to the Doppler
shift of Eq.~\eqref{eq:doppler}.  A differential angular kernel may be supplied in place of the
isotropic default; it sets only the scattered direction, and the comoving frequency is preserved
in either case.

Optionally, the solver tracks per-group time-averaged radiation energy densities
$\langle E_g \rangle$ for each cell, which can be used for diagnostic purposes or for coupling
to multigroup diffusion solvers (Section~\ref{sec:estimators}).

\subsection{Thermal Emission}\label{sec:emission}

At the beginning of each \IMC{} step, every cell emits new photon packets representing the thermal
source.  Under LTE the emission rate per unit volume into group $g$ is
$\kappa_{a,g}\, b_g(T)\, c\, a_R T^4$, and summing over groups gives the frequency-integrated rate
$\kappa_P c\, a_R T^4$ of Eq.~\eqref{eq:kappa_planck}.  The
\IMC{} source term of Eq.~\eqref{eq:imc_transport} carries only the fraction $f$ of this
emission, the remaining $1-f$ being represented by effective scattering, so the Fleck-reduced
energy emitted by cell~$i$ during a time step $\Delta t$ is
\begin{equation}\label{eq:emission_energy}
    \begin{aligned}
    E_{\mathrm{emit},i} &= f_i\, V_i\, a_R\, T_i^4\, \kappa_{P,i}\, c\, \Delta t, \\
    E_{\mathrm{emit},i,g} &= p_g(T_i)\, E_{\mathrm{emit},i},
    \end{aligned}
\end{equation}
where $f_i$ and $T_i$ are the cell Fleck factor and temperature, $V_i$ is the cell volume,
$E_{\mathrm{emit},i,g}$ is the emitted energy in group $g$, and $\kappa_{P,i}$ is the cell Planck mean; all material quantities are
evaluated at the beginning of the step.  This energy, boosted to the laboratory frame, is removed
from the cell together with its momentum through Eq.~\eqref{eq:material_exchange} so that emission
and later absorption remain consistent.

The number of thermal packets is controlled by a single parameter $N_{\mathrm{new}}$, quoted in
Section~\ref{sec:results} as the number of new packets ``per cell per step''.  It is a per-cell
floor rather than a uniform count: a global budget of
$N_{\mathrm{total}} = 10\, N_{\mathrm{new}}\, N_{\mathrm{cells}}$ packets, with $N_{\mathrm{cells}}$
the global cell count, is distributed among cells in proportion to $E_{\mathrm{emit},i}$, and each
cell's share is truncated to an integer and clamped to $[N_{\mathrm{new}},\, 20\,N_{\mathrm{new}}]$.
Every emitting cell therefore receives at least $N_{\mathrm{new}}$ packets, strongly emitting cells
up to twenty times as many, and the target mean over the domain is about $10\,N_{\mathrm{new}}$;
the floor and cap make the actual mean differ somewhat; $N_i$
denotes the resulting number emitted by cell $i$.  Each emitted packet is placed at a uniformly random position within
its cell, launched in a random isotropic \emph{comoving} direction at speed~$c$, and assigned a
random time-of-emission uniformly distributed in $[0, \Delta t]$.  The comoving packet frequency is
drawn by inverting the tabulated cumulative opacity-weighted Planck spectrum
$\sum_{g'\le g}\kappa_{a,g'} b_{g'}(T_i)$ at the local temperature, so that group~$g$ is selected
with the probability $p_g(T_i)$ of Eq.~\eqref{eq:emission_weights} and the group split of
Eq.~\eqref{eq:emission_energy} is reproduced.  The packet is then boosted to the laboratory frame,
\begin{equation}\label{eq:emit_boost}
    \nu_{\mathrm{lab}} = \frac{\nu_{\mathrm{co}}}{\Ddop},
    \qquad
    w = \frac{E_{\mathrm{emit},i}}{N_i\, \Ddop},
\end{equation}
with $\Ddop$ evaluated for its laboratory emission direction after aberration.  The packet
starts with comoving energy $E_{\mathrm{emit},i}/N_i$ and is boosted once; there is no additional
factor of $\gamma_i$ in that initial weight.  For the finite sample of emitted packets define
\begin{equation}\label{eq:emission_ledger}
    E_{\mathrm{src},i}=\sum_{k=1}^{N_i}w_k,\qquad
    \mathbf{P}_{\mathrm{src},i}=\sum_{k=1}^{N_i}\frac{w_k\hat{\mathbf{n}}_k}{c}
\end{equation}
The cell loses this actual laboratory energy and momentum through
Eq.~\eqref{eq:material_exchange}, with
$(\Delta E_i,\Delta\mathbf{p}_i)=(-E_{\mathrm{src},i},-\mathbf{P}_{\mathrm{src},i})$.
Isotropy gives the ensemble means
$\langle E_{\mathrm{src},i}\rangle=\gamma_i E_{\mathrm{emit},i}$ and
$\langle\mathbf{P}_{\mathrm{src},i}\rangle=\gamma_i E_{\mathrm{emit},i}\mathbf{v}_i/c^2$;
using those means as the debit would leave the finite-sample recoil uncompensated.

\subsection{Energy Deposition and Weight Attenuation}\label{sec:energy_dep}

Rather than analog absorption, the implementation uses continuous energy deposition: a packet's
weight attenuates exponentially along its trajectory, and the lost energy is deposited in the
current cell.  If a packet of laboratory-frame energy weight $w$ lying in group $g$ travels for a time $\delta t$ through a
cell with effective absorption opacity $\kappa_{a,g}^{\mathrm{eff}}$,
\begin{equation}\label{eq:energy_dep}
    \Delta E = w \bigl(1 - e^{-\kappa_{a,g}^{\mathrm{eff}} c\, \delta t\, \Ddop}\bigr),
\end{equation}
and
\begin{equation}\label{eq:weight_update}
    w \leftarrow w\, e^{-\kappa_{a,g}^{\mathrm{eff}} c\, \delta t\, \Ddop},
\end{equation}
where $\Ddop$ is the Doppler factor of Eq.~\eqref{eq:doppler}.  A packet co-propagating with the
flow has $\Ddop < 1$, interacts less frequently, and loses weight more slowly; a packet moving
against the flow has $\Ddop > 1$ and the opposite holds.  $\Delta E$ is the laboratory-frame energy
the packet loses, and that is what the cell is credited with; the split between internal and
kinetic energy is not made here but by Eq.~\eqref{eq:material_exchange}, once the accompanying
momentum $\Delta E\,\hat{\mathbf{n}}/c$ is known.  For an isolated absorption increment evaluated
at the material velocity before the kick, Eq.~\eqref{eq:material_exchange} gives
\begin{equation}\label{eq:absorption_internal}
    \Delta(U_mV)=\Delta E\,\Ddop/\gamma-\frac{(\Delta E)^2}{2mc^2}
\end{equation}
The first term is the comoving deposit divided by $\gamma$; the second accounts for the finite
momentum kick.  In the implementation, energy and momentum from all transport events are
accumulated before applying the kinetic-energy correction to their net sum, rather than assigning
an internal-energy increment to each event.  When a packet's weight falls below $10^{-3}$ of its weight at the start of the step
(or at its creation or last comb), the
packet is terminated and its residual energy $w$ and momentum $w\,\hat{\mathbf{n}}/c$ are deposited
locally.

\subsection{Geometry and the Per-Packet Event Loop}\label{sec:event_loop}

On a Voronoi mesh each cell is a convex polyhedron.  Before transport, \STORM{} caches, for every
cell, the inward-pointing unit normal $\hat{\mathbf{m}}_f$ and a point $\mathbf{x}_f$ on each
face (the face centroid).  A packet at position $\mathbf{x}$ with velocity $c\hat{\mathbf{n}}$
reaches face $f$ (here $f$ is a face index, not the Fleck factor) at time
\begin{equation}\label{eq:tface}
    t_f = \frac{(\mathbf{x}_f - \mathbf{x})\cdot\hat{\mathbf{m}}_f}{c\,\hat{\mathbf{n}}\cdot\hat{\mathbf{m}}_f},
\end{equation}
provided the denominator is negative (the packet is flying toward the face) and $t_f > 0$.  The
cell-boundary time is $t_{\mathrm{face}} = \min_f t_f$.  Convexity of the Voronoi cell guarantees
that this nearest-plane hit is the true exit face.  Emission sites are drawn uniformly in the
cell by decomposing it into tetrahedra from the generating point to each face triangle and
sampling a tetrahedron with probability proportional to its volume.

At each step of the Monte Carlo transport, a photon packet in cell~$i$ faces three competing
events:
\begin{enumerate}
    \item \textbf{Cell-boundary intersection} at time $t_{\mathrm{face}}$, from Eq.~\eqref{eq:tface}.
    \item \textbf{Scattering} at time $t_{\mathrm{scat}}$, sampled from an exponential distribution
        with rate $\kappa_{s,g}^{\mathrm{eff}}\, c\, \Ddop$, evaluated in the packet's own
        group $g$.  A uniform random number then decides whether the event is an elastic scatter
        (probability $\kappa_{s,g} / \kappa_{s,g}^{\mathrm{eff}}$) or an effective scatter.
    \item \textbf{Census} at time $t_{\mathrm{left}}$, the remaining time in the current Monte
        Carlo step.
\end{enumerate}
The event with the smallest time is selected,
\begin{equation}
    \delta t = \min(t_{\mathrm{face}},\; t_{\mathrm{scat}},\; t_{\mathrm{left}})
\end{equation}
In all three cases the packet first deposits energy and attenuates its weight according to
Eqs.~\eqref{eq:energy_dep}--\eqref{eq:weight_update}, and its position is advanced by
$\delta \mathbf{x} = \hat{\mathbf{n}}\, c\, \delta t$.  Then:
\begin{itemize}
    \item If the cell boundary is reached first, the packet moves to the adjacent cell.  If that
        neighbor lives on a different MPI rank, \STORM{} migrates the packet.
        If the face is a domain boundary, the problem-specific boundary
        condition is applied.
    \item If a scatter occurs, the packet is transformed to the comoving frame, scattered there,
        and transformed back.  The lab-to-comoving step multiplies frequency and weight by
        $\Ddop$, evaluated with the incoming direction.  Because the default scattering is
        isotropic in the comoving frame, the incoming comoving direction is not needed, and a new isotropic
        comoving direction is sampled directly: an elastic scatter preserves the comoving
        frequency, while an effective scatter redraws it from the local opacity-weighted Planck
        spectrum $p_g(T^{(n)})$, so that the packet may change group.  The inverse Lorentz boost,
        taken with the new direction, aberrates it into the laboratory frame and returns frequency
        and weight to laboratory values.  The scatter is elastic in the comoving frame but
        generally changes the laboratory weight.  Both elastic and effective scattering preserve
        the comoving packet energy, although effective scattering can change the photon frequency.
        Consequently $w_{\mathrm{before}}-w_{\mathrm{after}}=
        \mathbf{v}_{\mathrm{cell}}\cdot\Delta\mathbf{p}$ at the velocity used for the boosts.
        This cancels the linear work term when the same velocity is used in the material update;
        it does not cancel the finite $|\Delta\mathbf{p}|^2/(2m)$ kinetic-energy correction.
        The code tallies the laboratory energy and momentum differences and applies
        Eq.~\eqref{eq:material_exchange} to the complete cell tally.
    \item If the census time is reached, the packet is stored for the next time step.
\end{itemize}
In optically thick cells this three-way loop may be replaced by a \DDMC{} step
(Section~\ref{sec:ddmc}).

\subsection{Radiation-Field Estimators}\label{sec:estimators}

Two cell estimators of the radiation energy density are available.  The \emph{time-averaged}
(path-length) estimator integrates each packet's weight over the time it spends in the cell during
the step; for a segment of duration $\delta t$ with exponential attenuation at rate
$\lambda = \kappa_{a,g}^{\mathrm{eff}} c\, \Ddop$ the contribution is
$\int_0^{\delta t} w\, e^{-\lambda t'}\, dt' = w\,(1-e^{-\lambda\,\delta t})/\lambda$, and
\begin{equation}\label{eq:erad_time_avg}
    \langle E_{\mathrm{rad}} \rangle_i
      = \frac{1}{V_i\,\Delta t} \sum_{\mathrm{segments\ in}\ i}
        \int_0^{\delta t} w(t')\, dt' ,
\end{equation}
with the per-group $\langle E_g\rangle_i$ obtained by binning the same contributions by the packet's
group.  The \emph{census} estimator is the summed weight of the packets present in the cell at the
end of the step divided by its volume; it is instantaneous rather than time-averaged and therefore
noisier.  Radiation temperatures are defined from either estimator as
$T_{\mathrm{rad}} = (E_{\mathrm{rad}}/a_R)^{1/4}$.

\subsection{Boundary Conditions}\label{sec:boundary}

Domain-boundary packets are absorbed (vacuum), reflected specularly, or allowed to escape,
according to the problem.  Driven boundaries inject new packets each time step from a specified
blackbody at temperature $T_b$.  A face of area $A_f$ receives the blackbody energy
$\sigma_{\mathrm{SB}} T_b^4 A_f \Delta t$ per step, divided equally among a prescribed number of
packets.  Each packet starts at a uniformly sampled point of the face, at a uniformly sampled time
in the step, with an inward direction drawn from the Lambertian (flux-weighted) distribution,
$\mu = \sqrt{\xi}$ relative to the inward normal with uniform azimuth, and, in a multigroup or
continuous-frequency run, with an energy sampled from the Planck spectrum at $T_b$.  Each benchmark supplies its own boundary class (single-side blackbody for
the Marshak wave, two-side temperatures for radiative shocks, and shaped-geometry boundaries for
the hohlraum and crooked pipe).

%%%%%%%%%%%%%%%%%%%%%%%%%%%%%%%%%%%%%%%%%%%%%%%%%%
\section{Radiation--Hydrodynamics Coupling}\label{sec:radhydro}

\RICH{} couples radiation to hydrodynamics by operator splitting.  Following the same order as
the FLD solvers \citep{Giron2026}, one cycle (Fig.~\ref{fig:coupling}) (i)~advances the Euler
equations with a second-order Godunov scheme, (ii)~moves the mesh-generating points and
reconstructs the Voronoi tessellation with \MadVoro{}, and (iii)~holds that mesh fixed while
\STORM{} performs one Monte Carlo radiation substep.  Hydrodynamic work ($p\,\mathrm{d}V$, with $p$ the material pressure and $V$ the cell volume) and
advection therefore act on the material before the radiation exchange of the same cycle.  The
split is first-order in time; radiation and hydrodynamics share a common time step.

The radiation substep itself is:
\begin{enumerate}
    \item \textbf{Census reassignment.}  Mesh reconstruction can move cell faces past packets that
        were stored at the end of the previous step.  In a serial run each surviving packet is
        tested against its previous cell and, if it no longer lies inside that polyhedron, located
        with a containing-cell query on the new tessellation.  Under MPI the packet's cell index is
        first remapped through the persistent identifier carried by each generating point.  A
        packet lying within half the smallest neighbor distance of its cell's generating point is
        accepted directly; otherwise it walks across the faces it violates to the containing cell,
        for at most 64 hops.  Packets whose destination lies on another rank are sent directly to
        the owner of the corresponding ghost cell, and any that remain unresolved are located by an
        octree search and an iterative exchange between ranks.  Packets are not advected with the
        generating points: only their cell indices and ranks are updated.
    \item \textbf{Monte Carlo transport.}  \STORM{} creates thermal and boundary-source packets,
        propagates all packets, applies population control, and updates the material state.
        Inter-rank migration is handled by the communication layer.
\end{enumerate}

\subsection{Radiation Momentum Coupling}\label{sec:rad_momentum}

The frame transformations that let a packet interact with moving material were given with the
transport machinery they belong to in Sections~\ref{sec:physical_model}
and~\ref{sec:implementation}.  The remaining
element, treated in this section, is the back-reaction of the radiation field on the fluid, namely
the energy and momentum transferred from the radiation to the gas.  Both are evaluated in the laboratory frame, cell by cell, and equal the four-momentum
lost by the packets: the energy transfer is the net deposit of Eqs.~\eqref{eq:energy_dep}--\eqref{eq:weight_update}
less the sampled laboratory emission $E_{\mathrm{src},i}$ of Eq.~\eqref{eq:emission_ledger}, and the
momentum transfer is given by the increments listed below.
The fluid is Newtonian with a fixed cell mass $m_i$, so the momentum increment fixes its
kinetic-energy change, and the internal energy takes whatever remains,
\begin{align}
    m_i \mathbf{v}_i &\;\leftarrow\; m_i \mathbf{v}_i + \Delta\mathbf{p}_i , \nonumber\\
    U_{m,i} V_i &\;\leftarrow\; U_{m,i} V_i + \Delta E_i
        - \mathbf{v}_i\cdot\Delta\mathbf{p}_i - \frac{|\Delta\mathbf{p}_i|^2}{2 m_i},
    \label{eq:material_exchange}
\end{align}
with $\mathbf{v}_i$ the velocity before the kick, $V_i$ the cell volume, $U_{m,i}$ its internal
energy density, and $(\Delta E_i,\Delta\mathbf{p}_i)$ the net laboratory-frame energy and
momentum transferred to it.  The material total energy
$U_{m,i}V_i + m_i|\mathbf{v}_i|^2/2$ then changes by exactly $\Delta E_i$, and the sum of packet
weights and material total energy is conserved to roundoff by these source and interaction
updates, independently of the frame used to evaluate an interaction.  This statement assumes
direct \IMC{} feedback.  With momentum feedback disabled,
the internal-energy increment is simply $\Delta E_i$.  Prescribed-background transport can disable
material feedback altogether.

The linear work term has the expected low-velocity limit: for continuous absorption
$\Delta E - \mathbf{v}\cdot\Delta\mathbf{p} = \Delta E\,(1 - \hat{\mathbf{n}}\cdot\mathbf{v}/c)$,
which differs from $\Ddop\,\Delta E$ by the factor $\gamma$ alone.  The full internal-energy
increment also contains the finite-kick term in Eq.~\eqref{eq:material_exchange}; it must not be
identified exactly with the comoving deposit.  Equation
\eqref{eq:material_exchange} is applied once per cell after transport with the accumulated
increments, and once at emission with the four-momentum of Section~\ref{sec:emission}; the two
applications compose exactly because each uses the velocity current at the time.

\paragraph{Direct per-event momentum deposition.}
Momentum is accumulated from every packet interaction.  All increments below
are written as the momentum gained by the fluid.  During propagation the packet transfers
\begin{equation}\label{eq:mom_propagation}
    \Delta \mathbf{p} = \bigl(1 - e^{-\kappa_{a,g}^{\mathrm{eff}}\, c\, \delta t\, \Ddop}\bigr)\,
    w\, \frac{\hat{\mathbf{n}}}{c} = \Delta E\, \frac{\hat{\mathbf{n}}}{c},
\end{equation}
with $\Delta E$ the energy deposit of Eq.~\eqref{eq:energy_dep}.  At scattering the transfer is
$\Delta \mathbf{p} = (w_{\mathrm{before}}\, \hat{\mathbf{n}}_{\mathrm{old}}
- w_{\mathrm{after}}\, \hat{\mathbf{n}}_{\mathrm{new}}) / c$, tallied together with its energy
partner $w_{\mathrm{before}} - w_{\mathrm{after}}$ (Section~\ref{sec:event_loop}); a packet
terminated by the weight cutoff deposits its residual $w$ and $w\,\hat{\mathbf{n}}/c$.  Thermal emission is isotropic in the
comoving frame, but a finite sample has a nonzero random recoil.  The full sampled momentum
$\mathbf{P}_{\mathrm{src},i}$ is removed once per emitting cell, together with
$E_{\mathrm{src},i}$ (Section~\ref{sec:emission}).  Its expectation value is the physical bulk term
$\gamma_i E_{\mathrm{emit},i}\mathbf{v}_i/c^2$; the fluctuation about that value is the
finite-sample recoil.  This estimator retains the full anisotropy of the radiation field but can
be noisy in thick cells.

\subsection{Load Balancing}\label{sec:load_balance}

Photon work concentrates in regions of high emissivity and opacity rather than following the
mesh-point density.  When \STORM{} is run under MPI, \RICH{} keeps a separate domain
decomposition for the radiation substep and rebalances it with a user-supplied per-cell cost; in
the absence of one every cell has unit weight.  Typical costs are built from the number of
packet-step events recorded in each cell and, where packet storage dominates, the number of packets
present in it.  The partition is reconsidered periodically and is rebuilt only when the weighted load
imbalance exceeds a threshold, so that the cost of repartitioning is paid only when it is
recovered; problems in which the computational hot spot moves quickly can reconsider it more
frequently.  The communication mechanism is described in the companion \STORM{} paper.

%%%%%%%%%%%%%%%%%%%%%%%%%%%%%%%%%%%%%%%%%%%%%%%%%%
\section{Discrete Diffusion Monte Carlo}\label{sec:ddmc}

In optically thick cells, where the total optical depth
$\tau_{\mathrm{cell}} = \kappa_t \bar{\ell} \gg 1$ (with $\kappa_t=\kappa_a+\kappa_s$ the total
opacity and $\bar{\ell} = 4V/A_{\mathrm{surface}}$ the mean chord length, where $V$ and
$A_{\mathrm{surface}}$ are the cell volume and total surface area), event-by-event \IMC{} transport becomes expensive: each packet undergoes many
effective-scattering events before escaping.  \DDMC{} \citep{Densmore2007,Densmore2012} replaces
that random walk with a diffusion jump to a cell face.  During its residence time the packet
deposits energy analytically via the Fleck factor; upon leaking it is placed at the face, where it
either continues to diffuse in the neighboring cell or re-enters the standard \IMC{} loop,
depending on whether the neighbor is itself optically thick.

Each packet carries flags recording whether it is currently a \DDMC{} packet, resident in a cell,
and represented in the comoving frame.  These persist from event to event within a radiation
step and migrate with the packet between ranks, but they are cleared at census, so every packet
begins the next step as an ordinary \IMC{} packet and cell eligibility is re-evaluated on the
new mesh and material state.

\subsection{Cell Eligibility and Band Averages}\label{sec:ddmc_eligibility}

Eligibility is decided per group.  A group is diffusive in cell~$i$ when its total optical depth
across the mean chord exceeds a threshold,
\begin{equation}\label{eq:ddmc_cell_elig}
    \kappa_{t,g}\, \bar{\ell} \;\geq\; \tau_{\min}^{\mathrm{cell}},
    \qquad \kappa_{t,g} = \kappa_{a,g} + \kappa_{s,g},
\end{equation}
with $\bar{\ell} = 4V/A_{\mathrm{surface}}$ and a default threshold
$\tau_{\min}^{\mathrm{cell}} = 15$.  The groups are scanned from low to high energy and admitted
while Eq.~\eqref{eq:ddmc_cell_elig} holds; the first group that fails closes the \emph{diffusive
band} $g < G_D(i)$, where $G_D(i)$ is the number of contiguous eligible groups starting at
$g=0$ in cell $i$.  Photons below this group-index cutoff diffuse, photons above it are
transported by standard \IMC{} in the same cell.

The \DDMC{} coefficients are band averages over $g < G_D$, formed with two distinct means.  The
energy exchange with the material is an absorption rate and is averaged arithmetically with the
Planck weights, while the spatial transport is a flux and is averaged harmonically:
\begin{align}
    \kappa_A &= \frac{\sum_{g<G_D} b_g\, \kappa_{a,g}}{\sum_{g<G_D} b_g},
    &
    \kappa_T &= \frac{\sum_{g<G_D} b_g\, \kappa_{t,g}}{\sum_{g<G_D} b_g},
    \label{eq:ddmc_planck_means} \\
    \kappa_R &= \frac{\sum_{g<G_D} b_g}{\sum_{g<G_D} b_g/\kappa_{t,g}},
    &
    \Ddiff &= \frac{c}{3\,\kappa_R}
    \label{eq:ddmc_diff_coeff}
\end{align}
Here $\kappa_A$ and $\kappa_T$ are the band Planck means of absorption and total opacity,
$\Ddiff$ is the diffusion coefficient, $b_g$ is the local Planck weight of
Eq.~\eqref{eq:group_defs}, and $\kappa_R$ is a Rosseland-type (flux) mean, weighted with $b_g$ rather than with
$\partial B_g/\partial T$.  The harmonic mean appropriately emphasizes the least opaque groups,
which carry most of the diffusive flux; an arithmetic opacity average would over-weight the opaque
groups and underestimate the band diffusion coefficient.  In the grey limit
$G_D = 1$, $\kappa_A = \kappa_P$ and $\kappa_R = \kappa_T = \kappa_P + \kappa_s$, so
Eq.~\eqref{eq:ddmc_diff_coeff} reduces to $\Ddiff = c/(3\kappa_t)$.  Two further band quantities
enter below: the effective-scattering albedo in the sense of \citet{Densmore2006},
\begin{equation}\label{eq:ddmc_albedo}
    \omega = 1 - \frac{f\, \kappa_A}{\kappa_R},
\end{equation}
and the fraction of the cell's Planck emission that falls inside the diffusive band,
\begin{equation}\label{eq:ddmc_bandmass}
    \Gamma = \frac{\sum_{g<G_D} b_g\, \kappa_{a,g}}{\sum_{g} b_g\, \kappa_{a,g}},
\end{equation}
which drives the group-exit channel of Section~\ref{sec:ddmc_step}.

A cell is declared \DDMC{}-eligible when $G_D > 0$, $\Ddiff > 0$, and the band-averaged optical depth
$\kappa_T \bar{\ell}$ satisfies Eq.~\eqref{eq:ddmc_cell_elig}.  In addition, every domain-boundary
face of the cell must be one the diffusion discretization can represent, namely a rigid reflecting
wall or a thermal source (Section~\ref{sec:boundary}).  Cells touching a vacuum or escape boundary are excluded, so that a packet cannot leave the
problem through such a boundary by a \DDMC{} jump and must first return to standard \IMC{}.

\subsection{Face Leakage Rates}\label{sec:ddmc_leakage}

For each face~$f$ of an eligible cell the leakage rate follows a finite-volume discretization of
the diffusion equation.  Between two eligible cells the discretization is a series resistance
across the face, with $d_1$ and $d_2$ the perpendicular distances from the two cells' generating
points to the face plane:
\begin{equation}\label{eq:ddmc_leak_rate}
    \lambda_f^{\mathrm{int}}
    = \frac{A_f}{V}\left[\frac{d_1}{\Ddiff_1} + \frac{d_2}{\Ddiff_2}\right]^{-1},
\end{equation}
where $A_f$ is the face area and $V$ the volume of the cell the packet is leaving.  Subscripts
$1$ and $2$ label the source and neighboring cells, respectively, and $\Ddiff_1$, $\Ddiff_2$
are their diffusion coefficients; $\lambda_f^{\mathrm{int}}$ is the leakage rate through face $f$.  Equation
\eqref{eq:ddmc_leak_rate} reduces to $A_f/V$ times the harmonic mean $2\Ddiff_1\Ddiff_2/(\Ddiff_1+\Ddiff_2)$
divided by $2d_f$, where $d_f$ is the common half-distance when $d_1 = d_2 = d_f$.  On a Voronoi
mesh this is always the case, since every face lies on the perpendicular bisector plane of its two
generating points; the two distances differ on non-uniform Cartesian grids, where neighboring
cells have different widths and their shared face is not midway between the two cell centers.  Writing the rate in terms of both distances keeps one formula for every grid and
preserves reciprocity, $V_1 \lambda_{1\to2} = V_2 \lambda_{2\to1}$, where $V_j$ is the volume of
cell $j$ and $\lambda_{j\to k}$ the rate from cell $j$ to cell $k$, which \STORM{} monitors as a
diagnostic.
The two diffusion coefficients are re-evaluated at a common face temperature
\begin{equation}\label{eq:ddmc_face_temperature}
    T_f = \left[\tfrac{1}{2}\bigl(T_1^4 + T_2^4\bigr)\right]^{1/4},
\end{equation}
where $T_1$ and $T_2$ are the temperatures of the adjacent cells, so that the group opacities and
Planck weights entering the Rosseland means on the two sides are evaluated at the same temperature
and the two coefficients differ only through the local material.  Each side keeps its own band limit $G_D$,
set by its own cell temperature, and if either face-temperature coefficient vanishes the
cell-center coefficients are used instead.  Flux is then continuous at material interfaces
\citep{Densmore2012}.

When the neighbor is not eligible the packet is handed back to transport, and the rate is the
asymptotic diffusion-limit interface rate rather than a diffusion conductance.  \STORM{} uses the
emissivity-preserving conversion coefficient of \citet{Densmore2006} (their Eq.~48),
\begin{equation}\label{eq:ddmc_conversion}
    \begin{aligned}
    \mathcal{C}(\tau, \omega) &= \frac{\epsilon'\beta}{\beta - \tfrac{3}{4}\epsilon'\tau},
    \qquad \tau = 2\,\kappa_R\, d_1, \\
    \epsilon' &= \frac{4}{3}\,\frac{\sqrt{3(1-\omega)}}{1+\lambda_E\sqrt{3(1-\omega)}}, \\
    \beta &= \tfrac{3}{2}(1-\omega)\tau^2
      + \sqrt{3(1-\omega)\tau^2 + \tfrac{9}{4}(1-\omega)^2\tau^4},
    \end{aligned}
\end{equation}
where $\tau$ is the optical thickness of a planar cell of width $2d_1$, $\omega$ is the
single-scatter albedo of Eq.~\eqref{eq:ddmc_albedo}, $\epsilon'$ is the emissivity of the
analytic diffusion approximation for a half-space, and $\lambda_E = 0.7104$ is the extrapolation
length in optical-depth units.  $\mathcal{C}$ replaces the standard asymptotic conversion
probability $8/(3\tau+6\lambda_E)$, to which it reduces in the scattering-dominated limit
$\omega\to 1$ for any $\tau$, so that the discretized interface reproduces $\epsilon'$ for cells
of any optical thickness, and $\mathcal{C}\to\epsilon'$ as $\tau\to\infty$.  The leakage rate into the transport region is then
\begin{equation}\label{eq:ddmc_boundary_rate}
    \lambda_f^{\mathrm{ext}} = \frac{c\, A_f\, \mathcal{C}}{4\, V}
\end{equation}
Here $\lambda_f^{\mathrm{ext}}$ is the rate for leakage into a transport region and
$\mathcal{C}=\mathcal{C}(2\kappa_R d_1,\omega)$.  The coefficient is admissible when
$0 \le \mathcal{C} \le 2/(0.91+1.635)$, the bound that keeps the admission probability of
Eq.~\eqref{eq:ddmc_admission} below unity at normal incidence; outside that range the code falls back to the
\citet{Densmore2012} form $\lambda_f^{\mathrm{ext}} = c A_f/[3V(\kappa_R d_1 + \lambda_E)]$.

In multigroup problems a face may be partly diffusive, the neighbor supporting a narrower band
than the source cell, $G_D^{\mathrm{nb}} < G_D$.  The face rate is then split by the fraction of
the source cell's band emission that the neighbor can also diffuse,
\begin{equation}\label{eq:ddmc_face_split}
    \phi_f = \frac{\sum_{g<G_D^{\mathrm{nb}}} b_g}{\sum_{g<G_D} b_g},
    \qquad
    \lambda_f = \phi_f\, \lambda_f^{\mathrm{int}}
              + (1-\phi_f)\, \lambda_f^{\mathrm{ext}},
\end{equation}
so that the low-energy part of the band continues to diffuse while the remainder converts to
\IMC{}.  A face with $\phi_f = 1$ is purely internal and one with $\phi_f = 0$ is a pure \IMC{}
interface.  This split is an approximation: $\lambda_f^{\mathrm{int}}$ is formed from the
full-band diffusion coefficients of the two cells rather than recomputed over the shared groups
$g < G_D^{\mathrm{nb}}$, and $\phi_f$ only apportions the face's emission between the two
channels.  It is exact when the two bands coincide ($\phi_f = 1$) and for grey transport.

Rigid reflecting boundary faces carry no leakage channel, which is the zero-flux diffusion
condition.  Thermal-source boundary faces carry a channel of the interface
form~\eqref{eq:ddmc_boundary_rate}, which represents the boundary directly in the diffusion
discretization.  A \DDMC{} packet that leaks through such a face leaves the domain and is tallied
as escaping energy.  Conversely, a source packet injected through that face into an eligible cell
is converted into a \DDMC{} packet in that cell with the admission probability of
Eq.~\eqref{eq:ddmc_admission}; a source packet that is not converted is not injected at all,
which is equivalent to its being reflected straight back out by the diffusive region.  The total event rate is $\Lambda = \sum_f \lambda_f$.

The leakage rates complete the eligibility criteria of Section~\ref{sec:ddmc_eligibility}.  A cell
remains in transport unless its total rate $\Lambda$ is positive and the conversion coefficient
$\mathcal{C}$ of Eq.~\eqref{eq:ddmc_conversion}, evaluated on the mean chord, is admissible in the
sense above, since otherwise the asymptotic interface analysis does not close.

\subsection{Per-Packet DDMC Step}\label{sec:ddmc_step}

\DDMC{} imposes no minimum distance between the packet and the cell faces.  The face-based
discretization holds anywhere in the cell, and the transition is kept unbiased at the interface
(Section~\ref{sec:ddmc_transitions}).  A packet takes a \DDMC{} step when its cell is eligible, its
comoving frequency lies inside the diffusive band ($g < G_D$), and the event rate is positive, and
reverts to standard \IMC{} for that step otherwise.

If the packet qualifies, it is represented in the comoving frame and the step proceeds as follows.
\begin{enumerate}
    \item \textbf{Sample the event time.}  Leakage competes with a group-exit channel that exists
        only in multigroup runs.  Effective scattering inside the band can promote a photon out of
        it, at the rate
        \begin{equation}\label{eq:ddmc_upscatter}
            \lambda_{\uparrow} = c\,(1-f)\,\kappa_A\,(1 - \Gamma),
        \end{equation}
        with $\Gamma$ from Eq.~\eqref{eq:ddmc_bandmass}, so the total event rate is
        $\Lambda + \lambda_{\uparrow}$ and $t_{\mathrm{event}} = -\ln\xi/(\Lambda+\lambda_\uparrow)$,
        where $\Lambda=\sum_f\lambda_f$ is the total face-leakage rate and $\xi$ is a uniform
        random variate on $(0,1)$.
        The time advanced is $\delta t = \min(t_{\mathrm{event}},\, t_{\mathrm{left}},\,
        t_{\mathrm{cut}})$, where $t_{\mathrm{cut}}$ is the time at which a packet in a compressing
        cell would be blueshifted out of the band by the next item.
    \item \textbf{Comoving frequency drift.}  The packet is advanced in the comoving frame over a
        laboratory step, so the adiabatic term of the comoving transport equation acts on both the
        frequency and the weight,
        \begin{equation}\label{eq:ddmc_adiabatic}
            \nu \leftarrow \nu\, e^{-\nabla\cdot\mathbf{v}\,\delta t/3},
            \qquad
            w \leftarrow w\, e^{-\nabla\cdot\mathbf{v}\,\delta t/3},
        \end{equation}
        where $\nabla\cdot\mathbf{v}$ is the local fluid-velocity divergence and $\nu$ and $w$ are
        the comoving packet frequency and energy weight.  This is the \DDMC{} counterpart of the Doppler shift applied at each \IMC{} event.  The term
        vanishes for a static or uniformly translating medium.
    \item \textbf{Energy deposition.}  The weight attenuates at the Fleck effective-absorption
        rate built from the band Planck mean,
        \begin{equation}\label{eq:ddmc_abs}
            w \;\leftarrow\; w\, \exp(-\delta t\, \kappa_A\, f\, c),
        \end{equation}
        and $\Delta E = w_{\mathrm{before}} - w_{\mathrm{after}}$ is tallied for the cell together
        with the momentum $\Delta E\, \mathbf{v}/c^2$ of isotropic comoving radiation, both
        entering Eq.~\eqref{eq:material_exchange}.  Because a resident packet carries its comoving
        rather than laboratory weight, the laboratory energy budget of a \DDMC{} cell closes to
        $O(v^2/c^2)$, not to roundoff as in the \IMC{} path.
    \item \textbf{Census, group exit, or leak.}  If $t_{\mathrm{left}}$ expires first the packet is
        rebuilt as an \IMC{} census packet: it is placed at a position sampled uniformly in the
        cell, given an isotropic comoving direction and, in a multigroup run, a frequency drawn
        from the Planck spectrum of the diffusive band, and Lorentz-boosted to the laboratory
        frame.  If the group-exit channel is selected, the packet's frequency is
        redrawn from the part of the opacity-weighted Planck spectrum above the band and the
        packet resumes standard \IMC{} in the same cell.  Otherwise it leaks through a face chosen with
        probability $\lambda_f/\Lambda$, and within that face through the diffusive channel with
        probability $\phi_f\lambda_f^{\mathrm{int}}/\lambda_f$ and through the interface channel
        with probability $(1-\phi_f)\lambda_f^{\mathrm{ext}}/\lambda_f$, the two terms of
        Eq.~\eqref{eq:ddmc_face_split}.
    \item \textbf{Face placement.}  The packet is placed at the selected face centroid.  On the
        interface channel it re-enters standard \IMC{} with a cosine-distributed direction about
        the outward normal ($\mu = \sqrt{\xi}$, where $\mu$ is the cosine relative to that normal and $\xi$ is a
        fresh uniform variate on $(0,1)$, with uniform azimuth) and, in a multigroup run, with its
        frequency redrawn from the Planck spectrum of the part of the band that the neighbor cannot
        diffuse, $G_D^{\mathrm{nb}} \le g < G_D$ (the whole band when the neighbor is not
        eligible), consistent with the diffusive spectrum it carried.  On the
        diffusive channel it enters the neighbor as a \DDMC{} packet with
        direction drawn from the asymptotic probability density $p(\mu) \propto \mu\,(0.91 + 1.635\mu)$ of
        \citet{Densmore2012}, the distribution in detailed balance with the admission probability
        of Eq.~\eqref{eq:ddmc_admission}.
\end{enumerate}

\subsection{IMC--DDMC Transitions}\label{sec:ddmc_transitions}

For \IMC{}--\DDMC{} transport, the decision at each packet step is:
\begin{enumerate}
    \item If the cell is \DDMC{}-eligible, attempt a \DDMC{} step.
    \item If that attempt fails (frequency above the diffusive band, or a non-positive event
        rate), fall back to standard \IMC{}.
\end{enumerate}

The reverse transition, an \IMC{} packet arriving at the face of a \DDMC{} cell, requires more
care, a diffusion description being unable to accept a beam at grazing incidence.  \STORM{}
implements the asymptotic interface condition of \citet{Densmore2006,Densmore2012}: a packet
incident with direction cosine $\mu$ relative to the inward face normal is admitted into \DDMC{}
with probability
\begin{equation}\label{eq:ddmc_admission}
    P(\mu) = \tfrac{1}{2}\, \mathcal{C}\,(0.91 + 1.635\,\mu),
\end{equation}
with $\mathcal{C}$ the conversion coefficient of Eq.~\eqref{eq:ddmc_conversion} for the target
cell.  A packet that is not admitted is reflected back into the transport region with a
cosine-distributed direction and remains an ordinary \IMC{} packet.  Where
$\mathcal{C}$ is not admissible (Section~\ref{sec:ddmc_leakage}), the paired \citet{Densmore2012} form
$P(\mu) = 2(0.91+1.635\mu)/[3(\kappa_R d + \lambda_E)]$ is used instead, where $d$ is
the target-cell generating-point distance to the face centroid, $\kappa_R$ is that cell's band
transport mean, and $\lambda_E=0.7104$ is the extrapolation length in optical-depth units.  Grazing packets are
therefore rejected in the majority, and the $\mu$-weighting of the admitted population reproduces
that of the asymptotic diffusion solution.

When the interface moves with the material, the admitted weight is corrected by the
\citet{Wollaeger2013} factor
\begin{equation}\label{eq:ddmc_moving_interface}
    G(\mu, \beta_n) = 1 + 2\,\beta_n\, K(\mu),
    \qquad \beta_n = -\hat{\mathbf{m}}_f\cdot\mathbf{v}_f/c,
\end{equation}
where $\mathbf{v}_f$ is the interface velocity, $\hat{\mathbf{m}}_f$ is the target cell's inward
face normal, $\beta_n$ is the signed normal interface speed divided by $c$, $G$ is the
multiplicative weight correction, and $K(\mu)$ is the tabulated kernel of their Eq.~(60), built once and interpolated in
$\mu K(\mu)$ to preserve the grazing-angle limit.  Since $G$ can exceed unity, a packet admitted
into a cell on the same rank is split into up to $64$ equal-weight copies, so that no single
\DDMC{} packet carries more than twice the incident weight; a packet admitted into a cell on
another rank is sent as a single packet carrying the full corrected weight.  The splitting is
unbiased and serves only to control variance.  Crossings for which $|\beta_n|$ exceeds $0.1$,
$G$ is non-positive, non-finite or exceeds $10$, or more than $64$ copies would be needed, bypass
\DDMC{}: the packet continues in ordinary \IMC{}, which remains valid though slower, and \DDMC{}
stays disabled for it throughout the target cell until it leaves that cell through a face.

\subsection{Coupling to Hydrodynamics}\label{sec:ddmc_hydro}

Face areas, cell volumes, and the two face-to-center distances in Eq.~\eqref{eq:ddmc_leak_rate} are
recomputed from the current tessellation at the start of each radiation substep, so leakage rates
follow mesh motion.  Energy deposition uses the same Fleck factor as standard \IMC{}, and the
comoving frequency drift of Eq.~\eqref{eq:ddmc_adiabatic} supplies the $p\,\mathrm{d}V$ work that
event-by-event Doppler shifts supply in the \IMC{} path.

Standard \IMC{} follows a packet's direction through individual interactions, allowing the
momentum transferred to the fluid to be tallied at each event
(Section~\ref{sec:rad_momentum}).  \DDMC{} replaces those interactions by a residence time in a
cell followed by a jump through one of its faces.  It therefore does not resolve the individual
changes in direction needed for that tally.  Instead, we estimate the radiation energy flux
from the packet energy carried through the cell faces and use this flux to calculate the force
on the fluid.

Each time a \DDMC{} packet leaks through a face of cell $i$, or a packet is admitted into
\DDMC{} in cell $i$ through one of its faces, its energy weight is multiplied by the
outward unit normal of that face and added to a vector $\mathbf{R}_i$ for cell $i$, with a
positive sign when the packet leaves the cell and a negative sign when it enters, whether from
\IMC{} or from a neighboring \DDMC{} cell.  A packet crossing a face carries its full weight
across it whatever its direction, so accumulated over the radiation step $\mathbf{R}_i =
\sum_f E_f\,\hat{\mathbf{n}}_f$, where $E_f$ is the net energy that left through face $f$.
Converting it to a flux requires accounting for the areas and orientations of the cell faces.
We collect this geometric information in a matrix $\mathsf{M}_i$ and solve
\begin{equation}\label{eq:ddmc_flux_moment}
    \begin{aligned}
        \mathsf{M}_i &= \sum_f A_f\, \hat{\mathbf{n}}_f \otimes \hat{\mathbf{n}}_f, \\
        \mathsf{M}_i\, \mathbf{F}_i\, \Delta t &= \mathbf{R}_i,
    \end{aligned}
\end{equation}
where $A_f$ is the face area, $\hat{\mathbf{n}}_f$ is its outward unit normal, and $\otimes$
denotes the outer product.  Equation~\eqref{eq:ddmc_flux_moment} is the area-weighted
least-squares fit of $\mathbf{F}_i\cdot\hat{\mathbf{n}}_f = E_f/(A_f\Delta t)$ over the faces,
minimizing $\sum_f A_f\,(\mathbf{F}_i\cdot\hat{\mathbf{n}}_f\,\Delta t - E_f/A_f)^2$.  Packets
outside the diffusive band that cross the same faces as ordinary \IMC{} packets are not counted
here; their momentum is tallied by the per-event estimator of Section~\ref{sec:rad_momentum}.  The sum runs over the
leakage faces and over the cell's rigid reflecting walls; a wall carries no leakage channel, but its
zero-flux condition $\mathbf{F}_i\cdot\hat{\mathbf{n}}_f = 0$ is a genuine constraint, entering
with $E_f = 0$, and it keeps $\mathsf{M}_i$ well conditioned for cells whose leakage faces all lie
along one direction, as in the slab problems of Section~\ref{sec:results}.  The resulting
$\mathbf{F}_i$ is the estimated radiation energy flux in the cell, averaged over $\Delta t$.  With
the group fluxes $\mathbf{F}_g = -(c/3\kappa_{t,g})\nabla E_g$ and $E_g \propto b_g$ within the
band, the band force per unit volume is
$\sum_g \kappa_{t,g}\mathbf{F}_g/c = \kappa_R\, \mathbf{F}_i/c$, so the force is formed with the
same flux mean $\kappa_R$ as the diffusion coefficient of Eq.~\eqref{eq:ddmc_diff_coeff}.
Multiplying by the cell volume and the step duration gives the momentum transferred to the fluid,
\begin{equation}\label{eq:ddmc_force}
    \Delta\mathbf{p}_i = \frac{\kappa_R\, V_i\, \mathbf{F}_i\, \Delta t}{c}
\end{equation}
If $\mathsf{M}_i$ is nevertheless numerically singular, the code falls back to
$\mathbf{F}_i \Delta t = \mathbf{R}_i / \sum_f A_f$, summed over the leakage faces.  For a cell
whose leakage faces are a single opposed pair this is exactly the pseudo-inverse along their
common normal.

Packets cannot leave the domain through a vacuum or escape boundary by \DDMC{} leakage, because
cells touching such boundaries are not eligible; they must approach that face under standard
\IMC{} so that the problem boundary condition is applied.  The only boundaries a \DDMC{} packet
can cross directly are thermal-source faces, treated as described in
Section~\ref{sec:ddmc_leakage}, where source packets also enter \DDMC{} directly.

%%%%%%%%%%%%%%%%%%%%%%%%%%%%%%%%%%%%%%%%%%%%%%%%%%
\section{Population Control}\label{sec:pop_control}

After each transport step, and before the material state is updated, a comb redistributes the
census packet population so that statistical resolution is maintained without unbounded growth.
Let $W_i$ be the total packet weight in cell $i$ and $W$ the global weight.  The global target
population is
\begin{equation}\label{eq:comb_budget}
    N_{\mathrm{target}} = \phi_{\mathrm{comb}}\, N_{\min}\, N_{\mathrm{cells}},
\end{equation}
where $N_{\mathrm{cells}}$ is the global cell count, $N_{\min}$ the per-cell floor, and
$\phi_{\mathrm{comb}}$ a budget factor (default $2$), so that the target mean population is
$\phi_{\mathrm{comb}}N_{\min}$ packets per cell; the floor and cap below make the actual mean
differ somewhat.  It is converted into a per-cell target
\begin{equation}
    N_i = \min\!\Bigl(N_{\max},\; \max\!\Bigl(N_{\min},\;
        \Bigl\lfloor N_{\mathrm{target}}\,\frac{W_i}{W}\Bigr\rfloor\Bigr)\Bigr),
\end{equation}
where $N_{\max}$ is the per-cell cap.  The defaults are $N_{\min} = 20$ and
$N_{\max} = 20 N_{\min}$; the floor and cap can also be set independently of $N_{\min}$.  Cells
holding no weight receive no packets.

If cell $i$ already has no more than $N_i$ packets, overweight packets (weight greater than
$2 W_i/N_i$) are split into equal-weight copies; lighter packets are left unchanged.  If the cell
is over-occupied, the packets are randomly shuffled and a comb with random offset $r\in(0,1)$
walks through the cumulative weight, emitting a representative packet of weight $W_i/N_i$ each
time the comb teeth at $(k+r)W_i/N_i$ are crossed, with integer tooth index
$k=0,\ldots,N_i-1$.  The copies inherit the parent's position,
direction, and frequency.  By construction $\sum w = W_i$ in every cell, so the comb conserves
radiant energy to roundoff.  Resampling does not in general preserve the finite-sample directional
sum $\sum w\hat{\mathbf{n}}/c$ exactly; this statistical change is distinct from the balanced
material recoil of emission and transport interactions.  A stratified variant, grouping packets by frequency group before
combing, is available for multigroup problems.

%%%%%%%%%%%%%%%%%%%%%%%%%%%%%%%%%%%%%%%%%%%%%%%%%%
\section{Numerical Results}\label{sec:results}

We validate the \IMC{} implementation against a suite of benchmark problems that test different
aspects of the solver: pure radiative transport without hydrodynamics, coupled
radiation--hydrodynamics, single-group and multigroup opacities, and multidimensional geometry.
Unless
stated otherwise, coupled calculations use the \RICH{} Voronoi tessellation in three dimensions,
even for nominally one-dimensional problems, which are set up as narrow slabs.  Pure-transport
tests may equally be run in standalone \STORM{} on a Cartesian mesh; the physics kernel is the
same.  Table~\ref{tab:benchmarks} summarizes the geometric and algorithmic configuration of each
benchmark.

\begin{table*}
    \centering
    \caption{Configuration of the validation benchmarks.  ``Dim.'' is the physical dimensionality
    of the problem (a narrow three-dimensional slab is used for coupled radiation--hydrodynamics
    problems that are nominally one-dimensional).  ``Hydro'' indicates whether the material moves,
    so that mesh motion and, except in the RMTV problem (Section~\ref{sec:rmtv}), the frame
    transformations of Section~\ref{sec:transport_eq} are exercised;
    the remaining problems are pure radiative transport on a static background.  A dagger
    ($\dagger$) marks the moving slab, in which the material moves at a prescribed velocity and the
    transport is fully Doppler- and Lorentz-corrected, while no hydrodynamic equations are solved
    and the radiation does not feed back on the material (Section~\ref{sec:moving_slab}).  ``Groups'' is the single-group (grey) limit $G=1$, the
    number of frequency groups $G$, or ``cont.'' for continuous-frequency transport, in which
    opacities are evaluated at each packet's own energy rather than on a band structure.  ``DDMC'' indicates discrete diffusion Monte Carlo
    acceleration (Section~\ref{sec:ddmc}).  ``Ranks'' is the number of MPI ranks used for the run reported in
    Section~\ref{sec:results}.  ``Reference'' cites the publication that introduced the benchmark
    problem itself, as distinct from any publication cited only for comparison data.}
    \label{tab:benchmarks}
    \scriptsize
    \setlength{\tabcolsep}{2.5pt}
    \begin{tabular}{lccccccl}
        \hline\hline
        Benchmark & Dim. & Mesh & Hydro & Groups & DDMC & Ranks & Reference \\
        \hline
        Marshak waves 1--3 (\S\ref{sec:marshak}) & 1D & Cartesian & \no & grey & \no & 1 & \citet{KriefMcClarren2024} \\
        Marshak wave 4 (\S\ref{sec:marshak}) & 1D & Cartesian & \no & grey & \no & 16 & \citet{Derei2024} \\
        Cylindrical hohlraum (\S\ref{sec:hohlraum}) & 3D & Voronoi & \no & grey & \no & 512 & \citet{McClarrenUrbatsch2009} \\
        Mach~2 shock (\S\ref{sec:mach2}) & 1D (slab) & Voronoi & \yes & grey & \no & 16 & \citet{LowrieEdwards2008} \\
        Mach~45 shock (\S\ref{sec:mach45}) & 1D (slab) & Voronoi & \yes & grey & \no & 64 & \citet{LowrieEdwards2008} \\
        RMTV radiative shock (\S\ref{sec:rmtv}) & 3D (octant) & Voronoi & \yes & grey & \yes & 4--256 & \citet{ReinickeMeyerterVehn1991} \\
        Densmore step-opacity, IMC (\S\ref{sec:densmore}) & 1D & Cartesian & \no & 32 & \no & 32 & \citet{Densmore2012} \\
        Densmore step-opacity, DDMC (\S\ref{sec:densmore}) & 1D & Cartesian & \no & 32 & \yes & 32 & \citet{Densmore2012} \\
        Olson 2D multimaterial (\S\ref{sec:olson}) & 2D & Cartesian & \no & cont. & \no & 256 & \citet{Olson2020} \\
        Crooked pipe (\S\ref{sec:crooked_pipe}) & 3D & Voronoi & \no & grey & \yes & 480 & \citet{GrazianiLeBlanc2000} \\
        E3D stochastic medium (\S\ref{sec:e3d}) & 3D & Voronoi & \no & grey & \no & 160 & \citet{Olson2007} \\
        Moving slab (\S\ref{sec:moving_slab}) & 1D (slab) & Voronoi & \yes$^\dagger$ & 124 & \no & 48 & \citet{McClarrenGentile2021} \\
        \hline
    \end{tabular}
\end{table*}

\subsection{Marshak Wave}\label{sec:marshak}

The Marshak wave is a fundamental test of radiation diffusion into a cold, absorbing medium and is
one of the most widely used benchmarks for thermal radiative-transfer codes
\citep{SteinbergHeizler2022}.

We run four variants, drawn from two recent families of self-similar Marshak-wave solutions:
tests~1 and~2 are Tests~2 and~3 of \citet{KriefMcClarren2024}, and tests~3 and~4 are Tests~1
and~3 of \citet{Derei2024}.  Together they span equilibrium and non-equilibrium
matter--radiation coupling, uniform and spatially varying density, and uniform and geometrically
stretched meshes.

\subsubsection{Problem setup}
In every case a thermal radiation wave is driven into cold material by a time-dependent Planck
source at $x = 0$ whose temperature follows a power law in time, and the calculation runs to
$t_{\mathrm{final}} = 1$~ns.  The material model is a power-law equation of state
$u(T, \rho)$, denoting material internal energy per unit volume (the $U_m$ of
Section~\ref{sec:transport_eq}), together with power-law Rosseland and Planck opacities; the defining parameters of
each variant are collected in Table~\ref{tab:marshak_variants}.  The elastic scattering opacity is
set to
$\kappa_R - \kappa_P$ so that the transport mean free path reproduces the Rosseland mean, and the
nonlinear equation of state is inverted for $T$ from the internal energy after each step.

\begin{table*}
    \centering
    \caption{The four Marshak-wave variants.  Temperatures $T$ are in keV, times $t$ in ns, and
    densities $\rho$ in g\,cm$^{-3}$; opacities are in cm$^{-1}$ and energy densities in
    erg\,cm$^{-3}$.  Here $x$ is distance in cm from the driven boundary,
    $T_{\mathrm{bath}}$ is its blackbody temperature, $u$ is material internal energy density, and
    $\kappa_R$ and $\kappa_P$ are the Rosseland and Planck opacities.  Tests~1 and~2 are Tests~2 and~3 of \citet{KriefMcClarren2024}; tests~3 and~4
    are Tests~1 and~3 of \citet{Derei2024}.  All four run to $t_{\mathrm{final}} = 1$~ns.}
    \label{tab:marshak_variants}
    \begin{tabular}{clllllc}
        \hline\hline
        Test & $\kappa_R$ & $\kappa_P/\kappa_R$ & $u(T,\rho)$ & $\rho(x)$ &
        $T_{\mathrm{bath}}(t)$ & Domain (cm) \\
        \hline
        1 & $100\, T^{-3}$ & $10^{-3}$ & $6.860085\times10^{14}\, T^4$ & $1$ &
          $1.008038\, t^{1/3}$ & $[0,\, 0.2]$ \\
        2 & $100\, T^{-3}$ & $1$ & $6.860085\times10^{14}\, T^4$ & $1$ &
          $1.014565\, t^{1/3}$ & $[0,\, 0.2]$ \\
        3 & $40\, T^{-1.5} \rho^{1.2}$ & $2.5\times10^{-3}$ &
          $10^{14}\, T^{3.4} \rho^{0.86}$ & $x^{20/19}$ &
          $1.0470478\, t^{86/57}$ & $[0,\, 1]$ \\
        4 & $2\, T^{-4.5} \rho^{1.9}$ & $5\times10^{-4}$ &
          $10^{14}\, T^{6} \rho^{0.7}$ & $x^{-40/139}$ &
          $1.01008116\, t^{14/139}$ & $[10^{-5},\, 1+10^{-5}]$ \\
        \hline
    \end{tabular}
\end{table*}

Tests~1 and~2 share the Rosseland opacity, the equation of state, the density, and the domain, and
differ in the ratio $\kappa_P/\kappa_R$ (and hence in the normalization of the drive temperature):
test~2 has $\kappa_P = \kappa_R$, so matter and radiation stay in
equilibrium and a single temperature describes the wave, whereas test~1 has
$\kappa_P = 10^{-3} \kappa_R$, which decouples the two and produces a material temperature well
below the radiation temperature throughout the heated region.  Tests~3 and~4 add a spatially
varying density: test~3 has an increasing profile $\rho \propto x^{20/19}$ on a uniform mesh,
while test~4 has a divergent profile $\rho \propto x^{-40/139}$ and is run on a geometrically
stretched mesh, with the domain offset by $10^{-5}$~cm to avoid the singularity at the origin.

Tests~1 and~2 use 256 uniform cells and test~3 uses 512; test~4 uses 512 cells whose widths grow
geometrically from $7\times10^{-5}$~cm at the inner boundary by a factor $1.00969$ per cell.
Tests~1--3 run in serial and test~4 on 16 MPI ranks.  All four emit 15 thermal packets per cell
per step and 100 boundary-source packets per cell.  The material starts at $10$~eV in tests~1--3
and at $1$~eV in test~4.  The time step follows $\Delta t = \max(\Delta t_0,\,10^{-3}t)$, with
$\Delta t_0 = 10^{-15}$~s and a ceiling of $5\times10^{-11}$~s for tests~1--3 (both scaled down by
a factor of four for test~3), and $\Delta t_0 = 10^{-17}$~s with a ceiling of $5\times10^{-13}$~s
for test~4; over the run to $1$~ns the ceiling binds only in test~4.  Comb population control
(Section~\ref{sec:pop_control}) uses a floor $N_{\min} = 15$ packets per cell with budget factor
$\phi_{\mathrm{comb}} = 6$ in tests~1--3, and $N_{\min} = 100$ with $\phi_{\mathrm{comb}} = 1$ in
test~4.

\begin{figure*}
    \gridline{\fig{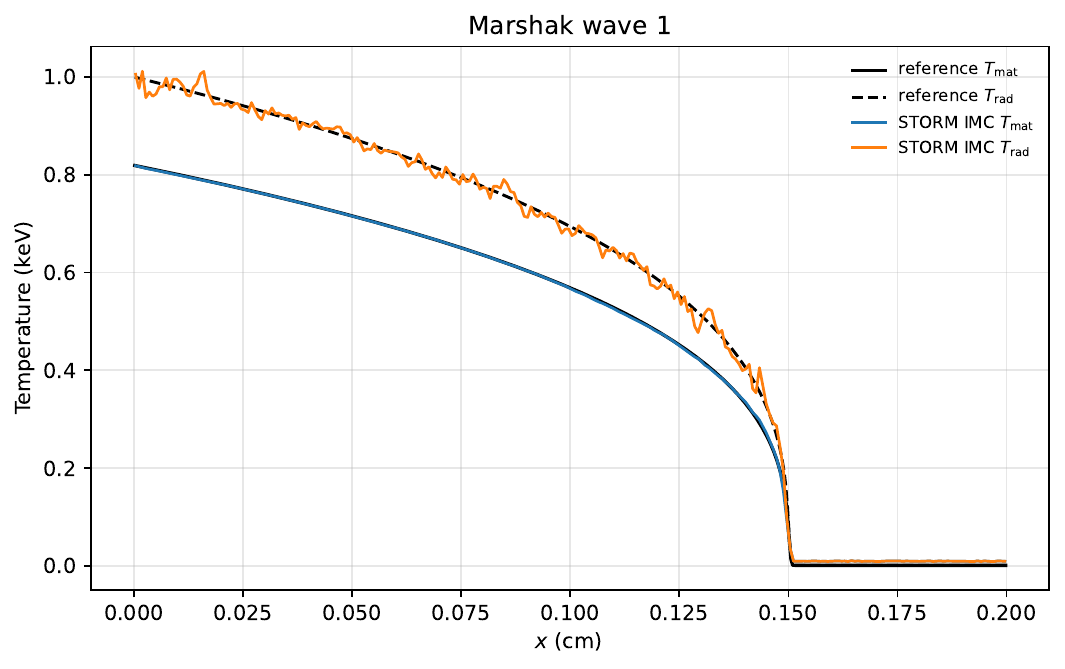}{0.47\textwidth}{(a) Test~1, non-equilibrium}
              \fig{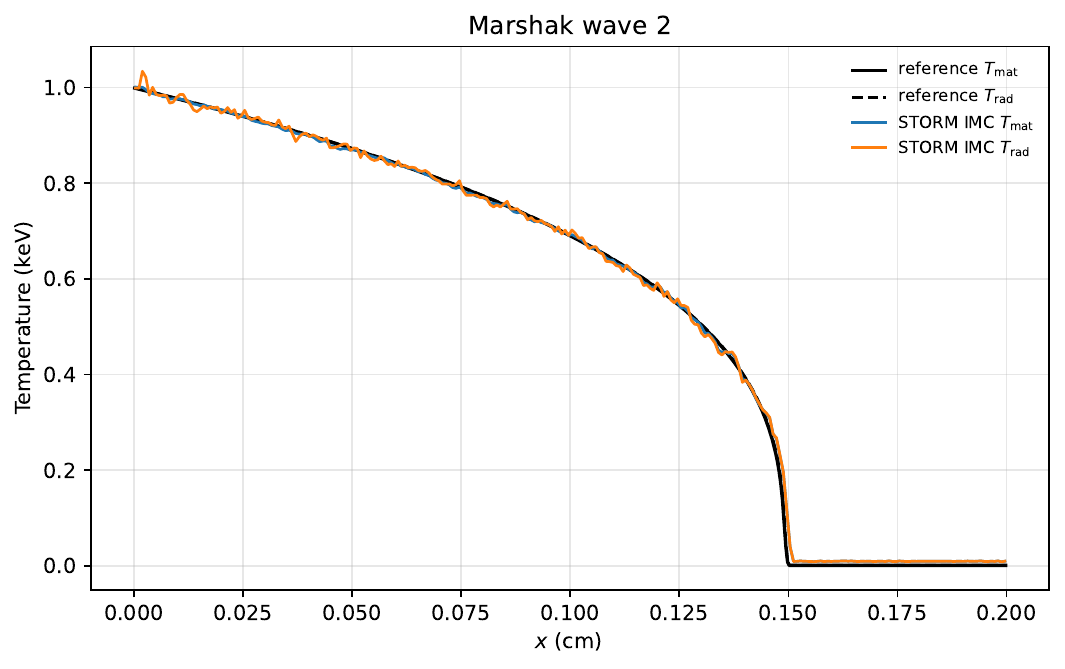}{0.47\textwidth}{(b) Test~2, equilibrium}}
    \gridline{\fig{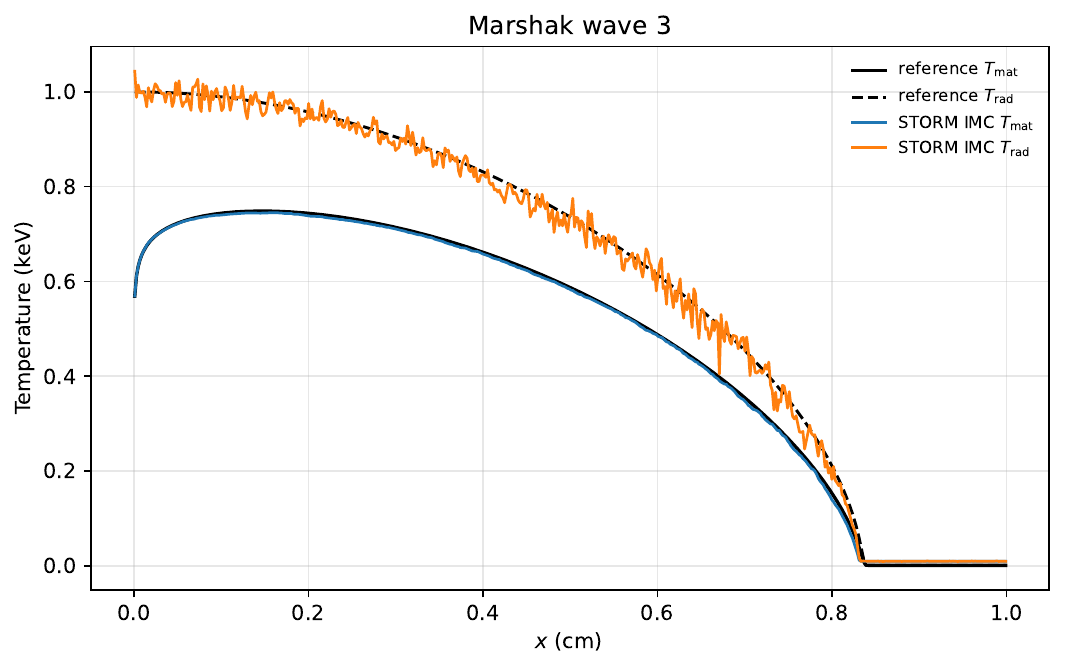}{0.47\textwidth}{(c) Test~3, $\rho \propto x^{20/19}$}
              \fig{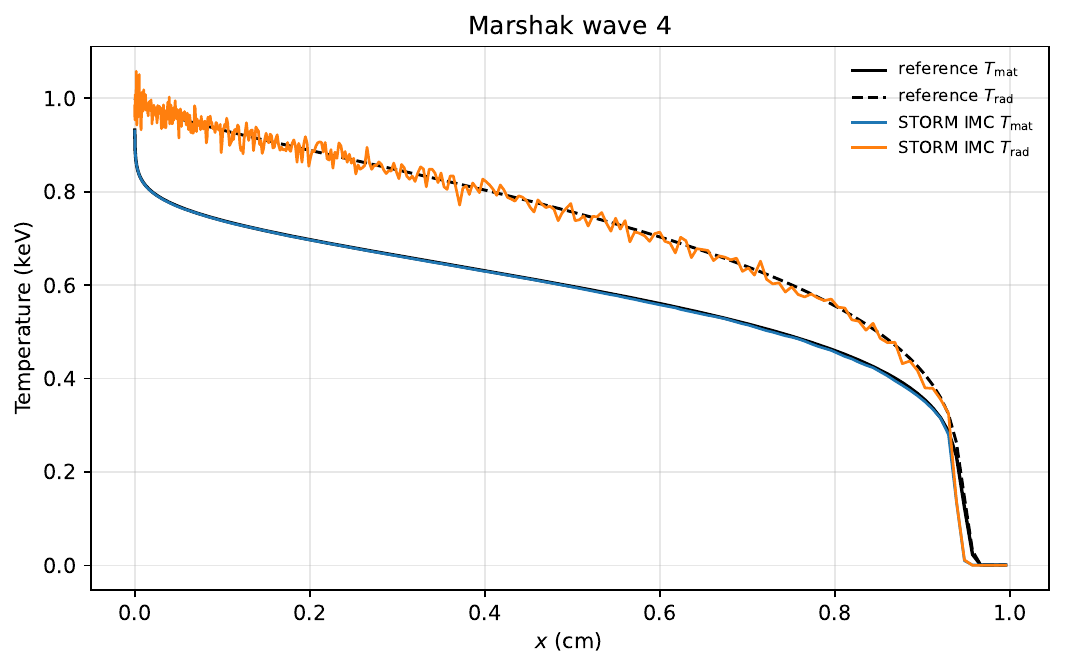}{0.47\textwidth}{(d) Test~4, $\rho \propto x^{-40/139}$}}
    \caption{Material and radiation temperature profiles at $t = 1$~ns for the four Marshak-wave
    variants of Table~\ref{tab:marshak_variants}.  Black curves are the published
    semi-analytic self-similar solutions of \citet{KriefMcClarren2024} (panels a and b) and
    \citet{Derei2024} (panels c and d); coloured curves are the \STORM{} \IMC{} results.  Panel~(b)
    is the equilibrium case, in which the two temperatures coincide; the other three are
    non-equilibrium, and the separation between the material and radiation temperatures is
    itself part of the solution being tested.  The radiation temperature is the noisier of the
    two because it is reconstructed from the instantaneous packet census.}
    \label{fig:marshak_results}
\end{figure*}

\subsubsection{Results}
Figure~\ref{fig:marshak_results} compares the \IMC{} profiles at $t = 1$~ns with the
semi-analytic self-similar solutions published for each problem, those of
\citet{KriefMcClarren2024} for tests~1 and~2 and of \citet{Derei2024} for tests~3 and~4.

Averaged over the heated region---defined here as the points where the reference temperature
exceeds $5\%$ of its peak, which excludes the cold material ahead of the front---the mean relative
difference in material temperature is $0.4\%$, $1.1\%$, $1.6\%$, and $0.4\%$ for tests~1 to~4,
and $1.6\%$, $1.3\%$, $3.1\%$, and $1.7\%$ in radiation temperature.  The wave front, taken where
the material temperature falls to half its peak, is reproduced to within $0.4$, $0.4$, $2.5$, and
$4.9\times10^{-3}$~cm respectively, in every case below one per cent of the distance travelled.
The solver therefore reproduces both the propagation speed and the two-temperature structure of
the wave, including the strong material--radiation decoupling of tests~1, 3 and~4 and the
equilibrium limit of test~2.  Pointwise relative differences remain large in the immediate
vicinity of the front, where a sub-cell offset in front position divides by a steeply falling
reference value, and in the low-temperature tail, where the census-based radiation temperature is
dominated by Monte Carlo noise.

\subsection{Cylindrical Hohlraum}\label{sec:hohlraum}

The hohlraum benchmark is a two-dimensional axisymmetric problem that tests radiation transport
through a complex, heterogeneous geometry with large opacity contrasts
\citep{McClarrenUrbatsch2009,SteinbergHeizler2022}.

\subsubsection{Problem setup}
The problem is specified in $RZ$ cylindrical geometry and consists of an evacuated cylindrical
cavity (the hohlraum) with absorbing walls and internal structures.  We implement this as a full
three-dimensional simulation by revolving the $RZ$ cross-section around the symmetry axis ($x$).
The problem occupies $x \in [0, 1.4]$~cm, $r = \sqrt{y^2 + z^2} \leq 0.65$~cm; the mesh fills
the box $x \in [0, 1.46]$~cm, $y, z \in [-0.71, 0.71]$~cm, the padding outside the problem being
vacuum, so that no material interface coincides with a domain face.

Four material regions define the absorbing structures:
\begin{itemize}
    \item Left wall: $x \in [0.10, 0.15]$, $r \leq 0.45$~cm.
    \item Capsule: $x \in [0.55, 0.95]$, $r \leq 0.45$~cm.
    \item Right end cap: $x \in [1.35, 1.40]$, $r \leq 0.65$~cm.
    \item Outer cylindrical wall: $x \in [0.10, 1.40]$, $r \in [0.60, 0.65]$~cm.
\end{itemize}
The absorbing material has opacity $\kappa_a = 300\, (T/\mathrm{keV})^{-3}$~cm$^{-1}$ and volumetric heat
capacity $\cV = 3 \times 10^{15}$~erg\,keV$^{-1}$\,cm$^{-3}$.  The vacuum regions have a negligible
opacity of $10^{-20}$~cm$^{-1}$ and a heat capacity
$\cV = 10^{15}$~erg\,keV$^{-1}$\,cm$^{-3}$; with that opacity they exchange no appreciable energy
with the radiation, so their heat capacity plays no role.  A blackbody source at $T = 1$~keV drives the left $x$-boundary.  The
initial temperature is 300~K and the simulation runs to $t = 10$~ns.

\begin{figure*}
  \gridline{\fig{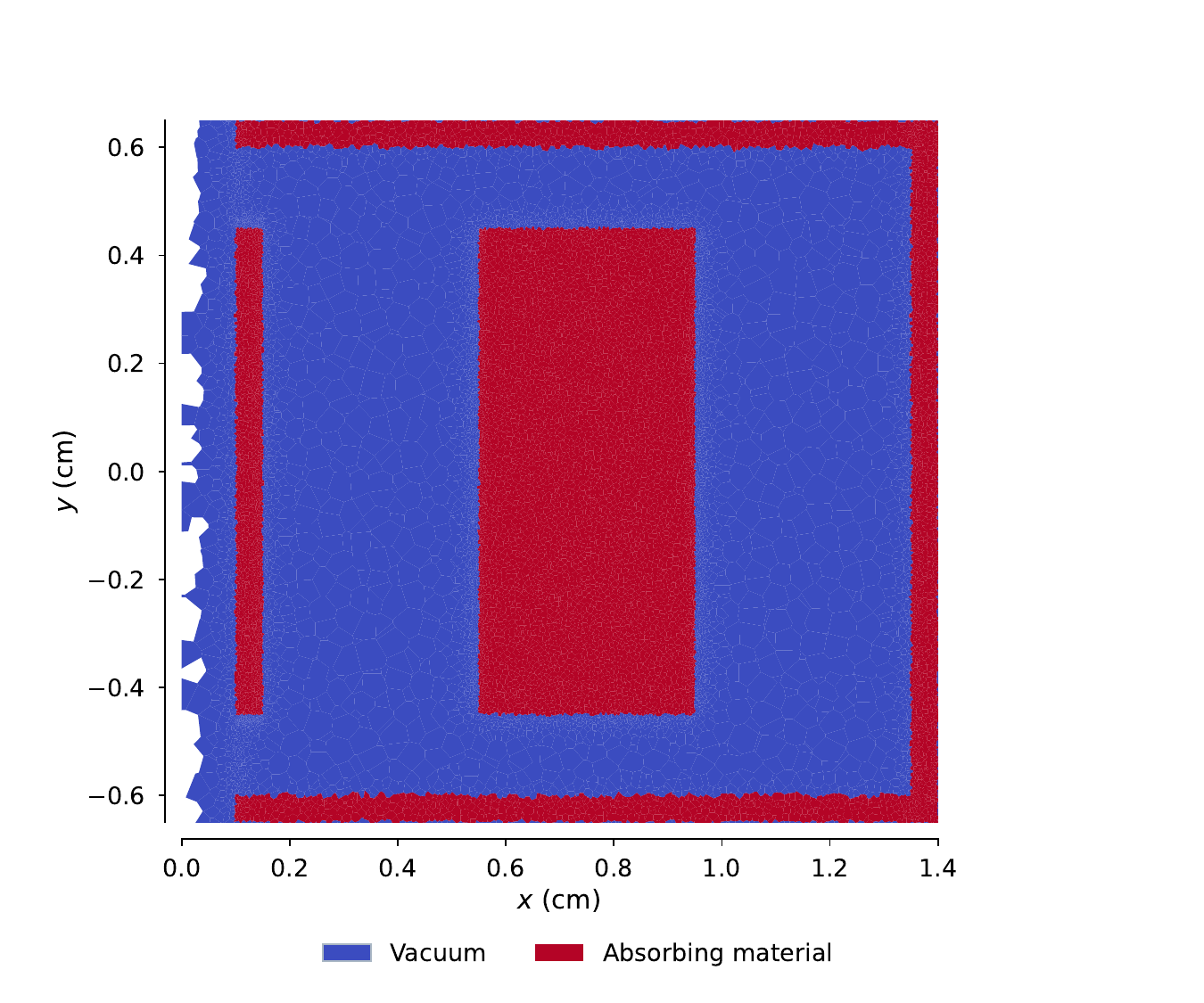}{0.46\textwidth}{(a) Material-type field}
            \fig{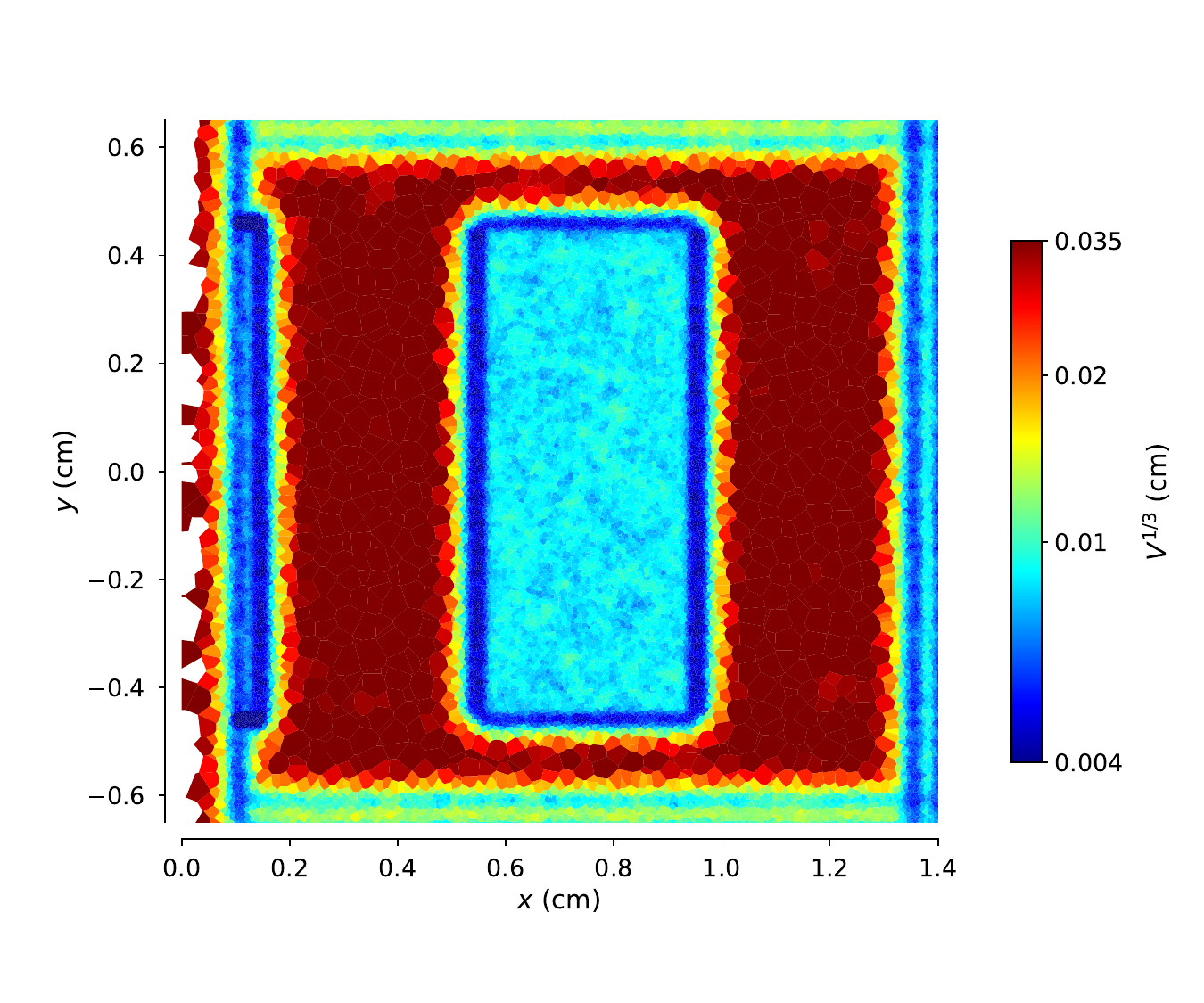}{0.46\textwidth}{(b) Voronoi-cell size}}
  \gridline{\fig{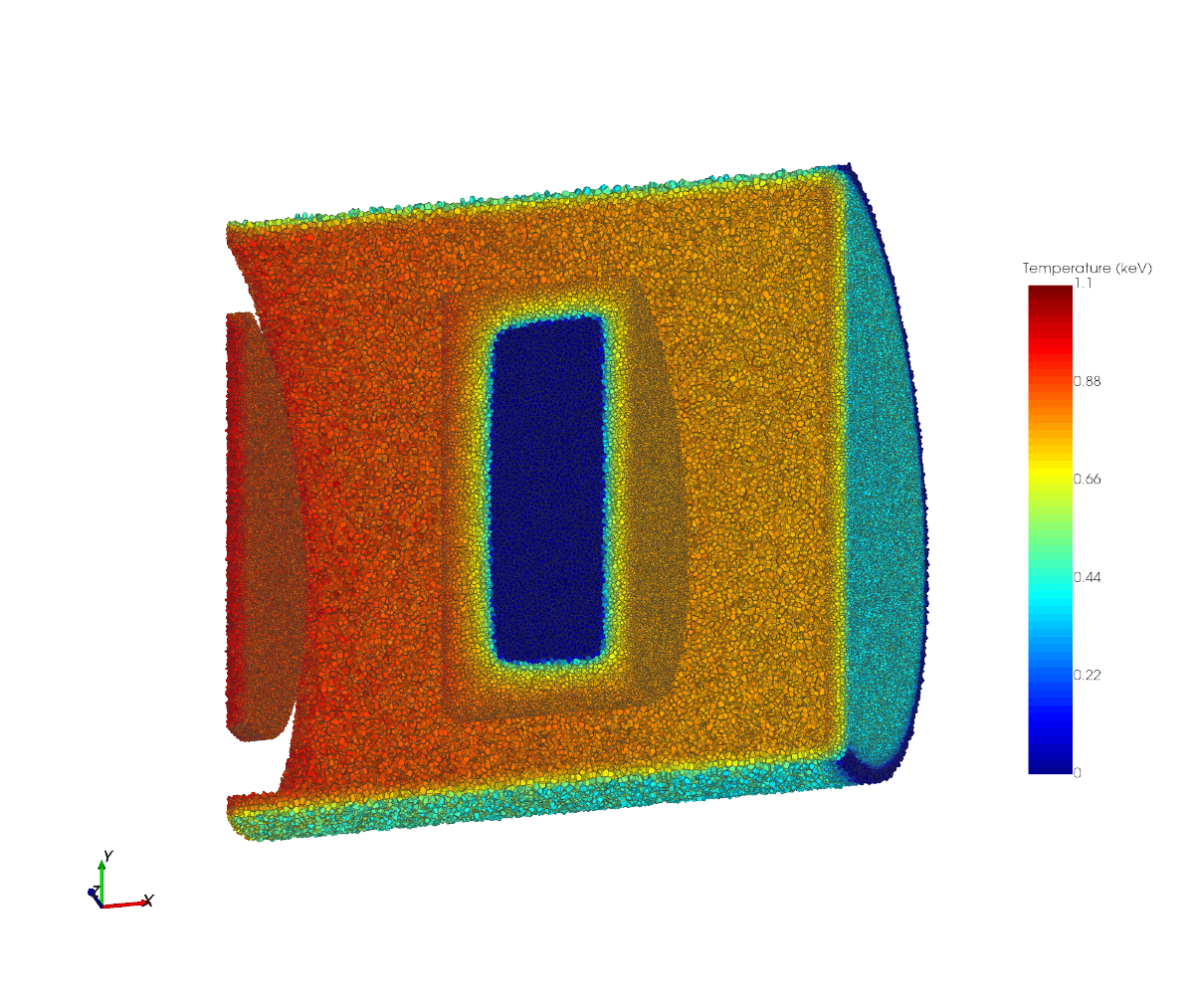}{0.46\textwidth}{(c) Three-dimensional tessellation}
            \fig{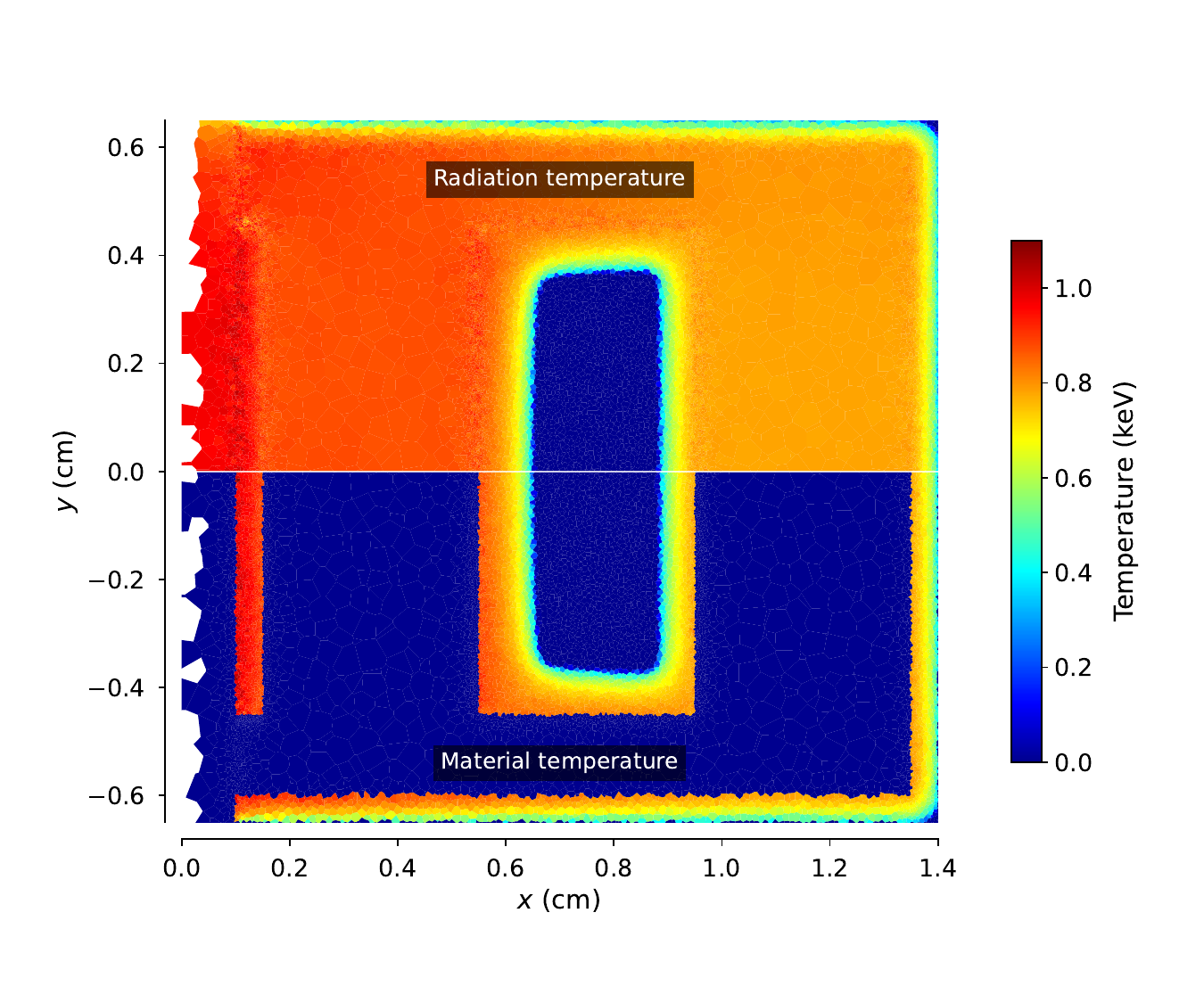}{0.46\textwidth}{(d) Radiation and material temperatures}}
  \caption{Setup of the cylindrical hohlraum benchmark of \citet{McClarrenUrbatsch2009}, shown
  for the region $x \in [0,\,1.4]$\,cm and $r = \sqrt{y^2+z^2} \le 0.65$\,cm, together with the
  temperature fields at the end of the run.  All panels use the $2{,}450{,}000$-cell mesh.
  (a)~Material layout in the central $z=0$ slice: the absorbing regions (left wall, capsule,
  right end-cap, and outer cylindrical shell) and the vacuum.
  (b)~Voronoi-cell size $V^{1/3}$ in the same slice on a logarithmic colour scale from $0.004$ to
  $0.035$~cm (values outside this range saturate at the end colours); the mesh is refined near
  the material--vacuum interfaces.
  (c)~Three-dimensional cutaway of the absorbing cells in $z<0$, with the cell edges drawn and
  the material temperature (keV) in colour.
  (d)~Radiation temperature (upper half) and material temperature (lower half) in the central
  slice, in keV.  Panels~(c) and~(d) are taken at the end of the run, $t=10$~ns; the
  radiation temperature is derived from the time-averaged radiation energy density over the
  final step.}
  \label{fig:hohlraum_setup}
\end{figure*}

The Voronoi mesh is generated with refinement near the material interfaces: the mesh-generating
points are concentrated at the boundaries between absorbing and vacuum regions to resolve the sharp
opacity gradients.  We use $N_{\mathrm{base}} = 50{,}000$ base points and interface refinement
for a total of $2{,}450{,}000$ Voronoi cells, distributed over $512$ MPI ranks.
Each time step samples $5$ new thermal photon packets per cell; comb population control uses a
floor of $N_{\min} = 15$ packets per cell and a budget factor $\phi_{\mathrm{comb}} = 6$
(Eq.~\ref{eq:comb_budget}), i.e.\ a mean of $90$ and a cap of $300$ packets per cell.  The driven boundary samples $1000$ packets per boundary cell.
The time step starts at $10^{-11}$~s and, after $t = 1$~ns, grows by a factor of $1.05$ per step
up to $5 \times 10^{-11}$~s.  Profiles and field snapshots are saved every $25$ cycles and at
the final time.
No hydrodynamics is applied.  Figure~\ref{fig:hohlraum_setup} shows the
material layout, the resulting mesh spacing, the three-dimensional tessellation, and a
representative temperature slice.

\subsubsection{Results}

\begin{figure}
    \centering
    \includegraphics[width=\columnwidth]{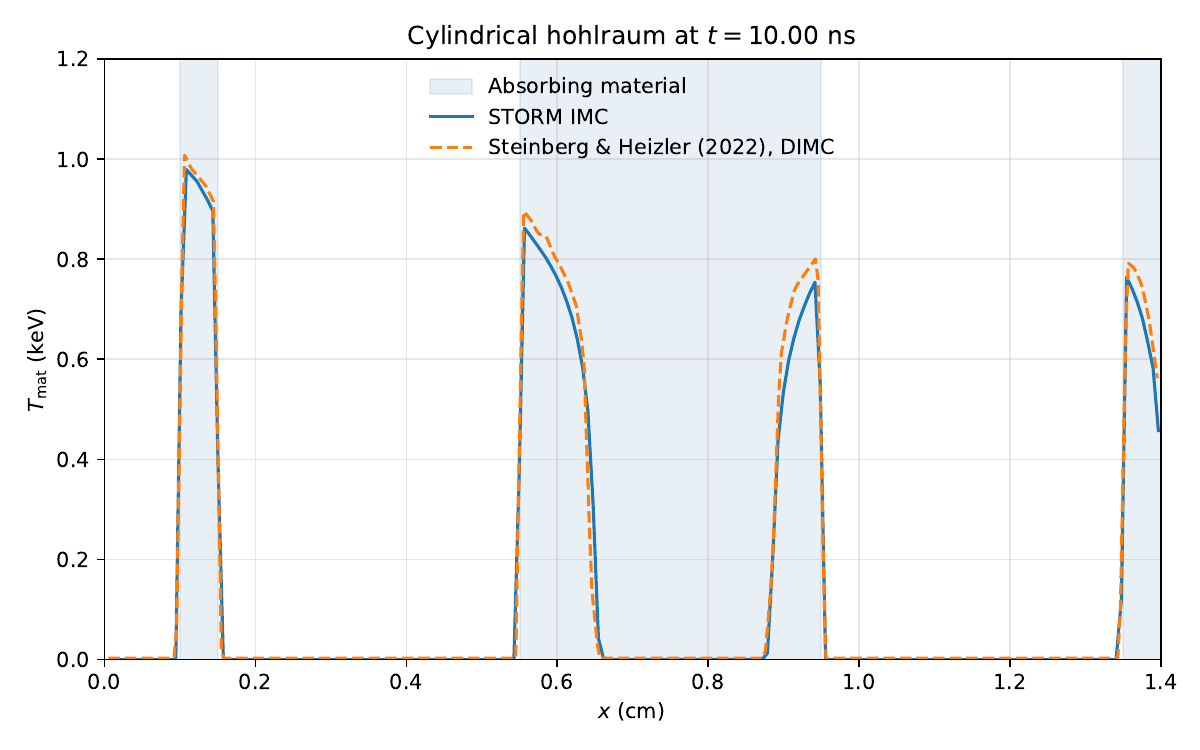}
    \caption{Material temperature in the cylindrical hohlraum at $t = 10$~ns
    from the $2{,}450{,}000$-cell \STORM{} \IMC{} run on $512$ MPI ranks.
    The blue solid curve shows the arithmetic mean over cells with
    $|r - 0.05\,\mathrm{cm}| < 0.03$~cm in $200$ uniform axial bins.
    The orange dashed curve is the digitized DIMC result of
    \citet{SteinbergHeizler2022}, Figure~5(e), at $r = 0.05$~cm and $t = 10$~ns.
    Shaded bands mark the absorbing-material regions: left wall, capsule, and right end-cap.
    The left boundary is driven by a $1$~keV blackbody source.}
    \label{fig:hohlraum_results}
\end{figure}

Fig.~\ref{fig:hohlraum_results} shows the material temperature near $r = 0.05$~cm at
$t = 10$~ns as a function of the axial coordinate~$x$, the radius at which
\citet{SteinbergHeizler2022} report their profile.  Each of $200$ uniform axial bins over
$x \in [0,1.4]$~cm takes the arithmetic mean of the temperatures of the cells whose generating
points lie within $|r - 0.05\,\mathrm{cm}| < 0.03$~cm; the mean is not volume weighted, so in bins
that straddle a material interface the more finely refined side carries more weight.
The radiation wave, driven by the $1$\,keV blackbody source at $x = 0$, propagates through the
vacuum channel and heats the absorbing walls and capsule progressively.  These profiles are in
good agreement with the two-dimensional results of \citet{SteinbergHeizler2022} and
\citet{McClarrenUrbatsch2009}, confirming that the 3D Voronoi-mesh IMC solver reproduces the
expected radiation transport physics.

\subsection{Mach~2 Radiative Shock}\label{sec:mach2}

The Mach~2 radiative shock is a coupled radiation--hydrodynamics benchmark that tests the
interaction between a moderately strong shock and the radiation field in a regime where the
radiation energy density is comparable to the material energy density
\citep{LowrieEdwards2008,SteinbergHeizler2022}.

\subsubsection{Problem setup}
The gas is ideal with ratio of specific heats $\gamma = 5/3$ (distinct from the Lorentz
factor of Section~\ref{sec:transport_eq}) and specific heat capacity $c_v = 1.91 \times 10^8$~erg\,g$^{-1}$\,K$^{-1}$.  The
absorption opacity follows a power law,
$\kappa_a = 0.362\, \rho\, (T/\mathrm{keV})^{-3.5}$~cm$^{-1}$, with no elastic scattering.  The
calculation is set up in the frame in which the unshocked gas is at rest.  The initial condition
is the relaxed Lowrie--Edwards semi-analytic profile itself, precursor included, interpolated onto
the cell centers with the shock at $x = 0$: shocked (downstream) gas fills $x < 0$ and moves
toward $+x$ behind the shock, and unshocked (upstream) gas fills $x > 0$.  The asymptotic states,
with $v_x$ the velocity component along $x$, are
\begin{center}
\begin{tabular}{lll}
    & Downstream ($x<0$) & Upstream ($x>0$) \\
    \hline
    $\rho$ (g/cc)   & 2.29 & 1.0 \\
    $T$ (keV)       & 0.253 & 0.122 \\
    $v_x$ (cm/s)    & $1.95 \times 10^7$ & 0 \\
\end{tabular}
\end{center}
The domain is $x \in [-0.21, 0.25]$~cm, resolved with 1024 cells, and the simulation runs to
$t = 5$~ns.  The jump conditions give a shock speed $D = \rho_2 u_2/(\rho_2-\rho_1) \simeq
3.46\times10^{7}$~cm\,s$^{-1}$, with $\rho_{1,2}$ the upstream and downstream densities and $u_2$ the
downstream velocity, so in this time the shock advances $D\,t \simeq 0.17$~cm, about $380$ cells
and $38\%$ of the domain.  The Eulerian run uses a fixed time step $\Delta t = 0.3\,\Delta x/\max(|v|+c_s)$,
evaluated from the two asymptotic states in the shock-front frame, where $c_s$ is the sound
speed; the Quasi Lagrangian run uses an adaptive hydrodynamic CFL step with Courant number $0.3$,
capped at that fixed step.  Boundary conditions inject blackbody radiation at the downstream and
upstream temperatures.  The runs use $16$ MPI ranks.  Each run emits $25$ new packets per cell per
step, starts with $50$ packets per cell, injects $50$ packets per boundary cell per step, and uses
comb population control with $N_{\min} = 100$ and $\phi_{\mathrm{comb}} = 10$
(Eq.~\ref{eq:comb_budget}).

The hydrodynamic step uses the HLLC Riemann solver with linear reconstruction, and the radiation
step is operator-split from it and solved with \IMC{}.  The problem is run on two meshes that
start from the same $1024$ cells and the same initial profile.

In the first, the Eulerian run, the mesh is fixed.  A captured hydrodynamic shock is then spread
over $6$--$7$ cells at every resolution we tested.  The gas radiates away part of its compressive
heating while the shock is still crossing a cell, before the jump is complete, which clips the
Zel'dovich spike.  This is a numerical effect of the smeared shock, not of the radiation solver.

In the second, the Quasi Lagrangian run, each mesh-generating point moves with the local fluid
velocity along $x$, with a small correction that keeps each mesh-generating point close to the center of mass of its cell.  The shock then
remains confined to a few cells instead of advecting through the grid, so the jump is resolved
before the gas has time to radiate its compressive heating away.  The outer domain walls stay fixed
while the interior points move, and the cell at the left wall therefore stretches as shocked gas
flows into the domain.  Adaptive mesh refinement splits this cell whenever its mass exceeds a few
times the initial cell mass, which increases the cell count from $1024$ to $1228$ by $t = 5$~ns.  At
the right wall the unshocked gas is at rest and no refinement is needed.

\begin{figure*}
    \centering
    \includegraphics[width=0.95\textwidth]{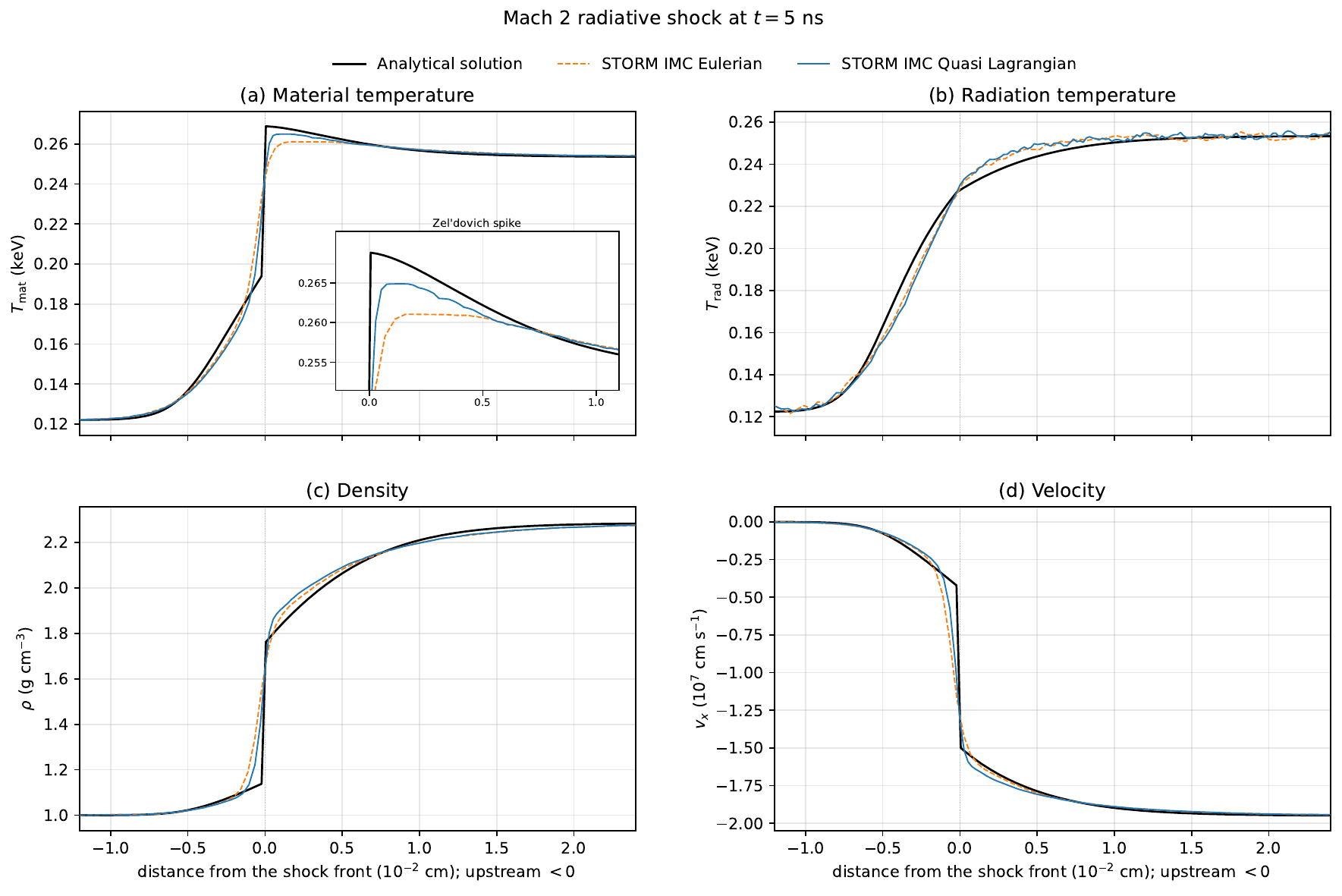}
    \caption{Mach~2 radiative shock at $t \simeq 5$\,ns, plotted against the distance $x_s - x$
    behind the shock position $x_s$ (upstream at negative distance): material temperature,
    radiation temperature, density, and $x$-velocity.  The velocity is that of the
    upstream-rest frame of the run, with its sign reversed to match the reversed axis.  The black curves show the Lowrie--Edwards semi-analytic solution
    \citep{LowrieEdwards2008}; orange dashed and blue solid curves show \STORM{} \IMC{}
    on Eulerian and Quasi Lagrangian meshes, respectively.  Both runs start with $1024$ cells
    and use $25$ new photon packets per cell per step, with comb population control at a floor of
    $N_{\min} = 100$ packets per cell and a budget factor $\phi_{\mathrm{comb}} = 10$
    (Eq.~\ref{eq:comb_budget}).  Boundary refinement increases the Quasi Lagrangian mesh to $1228$
    cells at the final time.  The inset in panel~(a) resolves the Zel'dovich spike: the
    Quasi Lagrangian run reaches $0.2649$\,keV and the Eulerian run $0.2611$\,keV, compared
    with the analytic peak of $0.2688$\,keV.  Relative to the $0.2530$\,keV downstream plateau,
    these correspond to $75\%$ and $51\%$ of the analytic spike amplitude.}
    \label{fig:mach2_results}
\end{figure*}

\subsubsection{Results}
Figure~\ref{fig:mach2_results} compares the Eulerian and Quasi Lagrangian \STORM{} \IMC{}
profiles at $t \simeq 5$\,ns to the Lowrie--Edwards semi-analytic solution, all replotted in
 the shock-front frame.  The upstream and downstream plateaus in material temperature,
radiation temperature, density, and velocity agree with the reference, and the solver reproduces the Zel'dovich spike---the local overshoot in material
temperature and density just ahead of the embedded hydrodynamic shock that is the hallmark of this
Mach number.  The inset quantifies why the Lagrangian mesh matters for this benchmark: at matched initial
$1024$-cell resolution and photon-sampling settings, the Eulerian mesh recovers only $51\%$ of the analytic spike amplitude
because the shock is spread over several cells and radiates part of its compressive heating away
before the jump completes, while advecting the mesh-generating points with the fluid confines the
shock to a few cells and recovers $75\%$ of the spike.  Exact agreement with the Lowrie--Edwards
curve is not expected: that solution is derived in the grey non-equilibrium diffusion
approximation, whereas \IMC{} solves the full transport equation, whose solution for the same
problem differs slightly from the diffusion one.  This illustrates the benefit of coupling \IMC{} to a moving
mesh: at the same initial cell count, quasi-Lagrangian motion sharpens hydrodynamic features that a
fixed grid smears.  The agreement also shows that the operator-split radiation--hydrodynamics
coupling correctly captures the interaction between the radiative precursor and the shock in a
regime where the radiation and material energy densities are comparable.

\subsection{Mach~45 Radiative Shock}\label{sec:mach45}

The Mach~45 radiative shock extends the previous test to a strongly radiation-dominated regime
in which the radiation precursor extends far ahead of the shock, heating the upstream gas well
above its initial temperature, and the post-shock radiation temperature exceeds the upstream
temperature by nearly two orders of magnitude \citep{LowrieEdwards2008,SteinbergHeizler2022}.

\subsubsection{Problem setup}
The gas has ratio of specific heats $\gamma = 5/3$ and specific heat capacity
$c_v = 1.45 \times 10^{15} / T_{\mathrm{1keV}}$
erg\,g$^{-1}$\,K$^{-1}$, where $T_{\mathrm{1keV}}=(1~\mathrm{keV})/k_B
\simeq 1.16045\times10^7$~K is the fixed temperature conversion factor.  The opacity includes both absorption and elastic scattering:
\begin{align}
    \kappa_a &= 0.0142\, \rho^2\, (T/\mathrm{keV})^{-3.5}~\mathrm{cm}^{-1}, \\
    \kappa_s &= 0.4006\, \rho~\mathrm{cm}^{-1}
\end{align}
The calculation is run in the shock rest frame and, as for the Mach~2 problem, is initialized
from the relaxed Lowrie--Edwards semi-analytic profile, precursor included, interpolated onto the
cell centers.  The asymptotic states, with $v_x$ the velocity component along $x$, are
\begin{center}
\begin{tabular}{lll}
    & Upstream & Downstream \\
    \hline
    $\rho$ (g/cc) & 1.0 & 6.43 \\
    $T$ (keV)     & 0.1 & 8.36 \\
    $v_x$ (cm/s)  & $5.71 \times 10^8$ & $0.89 \times 10^8$ \\
\end{tabular}
\end{center}
The domain is $x \in [1950, 2450]$~cm with 4000 cells ($\Delta x = 0.125$~cm), run on 64 MPI
ranks.  The shock is initially placed at $x = 2300$~cm and the simulation runs to
$t = 0.8$~$\mu$s.  The shock is stationary in this frame; the upstream gas crosses it at
$5.71\times10^{8}$~cm\,s$^{-1}$ and so travels $\simeq 457$~cm through it during the run, about
$91\%$ of the domain and nearly twice the $\sim 250$~cm extent of the radiation precursor.  The run emits $12$ new packets per cell per step, starts with $50$ packets per
cell, injects $50$ packets per boundary cell per step, and uses comb population control with
$N_{\min} = 100$ and $\phi_{\mathrm{comb}} = 5$ (Eq.~\ref{eq:comb_budget}).  The time step is
$0.3$ of the Courant step, starting at $10^{-3}$ of it and growing by $1\%$ per cycle from
cycle $750$.

The strong elastic scattering opacity makes this an excellent test of the effective-scattering
treatment and the interaction between elastic scattering and the Fleck linearization.

\begin{figure*}
    \centering
    \includegraphics[width=\textwidth]{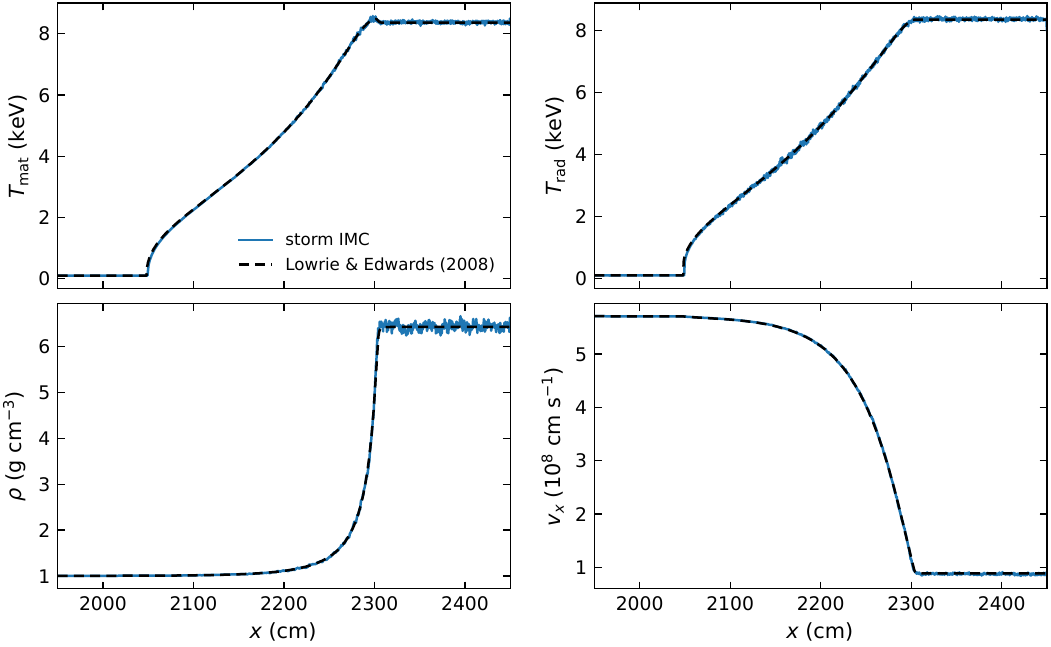}
    \caption{Mach~45 radiative shock at $t = 0.80$\,\textmu s, compared with the Lowrie--Edwards
    semi-analytic solution \citep{LowrieEdwards2008} for this benchmark's opacity law, as used by
    \citet{SteinbergHeizler2022} in their Figure~9(b).  Material temperature, radiation temperature,
    density, and $x$-velocity all lie on the analytic curve across the full radiation precursor,
    which extends roughly $250$\,cm ahead of the embedded shock, and through the strongly
    radiation-dominated post-shock plateau.  The self-similar solution is translation-invariant in
    $x$, and the \STORM{} profile (solid) is plotted shifted by the best-fit offset $A = +0.06$\,cm
    onto the unshifted analytic solution (dashed), less than half a cell ($\Delta x = 0.125$\,cm),
    i.e.\ within the resolution with which the captured shock can be located.}
    \label{fig:mach45_results}
\end{figure*}

\subsubsection{Results}
Figure~\ref{fig:mach45_results} shows the \STORM{} \IMC{} profile at $t = 0.8$\,\textmu s, compared
with the Lowrie--Edwards semi-analytic solution used
by \citet{SteinbergHeizler2022} in their Figure~9(b).  All four quantities lie on the analytic
solution across the precursor and through the shock jump, including the strongly
radiation-dominated plateau where the post-shock radiation temperature is nearly two orders of
magnitude above the upstream value.  Matching the two profiles requires only a small translational
offset, $A = 0.06$\,cm, along the self-similar coordinate, less than half a cell; the solution
started from the analytic profile therefore stays on it, with the shock drifting by less than half
a cell over the run and no change in the shape of the wave.
The level of agreement in a regime with both elastic scattering and absorption confirms that the
effective opacities of Eqs.~\eqref{eq:eff_abs}--\eqref{eq:eff_scat} remain accurate when elastic
scattering dominates.

\subsection{Reinicke--Meyer-ter-Vehn Radiative Shock}\label{sec:rmtv}

The Reinicke--Meyer-ter-Vehn (RMTV) problem \citep{ReinickeMeyerterVehn1991} is a self-similar,
spherically symmetric radiative shock driven by strongly temperature-dependent thermal
conduction. Unlike the planar Mach~2 and Mach~45 tests, it exercises the code in a genuinely
three-dimensional, spherical geometry, and the radiative heat exchange is purely conductive, so the
optically thick cells are treated with \DDMC{}.  It therefore provides an independent check of the
\DDMC{} coupling under hydrodynamic motion.

\subsubsection{Problem setup}
The problem is described by the Euler equations with nonlinear heat conduction,
\begin{multline}\label{eq:rmtv_target}
    \rho \frac{De}{Dt} + p\, \nabla\!\cdot\!\mathbf{v} = \nabla\!\cdot\!\bigl(\chi \nabla T\bigr), \\
    p = (\gamma-1)\rho e, \qquad e = C_{V\!,s} T,
\end{multline}
where $e=U_m/\rho$ is the specific material internal energy, $p$ is the gas pressure,
$D/Dt=\partial/\partial t+\mathbf{v}\cdot\nabla$ is the material derivative, $C_{V\!,s}$ is the
constant specific heat capacity per unit mass, and $\gamma=5/4$ is the adiabatic index.  The
thermal conductivity $\chi = \chi_0\, \rho^{-2} T^{13/2}$, with constant $\chi_0$, increases
steeply with temperature.

These equations contain no radiative force.  We therefore disable the radiation momentum
coupling and the moving-material frame corrections in the transport (the Doppler factor, the
Lorentz boosts, and the factor $\gamma$ in the Fleck factor), while the thermal exchange between
radiation and material remains fully active.  The hydrodynamic step evolves the density, velocity,
and pressure independently, and after each Monte Carlo step the gas total energy is synchronized
with the updated material internal energy at unchanged momentum.  In the diffusion limit this
thermal exchange reduces to Eq.~\eqref{eq:rmtv_target} for a grey, purely absorbing opacity
\begin{equation}\label{eq:rmtv_opacity}
    \Sigma = \frac{4 a_R c}{3 \chi_0}\, \rho^2\, T^{-7/2},
\end{equation}
where $\Sigma=\kappa_a$ is the macroscopic absorption opacity.  With this choice the radiation
energy flux $\mathbf{F} = -(c/3\Sigma)\nabla E_{\mathrm{rad}} = -[4 a_R c T^3/(3\Sigma)]\nabla T$
equals the conductive flux $-\chi\nabla T$ of Eq.~\eqref{eq:rmtv_target}.  Elastic scattering is switched off and a single energy
group is used.

The problem is solved in the octant $[0,1.2]^3$, discretized on a Cartesian Voronoi mesh of $n^3$
cells with $n \in \{16,24,32,48,96\}$ cells per coordinate direction.  The hydrodynamics and the
radiation are both reflected at the three symmetry planes and at the three outer walls, which lie
beyond the final heat front.  Lengths, times, and temperatures are scaled by units of $1$~cm,
$0.0256$~s, and $937.5$~K, respectively.  The heat front and the embedded shock are initially at
the spherical radii $r=0.45$ and $r=0.225$.  The initial mass, momentum, and total energy of each
cell follow the similarity solution in the heated interior and a cold, stationary ambient medium
with $\rho \propto r^{-19/9}$ outside it.  They are computed with an eight-point midpoint tensor
quadrature, recursively refined in the cells cut by either front.  The radiation is initialized in
equilibrium with the resulting cell temperature.  The initial material temperature is floored at
$1$~K, and the opacity is capped at $10^{8}$~cm$^{-1}$ in the cold exterior.  The reference solution is the
strong-conduction self-similar solution of \citet{ReinickeMeyerterVehn1991}, studied in detail by
\citet{Kamm2000}.  We obtain it by integrating the similarity equations using the RMTV solver in \textsc{ExactPack} \citep{ExactPack}, for $\gamma = 5/4$, the
conductivity $\chi \propto \rho^{-2}T^{13/2}$, and the ambient density $\rho \propto r^{-19/9}$
stated above, and we rescale it to the units used here.

The \DDMC{} treatment is applied in every cell with optical depth $\Sigma\,\bar{\ell} \ge 3$, where
$\bar{\ell}=4V/A_{\mathrm{surface}}$ is the cell mean chord (Section~\ref{sec:ddmc_eligibility}).
At each resolution the problem is run on a fixed (Eulerian) mesh and on a moving (Lagrangian) mesh
whose generating points move with the fluid velocity.  The time step, common to the hydrodynamic
and radiation steps, is the smaller of an acoustic CFL bound with Courant number $0.2$ and $0.005$
of the current physical age of the similarity solution.  We use an emission floor of
$N_{\mathrm{new}} = 16$ packets per cell, sample the initial radiation with $16$ packets per cell,
and apply comb population control with $N_{\min} = 32$ and $\phi_{\mathrm{comb}} = 1$
(Eq.~\ref{eq:comb_budget}).

The calculation starts at $t_0 = 4.82\times10^{-4}$~s, when the fronts are at the initial radii
given above, and runs to $t_{\mathrm{end}} = 1.312\times10^{-3}$~s, when the heat front and the
shock reach $r=0.90$ and $r=0.45$; the heat-front radius grows as $t^{9/13}$.  The error is measured as the volume-weighted $L_1$ norm relative to the analytic
solution, which is averaged conservatively over the volume of each cell so that the comparison
remains valid as the Lagrangian cells compress.

\begin{figure}
    \centering
    \includegraphics[width=0.95\columnwidth]{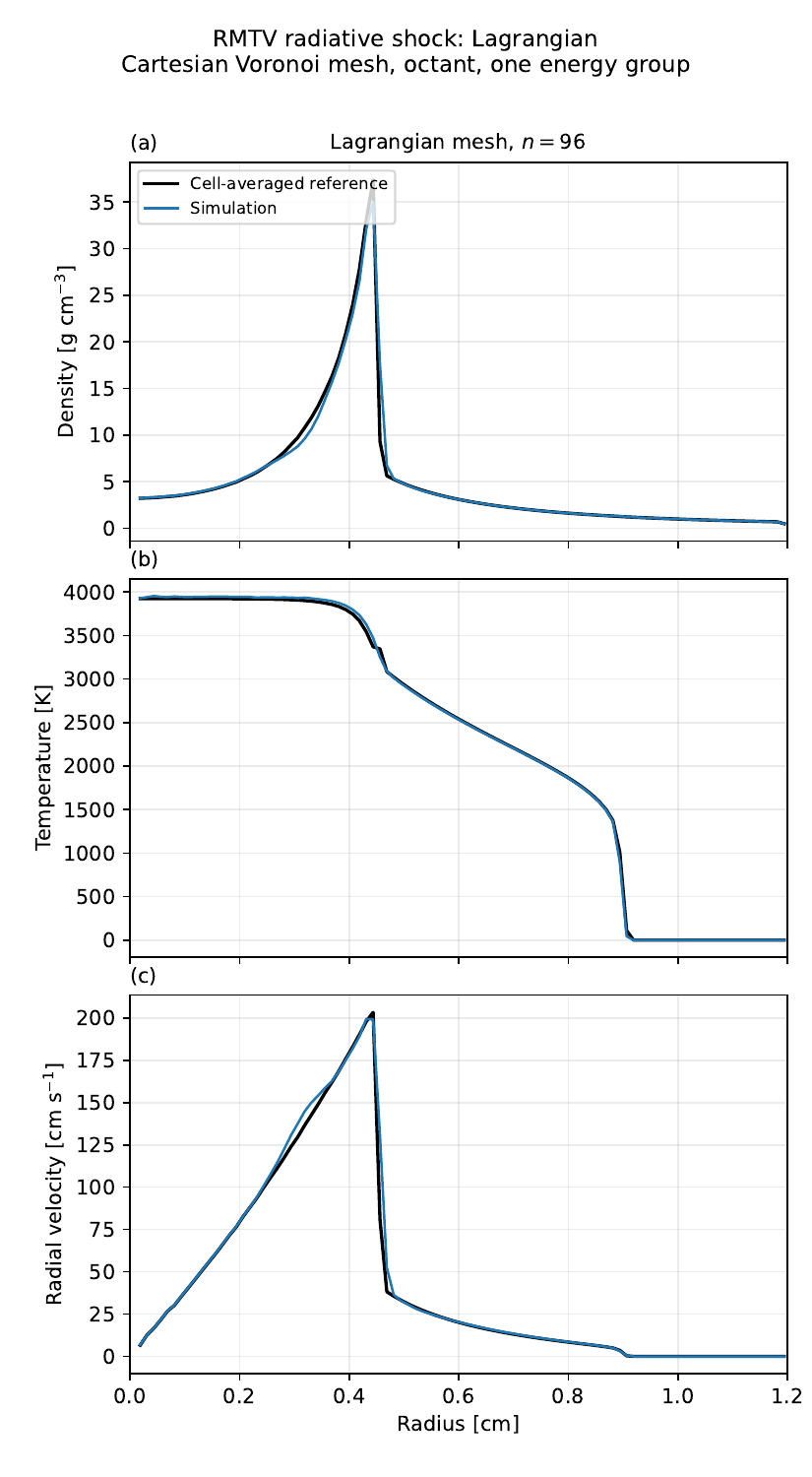}
    \caption{RMTV radiative shock at $t_{\mathrm{end}}$, Lagrangian mesh, the headline
    configuration of this benchmark: (a)~density, (b)~temperature, and (c)~radial velocity
    profiles at $n=96$ (blue) against the cell-averaged analytic reference (black).  Convergence
    with resolution and the Eulerian-mesh comparison, which is systematically worse at every
    resolution, are summarized in Table~\ref{tab:rmtv_l1} and discussed in
    Section~\ref{sec:rmtv}.}
    \label{fig:rmtv_results}
\end{figure}

\subsubsection{Results}
Table~\ref{tab:rmtv_l1} and Figure~\ref{fig:rmtv_results} summarize the volume-weighted $L_1$
errors at $t_{\mathrm{end}}$.  Temperature and radial velocity converge close to their expected
second and first order on both meshes (fitted orders $1.88$--$1.91$ and $0.98$--$1.24$
respectively), while density converges at only about half order on the Eulerian mesh and
about $0.7$ on the Lagrangian mesh.  This is not a scheme
deficiency: 85\% of the density error sits in the two cells straddling the shock, and it barely
improves under refinement because the reference itself is a \emph{cell-averaged} quantity that
sharpens as the cell size shrinks---between $n=48$ and $n=96$ the reference peak rises from
$31.96$ to $35.85$ (in code units) while its value one cell downstream falls from $9.75$ to
$5.64$. The captured numerical shock keeps a width of two to four cells regardless of resolution,
so the error volume shrinks while the local mismatch amplitude grows, giving the observed
$\sim 2^{0.45}$ improvement per doubling rather than first order; the shock and front radii, shell
mass, and global conservation are the more informative headline diagnostics for this benchmark.
Following the fluid helps throughout: Lagrangian $n=48$ ($L_1(\rho)=0.0499$) already beats
Eulerian $n=96$ ($L_1(\rho)=0.0557$) with one eighth of the cells; it ran in $13$~min on 64 cores
against $118$~min on 256 cores, about $14$ against $505$ core-hours, recovers the
Zel'dovich-like temperature spike that the Eulerian mesh clips, and places the shock peak on the
correct reference shell one cell earlier than the fixed mesh.  Mass and total energy are conserved
to round-off in every run, with relative drifts of $10^{-15}$--$10^{-14}$ and at most
$1.4\times10^{-13}$ (Eulerian $n=24$), confirming that the thermal-only
coupling used here (Eq.~\eqref{eq:rmtv_target}) does not introduce a spurious energy sink or
source under mesh motion.

\begin{table}
    \centering
    \caption{RMTV volume-weighted $L_1$ errors at $t_{\mathrm{end}}$ for density, temperature, and
    radial velocity on the Eulerian (E) and Lagrangian (L) meshes.}
    \label{tab:rmtv_l1}
    \small
    \setlength{\tabcolsep}{4pt}
    \begin{tabular}{ccccccc}
        \hline\hline
        & \multicolumn{2}{c}{$L_1(\rho)$} & \multicolumn{2}{c}{$L_1(T)$} & \multicolumn{2}{c}{$L_1(v_r)$} \\
        $n$ & E & L & E & L & E & L \\
        \hline
        16 & 0.139 & --    & 0.380 & --    & 0.431 & --    \\
        24 & 0.103 & 0.087 & 0.188 & 0.150 & 0.281 & 0.227 \\
        32 & --    & 0.064 & --    & 0.072 & --    & 0.159 \\
        48 & 0.078 & 0.050 & 0.048 & 0.019 & 0.135 & 0.088 \\
        96 & 0.056 & 0.034 & 0.013 & 0.011 & 0.075 & 0.041 \\
        \hline
    \end{tabular}
\end{table}

\subsection{Densmore Heterogeneous Step-Opacity}\label{sec:densmore}

The Densmore step-opacity problem tests multigroup transport through a heterogeneous medium with
sharply varying frequency-dependent opacities \citep{Densmore2012}.

\subsubsection{Problem setup}
The domain is $x \in [0, 3]$~cm with a Planck source at $T = 1$~keV on the left boundary.
The right boundary is reflective.  The opacity is frequency-dependent:
\begin{equation}
    \kappa_a(E) = \frac{\sigma_0}{\sqrt{k_B T}\, E^3},
\end{equation}
where $E=h\nu$ is photon energy, $k_B$ is Boltzmann's constant, $k_B T$ and $E$ are
expressed in keV, and the opacity normalization is $\sigma_0 = 10$~keV$^{3.5}$/cm for $x < 2$~cm and $\sigma_0 = 1000$~keV$^{3.5}$/cm for
$x \geq 2$~cm.  The equation of state is ideal gas with ratio of specific heats $\gamma = 1.4$ and
specific heat capacity $c_v = 10^{15} / T_{\mathrm{1keV}}$~erg\,g$^{-1}$\,K$^{-1}$, where
$T_{\mathrm{1keV}}=(1~\mathrm{keV})/k_B$ is the fixed temperature conversion factor.  The initial temperature is
$T_0 = 1$~eV, $\rho = 1$~g/cc.  The photon-energy range $[10^{-4}, 10^2]$~keV is divided into $G = 32$ logarithmically spaced
groups.  The domain is resolved with $256 \times 1 \times 1$
cells, $\Delta x = 0.01171875$~cm, and the simulation runs to $t = 1$~ns with $\Delta t = 5$~ps
(200 steps).  The problem is run twice on 32 MPI ranks, once with standard \IMC{} alone and
once with multigroup \DDMC{} acceleration at the default eligibility threshold
$\tau_{\min}^{\mathrm{cell}} = 15$ (Section~\ref{sec:ddmc_eligibility}); the two runs use the same
mesh, time step, and packet counts, so they isolate the effect of the acceleration.

No hydrodynamics is applied.  This benchmark validates the multigroup IMC implementation,
the frequency-dependent opacity evaluation, and the Planck-weighted emission and re-emission
sampling.  The sharp two-order-of-magnitude jump in $\sigma_0$ at $x = 2$~cm creates a strong
heterogeneity that tests the solver's ability to handle disparate optical depths within the same
simulation.

\begin{figure}
    \centering
    \includegraphics[width=\columnwidth]{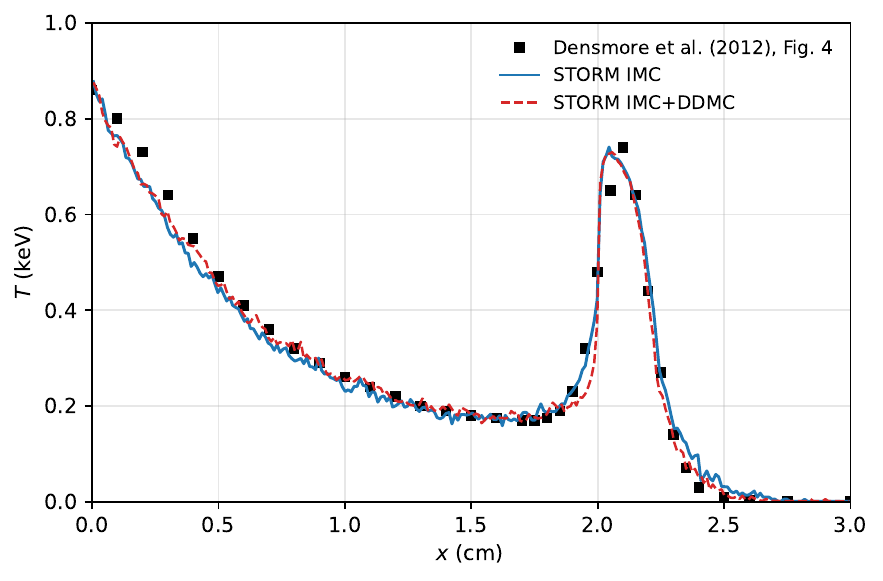}
    \caption{Material temperature at $t = 1$\,ns for the Densmore heterogeneous step-opacity
    problem with standard \IMC{} (solid) and with \DDMC{} acceleration (dashed), compared with
    the digitized Monte Carlo reference curve from Figure~4 of \citet{Densmore2012} (squares).
    The opacity jumps by two orders of magnitude at $x = 2$\,cm, producing the secondary heating
    bump just past the interface; the RMS temperature differences from the reference over the
    full profile are $0.032$\,keV for both runs.}
    \label{fig:densmore_results}
\end{figure}

\subsubsection{Results}
Figure~\ref{fig:densmore_results} compares the \STORM{} material-temperature profiles at
$t = 1$\,ns to the digitized Monte Carlo reference curve from Figure~4 of \citet{Densmore2012}.
The \IMC{} run reproduces the smooth cooling of the low-opacity region, the sharp heating bump
created as the wave crosses into the high-opacity material at $x = 2$\,cm, and the subsequent
falloff, with an RMS temperature difference of $0.032$\,keV over the full domain.  This confirms
that the multigroup Planck-weighted opacity evaluation (Section~\ref{sec:multigroup}) and the
frequency-resampling of effectively scattered packets correctly propagate a two-order-of-magnitude
opacity discontinuity.  With \DDMC{} enabled, the low-energy groups in the opaque cells are
diffused while the rest of the spectrum is transported, so the run exercises the partially
diffusive bands, group-exit channel, and \IMC{}--\DDMC{} interfaces of Section~\ref{sec:ddmc}.
The accelerated profile agrees with the reference to the same accuracy, an RMS difference of
$0.0315$\,keV against $0.0321$\,keV for \IMC{}, and with the \IMC{} profile to $0.022$\,keV RMS,
comparable to the Monte Carlo noise, while the wall-clock time on 32 ranks falls from $698$\,s
to $279$\,s.

\subsection{Olson Two-Dimensional Multimaterial Problem}\label{sec:olson}

The Olson complex two-dimensional problem \citep{Olson2020} drives a strongly
frequency-dependent, two-material geometry with a smooth interior volume source.  It exercises
continuous-frequency transport, sharp material interfaces, and a nonlinear ionizing equation of
state simultaneously, and reference solutions exist from both a filtered $P_N$ calculation
\citep{Olson2020} and a frequency-dependent DIMC calculation \citep{SteinbergHeizler2023}.

\subsubsection{Problem setup}
The domain is the first-quadrant square $x, y \in [0, 3.8]$~cm with reflecting boundaries on all
sides.  It is realized in \STORM{} as a Cartesian mesh extruded by $1$~cm in $z$ with reflecting
$z$ faces, so the solution is spatially two-dimensional while photon directions remain isotropic
over the full sphere rather than being confined to the $xy$ plane.  Three aluminum blocks are
embedded in a foam background:
\begin{itemize}
    \item $[0.5, 1.5] \times [0.5, 1.5]$~cm,
    \item $[1.5, 2.5] \times [1.5, 2.5]$~cm,
    \item $[1.5, 2.5] \times [0, 0.5]$~cm, a half-block cut by the reflecting symmetry axis.
\end{itemize}
Both materials have the same density, $\rho = 10^{-3}$~g\,cm$^{-3}$; the material identity is
carried explicitly in a cell tracer rather than inferred from the density.  Material and radiation
start in equilibrium at $T = 0.01$~keV.

An isotropic volume source is active throughout the calculation,
\begin{equation}\label{eq:olson_source}
    Q(r) = B(0.5~\mathrm{keV})\, e^{-18.7\, r^3},
\end{equation}
where $Q$ is the isotropic source amplitude, $r = \sqrt{x^2+y^2}$ is distance from the origin
in cm, and $B(T)=\int_0^\infty B_\nu(T)\,d\nu=c a_R T^4/(4\pi)$ is the frequency-integrated
Planck intensity.  This corresponds to an energy source density
$c\, a_R\, (0.5~\mathrm{keV})^4 e^{-18.7 r^3}$.  Source packets sample positions throughout each
cell, emission times uniformly within the step, and the Planck spectrum at $0.5$~keV; the spatial
profile is reproduced by stochastic allocation and packet weights rather than by a hard cutoff
radius.

The absorption opacity is a piecewise function of both photon energy and temperature with several
absorption edges, given by Eqs.~(5) and~(9) of \citet{SteinbergHeizler2023} restating
\citet{Olson2020}.  It is $\kappa_a(E,T) = \rho\, k(E,T)$, with $\rho$ the density and $k$ the mass
opacity in cm$^2$\,g$^{-1}$, where $E$ is the photon energy and $T$ the temperature, both in keV.
For aluminum
\begin{equation}\label{eq:olson_opacity_al}
    k_{\mathrm{Al}} =
    \begin{cases}
        \min\!\bigl(10^{7},\, 10^{8}\,T\bigr), & E < 0.01, \\
        a_{\mathrm{Al}}, & 0.01 \le E < 0.1, \\
        a_{\mathrm{Al}} + b_{\mathrm{Al}}, & 0.1 \le E < 1.5, \\
        a_{\mathrm{Al}}\sqrt{1.5/E} + c_{\mathrm{Al}}, & E \ge 1.5,
    \end{cases}
\end{equation}
where
\begin{equation}
    \begin{aligned}
        a_{\mathrm{Al}} &= \frac{10^{7}\,(0.01/E)^{2}}{1 + 20\,T^{3/2}}, \\
        b_{\mathrm{Al}} &= \frac{10^{6}\,(0.1/E)^{2}}{1 + 200\,T^{2}}, \\
        c_{\mathrm{Al}} &= \frac{10^{5}\,(1.5/E)^{5/2}}{1 + 1000\,T^{2}}
    \end{aligned}
\end{equation}
For the foam
\begin{equation}\label{eq:olson_opacity_foam}
    k_{\mathrm{foam}} =
    \begin{cases}
        \min\!\bigl(10^{7},\, 10^{9}\,T^{2}\bigr), & E < 0.008, \\
        a_{\mathrm{foam}}, & 0.008 \le E < 0.3, \\
        a_{\mathrm{foam}}\sqrt{0.3/E} + c_{\mathrm{foam}}, & E \ge 0.3,
    \end{cases}
\end{equation}
where
\begin{equation}
    \begin{aligned}
        a_{\mathrm{foam}} &= \frac{3\times10^{6}\,(0.008/E)^{2}}{1 + 200\,T^{3/2}}, \\
        c_{\mathrm{foam}} &= \frac{4\times10^{4}\,(0.3/E)^{5/2}}{1 + 8000\,T^{2}}
    \end{aligned}
\end{equation}
The opacity varies by
more than three orders of magnitude across the band.  The equation
of state is nonlinear and ionizing,
\begin{equation}\label{eq:olson_eos}
    U_m = a_R T_{\mathrm{1keV}}^4\, H \left[\theta + (\theta + \chi)\, \alpha(\theta, \chi)\right],
\end{equation}
where $U_m$ is material internal energy per unit volume,
$T_{\mathrm{1keV}}=(1~\mathrm{keV})/k_B$ is the fixed reference temperature,
$\theta=T/T_{\mathrm{1keV}}$ is dimensionless temperature, $H$ is a dimensionless heat-capacity
normalization, and $\chi$ is the ionization energy in units of $1$~keV.  The material parameters
are $(H, \chi) = (0.1, 0.1)$ for foam and $(0.5, 0.3)$ for aluminum.  The ionization fraction is
$\alpha(\theta,\chi)=2\sqrt{q}/(\sqrt{q+4}+\sqrt{q})$, with $q=\exp(-\chi/\theta)$.

The calculation is run as continuous-frequency transport rather than with banded groups: a single
bookkeeping band spans $10^{-7}$--$100$~keV, but every absorption is evaluated from the exact
$\kappa_a(E, T)$ at the packet energy.  Thermal emission and effective scattering sample
$\kappa_a(E,T) B(E,T)$, while the imposed source and the initial field sample $B(E,T)$, where
$B(E,T)=B_\nu(T)/h$ at $\nu=E/h$ is the Planck specific intensity per unit photon energy.  The
emission cumulative distributions and Planck means are tabulated with quadrature split at every
absorption edge, and verified independently against adaptive quadrature to a maximum relative
error of $1.0 \times 10^{-4}$.

The mesh is $380 \times 380$ cells---a multiple of $38$, so that every material interface at
$0.5$, $1.5$, and $2.5$~cm coincides with a cell face---and the time step is
$\Delta t = 10^{-13}$~s.  Snapshots are taken at $ct = 2$, $2.5$, and $3$~cm.  No hydrodynamics, elastic scattering, or \DDMC{}
is active, so this benchmark isolates the frequency-dependent \IMC{} kernel
itself.  The run used $256$ MPI ranks; because the source concentrates the work near the origin, a
static source-weighted cost function drives the domain decomposition, which changes cell ownership
only and never the mesh or the physical inputs.

\begin{figure*}
    \centering
    \includegraphics[width=0.98\textwidth]{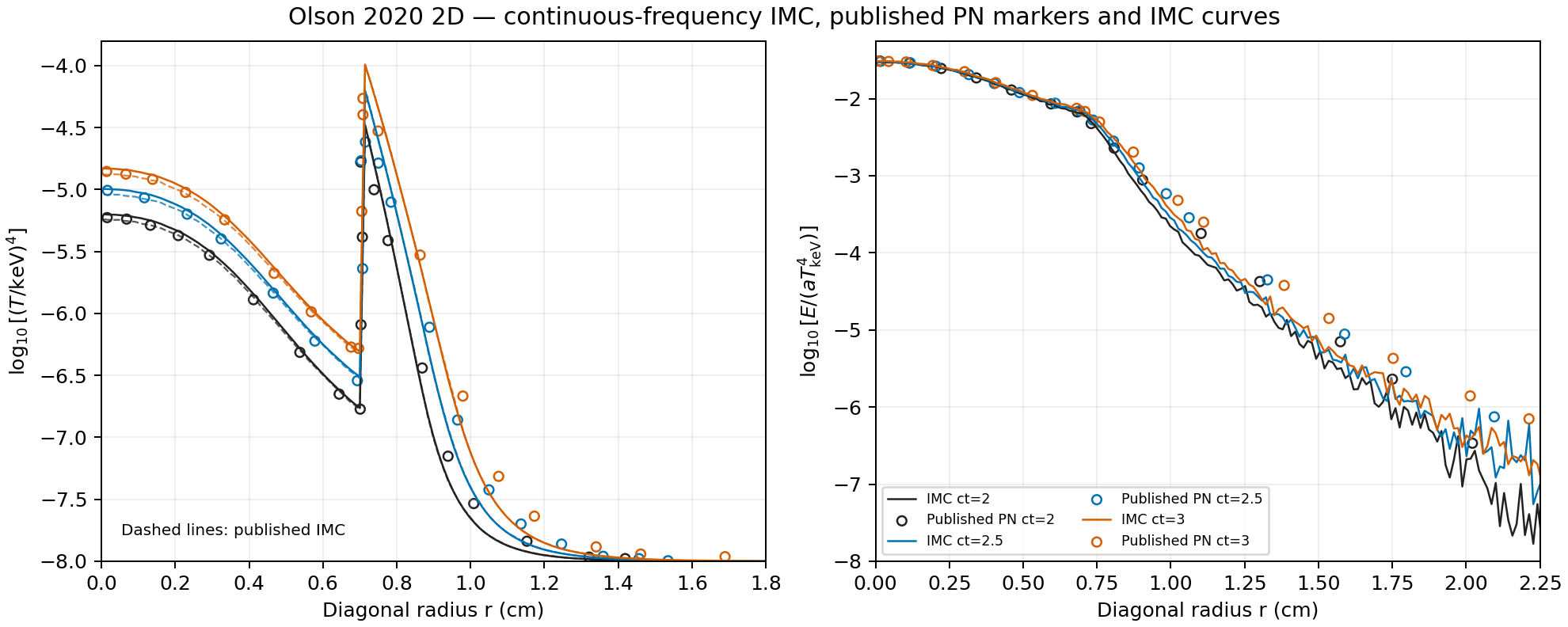}
    \caption{Diagonal profiles for the Olson two-dimensional problem at $ct = 2$, $2.5$, and
    $3$~cm, plotted against $r = \sqrt{x^2+y^2}$.  Left: material temperature as
    $\log_{10}(T/\mathrm{keV})^4$, with the \STORM{} continuous-frequency \IMC{} result (solid),
    the published $P_N$ markers of \citet{Olson2020} (open circles), and the published IMC curves
    of \citet{SteinbergHeizler2023} (dashed).  Right: scaled radiation energy density
    $\log_{10}[E_{\mathrm{rad}}/(a_R T_{\mathrm{1keV}}^4)]$, where $E_{\mathrm{rad}}$ is radiation
    energy density and $T_{\mathrm{1keV}}=(1~\mathrm{keV})/k_B$ is the fixed reference temperature.  The discontinuity at $r = \sqrt{0.5} \approx
    0.707$~cm is where the diagonal enters the first aluminum block; the solver captures the jump
    without smearing it across cells because the material interfaces coincide with mesh faces.}
    \label{fig:olson_profiles}
\end{figure*}

\begin{figure*}
    \centering
    \includegraphics[width=0.62\textwidth]{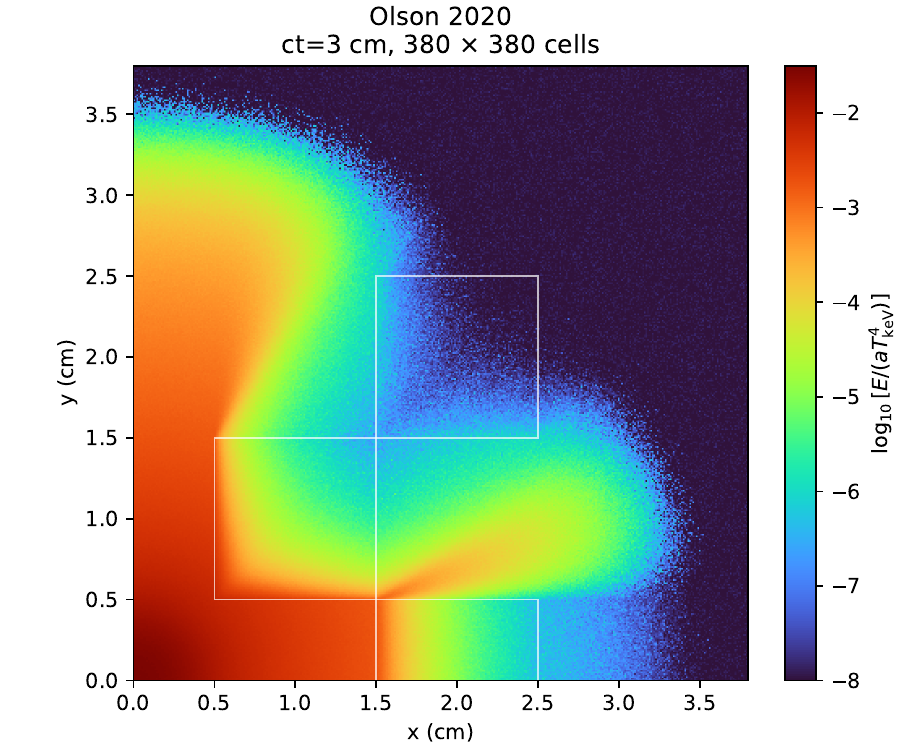}
    \caption{Scaled radiation energy density over the full domain at $ct = 3$~cm on the
    $380 \times 380$ mesh.  White outlines mark the three aluminum blocks.  The field spans six
    decades: radiation streams outward from the source at the origin, is strongly attenuated
    inside the blocks, and reaches the region beyond them mainly through the foam channels between
    and around them, producing the shadowed structure behind each block.}
    \label{fig:olson_map}
\end{figure*}

\subsubsection{Results}
Figure~\ref{fig:olson_profiles} shows the diagonal profiles at the three snapshot times, and
Figure~\ref{fig:olson_map} the corresponding two-dimensional radiation field at $ct = 3$~cm.
Along the diagonal $r = \sqrt{x^2+y^2}$, the material temperature agrees with the published IMC
solution of \citet{SteinbergHeizler2023} to a relative $L_1$ difference of $1.24\%$, $1.35\%$, and
$1.41\%$ at $ct = 2$, $2.5$, and $3$~cm respectively, evaluated for $r \le 1.4$~cm and excluding
one diagonal cell spacing around the material interfaces.  In the smooth foam region inside
$r = \sqrt{0.5}$~cm, where \citet{SteinbergHeizler2023} report $P_N$/Monte Carlo agreement, the
RMS difference from the $P_N$ reference is $0.034$, $0.032$, and $0.031$~dex in
$\log_{10}(T/\mathrm{keV})^4$ and $0.009$, $0.010$, and $0.005$~dex in the scaled radiation energy
density, against a digitization uncertainty of about $0.03$~dex; the material comparison is thus
at the level of the uncertainty in reading the reference curves.  Beyond that radius the reference
solutions themselves differ near the material interfaces.  The energy ledger---material plus
census energy against the initial energy plus the injected source energy, with all walls
reflecting---closes to a relative residual of $1.4 \times 10^{-14}$, and the sampled source
normalization matches the analytic integral of Eq.~\eqref{eq:olson_source} to
$1 \times 10^{-6}$.

The radiation profiles remain visibly noisy in the faint tail beyond $r \approx 1.5$~cm, where the
scaled energy density has fallen five decades below its peak; the published Monte Carlo reference
is noisy there as well, and agreement in that region should not be assumed to arbitrarily small
intensity.

\subsection{Crooked Pipe}\label{sec:crooked_pipe}

The three-dimensional crooked pipe is a challenging photon-transport benchmark that tests the
solver's ability to handle radiation streaming through a narrow, optically thin channel embedded
in a thick absorbing medium.

\subsubsection{Problem setup}
The domain is $x \in [0, 7]$~cm, $y, z \in [-2, 2]$~cm.  Two material regions are defined by
tracer fields:
\begin{itemize}
    \item \textbf{Thick material} (absorbing walls): $\kappa_a = 2000$~cm$^{-1}$,
        $\cV = 10^{16} / T_{\mathrm{1keV}}$~erg\,cm$^{-3}$\,K$^{-1}$.
    \item \textbf{Thin pipe} (vacuum channel): $\kappa_a = 0.2$~cm$^{-1}$,
        $\cV = 10^{13} / T_{\mathrm{1keV}}$~erg\,cm$^{-3}$\,K$^{-1}$.
\end{itemize}
Here $\cV$ is volumetric heat capacity and $T_{\mathrm{1keV}}=(1~\mathrm{keV})/k_B$ is the
fixed temperature conversion factor.  The radial coordinate is $r=\sqrt{y^2+z^2}$.
The pipe has a complex three-dimensional geometry: it consists of two straight cylindrical segments
(of radius $r = 0.5$~cm, running along the $x$-axis for $x \in [0, 2.5]$ and $x \in [4.5, 7]$)
connected by an annular elbow section ($r \in [0.5, 1.5]$~cm for $x \in [2.5, 4.5]$), with a
thick plug ($r < 1$~cm) blocking the direct line of sight through the elbow at $x \in [3, 4]$.
Radiation enters the thin pipe from the left $x$-boundary at a blackbody temperature of
$T = 0.5$~keV and must navigate through the elbow and around the plug to reach the exit.

The initial temperature is $T_0 = 0.05$~keV, no hydrodynamics is applied, and the simulation
runs to $t = 1$~$\mu$s.  The four-order-of-magnitude opacity contrast between pipe and walls
makes this a stringent test of the MC transport's geometric fidelity and the thick-cell
accelerations' ability to handle the absorbing regions.

\subsubsection{Numerical setup}
The mesh is generated from a uniform random background of $2\times10^{4}$ points, refined at the
material interfaces by structured anisotropic slabs and by $4\times10^{5}$ additional random points
filling the thin channel, for a total of $771\,345$ cells.  Each interface carries a stack of
concentric slabs whose lattice sites are shared by every slab of the stack, so that the Voronoi
bisector between a mirrored opaque/transparent pair lies exactly on the material interface; the
tangential site spacing is $0.06$~cm and the first slab on either side is $10^{-4}$~cm thick,
i.e.\ $0.2$ mean free paths in the opaque material, growing by a factor of two per slab out to
$0.8$~cm on the opaque side and $0.026$~cm on the transparent side.  This alignment matters: with
independently jittered slabs the bisectors tilt and the interface becomes a fuzzy layer of mixed
cells roughly $10^{-2}$~cm thick, which by itself delays the arrival at the far probes by factors
of three to five.

The time step follows \citet{SteinbergHeizler2022}: $\Delta t = 10^{-11}$~s growing by $10\%$ per
cycle and capped at $10^{-8}$~s, reaching $t = 1$~$\mu$s in $163$ cycles.  \DDMC{} is enabled for
cells whose mean chord is at least three mean free paths (Section~\ref{sec:ddmc_eligibility}),
which claims every opaque slab from the fourth outward while leaving the three sub-mean-free-path
surface slabs---where the boundary layer and the wall albedo are set---in transport.  The
probe temperatures are averaged over the ring of transparent cells within $0.1$~cm of each
$(r, x)$ location: the problem is axisymmetric, so this costs no physics, and it removes a
sampling noise of $\sim 0.09$~keV per cycle that a single cell of the low-heat-capacity channel
material would otherwise show.  The run emits $6$ new packets per cell per step with an
emission floor of $4$, uses comb population control with $N_{\min} = 6$ and census floor and cap
of $8$ and $2000$ packets per cell (Section~\ref{sec:pop_control}), and injects $100$ drive
packets per source face per step.

\begin{figure}
    \centering
    \includegraphics[width=\columnwidth]{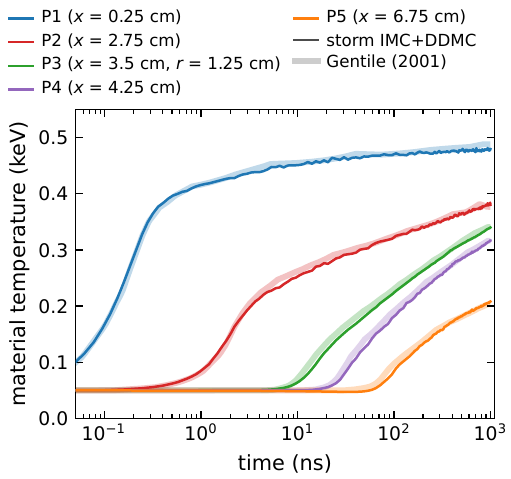}
    \caption{Crooked pipe: material temperature at the five reference probes of
    \citet{SteinbergHeizler2022}, Fig.~8(a), against the \citet{Gentile2001} reference solution
    (pale).  P1 sits just inside the inlet, P2 at the end of the first straight segment, P3 in the
    elbow bypass above the plug, P4 behind the plug and P5 near the exit; positions are given along
    the pipe axis $x$.  The \STORM{} curves are
    ring averages over the transparent cells within $0.1$~cm of each probe.  Deviations below
    $\sim 0.01$~keV are at the level of the digitization uncertainty of the reference curves.}
    \label{fig:crooked_pipe_probes}
\end{figure}

\subsubsection{Results}
Figure~\ref{fig:crooked_pipe_probes} compares the five probe histories with the
\citet{Gentile2001} reference solution.  The agreement is at the level of the digitization
uncertainty of the published curves: the time-averaged errors are $0.006$, $0.005$, $0.006$,
$0.005$ and $0.004$~keV at P1--P5, with worst-case deviations of $0.013$--$0.017$~keV.  Taking the
first crossing of $0.08$~keV as the heat-arrival time, the wave reaches the probes within $5\%$
(P1, P2), $16\%$ (P3), $10\%$ (P4) and $19\%$ (P5) of the reference times of $0.04$, $0.96$,
$11.1$, $27.9$ and $81.0$~ns, so both the direct filling of the first segment and the two
indirect paths---around the plug through the elbow bypass, and into the shadow behind it---are
reproduced.

The deep-wall treatment is what sets the late-time level.  Repeating the calculation with
\DDMC{} disabled, so that the opaque slabs are transported by \IMC{} alone, leaves the arrival
times almost unchanged but drops every probe below the reference by a margin that grows down the
pipe and with time, reaching $0.016$~keV at P2 and $0.037$--$0.039$~keV at P3 and P4 at
$t = 1$~$\mu$s (time-averaged errors $0.016$, $0.026$, $0.028$ and $0.020$~keV at P2--P5, against
$0.004$--$0.006$~keV here).  The reason is the familiar \IMC{} teleportation error: once the
thermal front leaves the resolved surface slabs and enters the geometrically growing ones, whose
optical depths run from $1.6$ to several hundred mean free paths, \IMC{} drives energy too deep
into the wall, and the energy so buried is energy the channel does not carry downstream.  Enabling \DDMC{} above three mean free paths removes it.  It is worth
noting that the \IMC{} curves published by \citet{SteinbergHeizler2022} for this problem lie below
the reference in the same manner as our \DDMC{}-disabled run.

\subsection{E3D Binary Stochastic Medium}\label{sec:e3d}

The E3D problem is a genuinely three-dimensional thermal-transport benchmark in a binary
stochastic medium: a low-opacity background seeded with many optically thick spherical inclusions
placed at random.  The benchmark was proposed by \citet{Olson2007} to test approximate treatments
of radiation transport through stochastic media against a fully resolved reference; we adopt the
constant-opacity, constant-heat-capacity parameterization of this problem used by
\citet{BrantleyNovellino2025}, who constructed a high-statistics explicit-geometry \IMC{} ensemble
solution for this configuration, together with results from the approximate stochastic-media
treatments---atomic mix, Levermore--Pomraning chord-length sampling (CLS~LP),
local-realization-preserving CLS (CLS~LRP), and LRP with ramp renewal statistics---that are
normally used in place of resolving the inclusions.  In the atomic-mix approximation, the two
materials are replaced by a single homogeneous medium with volume-averaged properties, which removes
the geometry of the inclusions entirely.  The results of these approximate treatments are taken
from \citet{BrantleyNovellino2025} and are not recomputed here.  Because the random geometry produces
transport paths that cannot be collapsed onto a one-dimensional slab, and because \RICH{}
represents each sphere explicitly on the Voronoi mesh rather than through a mixing model, this
problem tests the transport kernel on a strongly heterogeneous, fully three-dimensional
tessellation.

\subsubsection{Problem setup}
The domain is a unit cube, $0 \le x, y, z \le 1$~cm.  Two materials are present:
\begin{itemize}
    \item \textbf{Spheres:} $\kappa_a = 100$~cm$^{-1}$, $\rho = 1.0$~g\,cm$^{-3}$.
    \item \textbf{Background:} $\kappa_a = 1$~cm$^{-1}$, $\rho = 0.1$~g\,cm$^{-3}$.
\end{itemize}
Both have specific heat capacity $c_v = 10^{15}$~erg\,g$^{-1}$\,keV$^{-1}$.  The inclusions all have radius
$R = 0.0225$~cm, and $2096$ of them are placed without overlap, giving a sphere volume fraction
$p_1 = 0.1$ and mean chord lengths $\Lambda_1 = 4R/3 = 0.03$~cm in the spheres and
$\Lambda_2 = 0.27$~cm in the background.  Each sphere is optically thick across its diameter
($\kappa_a\, 2R = 4.5$) while the background is thin over the domain ($\kappa_a L = 1$, where $L=1$~cm is the cube side length), so the
radiation must find its way around the inclusions rather than through them.

A blackbody source at $T_s = 1$~keV drives the $z = 0$ face.  The $x$ and $y$ faces are
reflecting, $z = 1$~cm is a vacuum boundary where escaping packets are tallied as transmission,
and packets leaving through $z = 0$ are tallied as reflection.  The material starts at
$T_0 = 0.003$~keV with an empty radiation field, and the calculation runs to $t = 5$~ns.  There is
no hydrodynamics, no elastic scattering (only the Fleck effective scattering of
Section~\ref{sec:imc}), and the sphere realization and mesh are held fixed throughout.

The mesh-generating points are concentrated on the surfaces and in the interiors of the spheres,
so that every material interface is resolved by Voronoi faces.  The run reported here
used $8.0 \times 10^6$ generator points and $160$ MPI ranks, with 5 new thermal packets per cell
per step, 20 boundary packets per face, and a time step growing geometrically from $10^{-15}$~s to
a ceiling of $4 \times 10^{-12}$~s.  \DDMC{} is not enabled, so
the spheres are traversed by standard \IMC{} transport.

\begin{figure*}
    \centering
    \includegraphics[width=\textwidth]{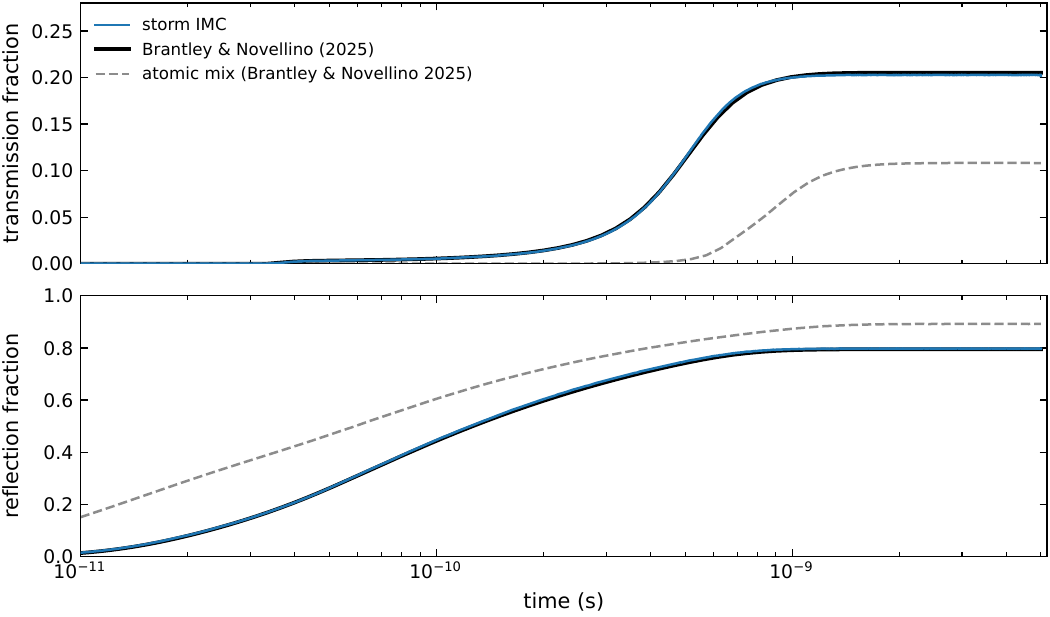}
    \caption{Photon transmission (top) and reflection (bottom) fractions versus time for the E3D
    binary stochastic medium.  The \STORM{} \IMC{} result (blue) is compared with the
    explicit-geometry ensemble benchmark of \citet{BrantleyNovellino2025} (black) and with the
    atomic-mix approximation reported in the same work (grey dashed).  Resolving the spheres
    explicitly on the Voronoi mesh reproduces the benchmark to an RMS difference of
    $1.3 \times 10^{-3}$ in transmission and $4.3 \times 10^{-3}$ in reflection, whereas replacing
    the geometry by a homogenized mixture roughly halves the transmitted fraction and
    correspondingly overpredicts the reflected fraction.}
    \label{fig:e3d_results}
\end{figure*}

\subsubsection{Results}
Figure~\ref{fig:e3d_results} shows the transmission and reflection histories.  The \STORM{} curves
follow the explicit-geometry benchmark of \citet{BrantleyNovellino2025} throughout the transient
and onto the late-time plateau, with an RMS difference of $1.3 \times 10^{-3}$ (maximum
$3.4 \times 10^{-3}$) in the transmitted fraction and $4.3 \times 10^{-3}$ (maximum
$7.2 \times 10^{-3}$) in the reflected fraction; the final transmitted fraction is $0.202$ against
the benchmark's $0.205$.  The atomic-mix approximation, plotted for contrast, reaches only about
half the transmitted fraction, since homogenizing the opacity removes the low-opacity channels
between the spheres through which most of the transmitted radiation actually travels.  The
chord-length-sampling treatments also reported by \citet{BrantleyNovellino2025} do better than
atomic mix but still overpredict the transmitted plateau.  That the explicit Voronoi
representation lands on the benchmark while the mixing models do not is the expected result, and
it confirms that the \IMC{} kernel handles a dense population of optically thick interfaces on an
unstructured three-dimensional mesh without systematic leakage or streaming errors.

We note two differences from the published reference calculation.  \citet{BrantleyNovellino2025}
average over ten independent geometry realizations, whereas the result shown here is a single
realization, so a part of the residual difference is realization noise rather than solver error.
The published run also used a substantially larger packet budget.  A full ensemble average over
realizations would be needed to quote a converged discrepancy.

\subsection{McClarren--Gentile Moving Slab}\label{sec:moving_slab}

The moving-slab benchmark tests the Doppler and Lorentz corrections for frequency-dependent
radiation transport in a moving medium \citep{McClarrenGentile2021}.  The material moves at
$v/c = 2.0\times10^{-2}$, the Voronoi mesh moves with it, and every emission, absorption and
scattering event is transformed between the comoving and laboratory frames, so the problem belongs
with the radiation--hydrodynamics tests rather than the pure-transport ones even though no
radiation force or radiation work acts on the gas.  It exercises the same machinery as the
radiative-shock problems --- mesh motion, census reassignment, and the Doppler factor in the
weight attenuation --- with the
radiation--matter feedback switched off, so that the emergent spectrum can be compared against a
semi-analytic solution.  It is listed accordingly in Table~\ref{tab:benchmarks}, the dagger
recording the absence of feedback.

\subsubsection{Problem setup}
An aluminum slab of density $\rho = 0.1$~g/cc and width $L = 0.4$~cm, at $T = 1$~keV, starts at
$0 \le x \le 0.4$~cm and moves along $+x$ at $v = 0.5994$~cm/ns ($v/c = 2.0\times10^{-2}$) toward
an observer at $x = 12$~cm.  The slab is given a volumetric heat capacity of
$10^{23}$~erg\,keV$^{-1}$\,cm$^{-3}$, so that its temperature stays at $1$~keV throughout, and the
region ahead of it is filled with vacuum of density $10^{-10}$~g/cc.  The opacity is given by a
124-group tabulated absorption coefficient, the table of the benchmark of
\citet{McClarrenGentile2021}.  The simulation uses multigroup \IMC{} without \DDMC{} or population control,
with the full Lorentz corrections of Sections~\ref{sec:physical_model}
and~\ref{sec:implementation}.  No hydrodynamic equations are solved: the material velocity is
prescribed, and the mesh-generating points are moved kinematically, those in the slab with the
slab and those ahead of it compressed uniformly between the slab front and the observer plane,
with the domain's trailing face moving with the slab.  The radiation does not act back on the
material, which isolates the transport error from any hydrodynamic error.  The transverse
extent is $0.2\times0.2$~cm with specularly reflecting $y$ and $z$ faces, and photons escape through
both $x$ faces.  The mesh has $80 \times 3 \times 3 = 720$ cells: 20 generating points across the
slab, a spacing of $0.02$~cm, and 60 in the vacuum ahead of it along the direction of motion.
The emission floor is $N_{\mathrm{new}} = 3333$ packets per cell (Section~\ref{sec:emission}).
The time step starts at $1$~ps and grows by a factor of
$1.1$ per step to a ceiling of $100$~ps, and the spectrum is recorded at the observation time
$t_O = 10$~ns; the run stops at $t_O$ plus half a maximum step, $10.05$~ns, after 139 steps.

This benchmark validates the frame transformations of Sections~\ref{sec:transport_eq},
\ref{sec:energy_dep}, \ref{sec:event_loop} and~\ref{sec:emission}: the observed
photon spectrum should be blue-shifted by the slab's velocity, and the total emergent energy should
be consistent with the Lorentz-boosted thermal emission.

\begin{figure}
    \centering
    \includegraphics[width=\columnwidth]{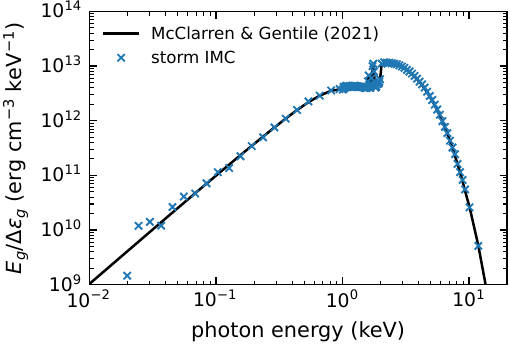}
    \caption{Emergent spectral radiation energy density $E_g/\Delta\varepsilon_g$, the group radiation
    energy density at the observer averaged over the last step (Section~\ref{sec:estimators}) and
    divided by the group width, seen by an observer toward whom
    the aluminum slab is moving at $v = 0.5994$\,cm/ns, compared with the McClarren--Gentile
    semi-analytic solution \citep{McClarrenGentile2021} for the 124-group tabulated opacity.
    \STORM{} Monte Carlo samples (crosses) follow the semi-analytic spectrum (black) across three
    decades in energy, including the opacity-edge features near $1$--$2$\,keV; the flux-averaged
    error is $2.4\%$.}
    \label{fig:moving_slab_results}
\end{figure}

\subsubsection{Results}
Figure~\ref{fig:moving_slab_results} compares the emergent multigroup spectrum to the
McClarren--Gentile semi-analytic solution.  The Monte Carlo spectrum tracks the semi-analytic curve
across the thermal peak, the exponential tail, and the opacity-edge structure inherited from the
124-group aluminum table, with a flux-averaged error of $2.4\%$ and an $L_1$ error of $2.4\%$.
Because the slab moves toward the observer, this tests Doppler-shifted transport and the
transformation of the emergent spectrum.  The implementation uses the full Lorentz boosts
described in Sections~\ref{sec:transport_eq}, \ref{sec:emission}, and~\ref{sec:event_loop}.
At $v/c\simeq0.02$, however, the $2.4\%$ spectral error does not resolve corrections of order
$(v/c)^2\simeq4\times10^{-4}$ and cannot distinguish the full boost from a first-order
approximation.  This prescribed-flow benchmark also does not test the recoil update of a
dynamically evolving gas.

%%%%%%%%%%%%%%%%%%%%%%%%%%%%%%%%%%%%%%%%%%%%%%%%%%
\section{Conclusion}\label{sec:conclusion}

We have presented an Implicit Monte Carlo radiation-transport solver in the mesh-agnostic \STORM{}
library, coupled to \RICH{} on a moving Voronoi mesh constructed by \MadVoro{}.  The solver
combines the Fleck--Cummings \IMC{} linearization, formulated in the fluid's proper time, with
multigroup and continuous-frequency transport, exact Lorentz transformations of photon packets in
moving media, multigroup \DDMC{} acceleration with an emissivity-preserving interface, comb-based
population control, and first-order operator-split coupling to \RICH{}'s hydrodynamics.  The same
kernel runs unchanged on Cartesian and Voronoi grids, and parallel packet migration is described
in the companion \STORM{} paper.

The benchmarks establish the solver quantitatively across transport, multigroup, and
radiation--hydrodynamics regimes.  Marshak-wave fronts are placed to better than one per cent of
the distance travelled, and the three-dimensional hohlraum and crooked-pipe problems follow the
published reference solutions through strongly heterogeneous, partly vacuum geometries.  Explicit transport through the E3D binary stochastic medium matches the benchmark
transmission to an RMS difference of $1.3\times10^{-3}$, whereas an atomic-mix treatment gives
only about half the transmitted fraction.
For coupled problems, the moving mesh recovers $75\%$ of the Mach~2 Zel'dovich spike against $51\%$
on a fixed mesh, the Mach~45 profile stays on the semi-analytic solution through a
radiation-dominated, scattering-dominated shock, and on the RMTV problem a Lagrangian mesh with one
eighth of the cells gives a lower density error than the Eulerian one while conserving mass and
energy to roundoff.  \DDMC{}
reduces the run time of the multigroup Densmore problem from $698$ to $279$~s at unchanged
accuracy, and the moving-slab spectrum agrees with the semi-analytic solution to $2.4\%$.

For direct \IMC{} feedback, energy and momentum are conserved by construction.  Thermal emission
is debited using the actual emitted packet energies and momenta, and transport transfers are
tallied in the laboratory frame.  The internal-energy change is the total-energy transfer minus the
exact Newtonian kinetic-energy change (Eq.~\eqref{eq:material_exchange}), so these source and
interaction updates balance packet and material energy and momentum to roundoff.

The method has several limitations.  The operator split is first-order, so radiation and
hydrodynamics exchange energy only once per cycle.  Standard \IMC{} can violate the maximum
principle on large time steps \citep{Wollaber2016}.  The \DDMC{} path uses comoving packet weights
and a flux-based force estimator that relies on the diffusion and low-velocity approximations
(Section~\ref{sec:ddmc_hydro}).  Finally, the conservation above is a discrete property of the
Newtonian material update, not a relativistic treatment of the fluid, and it does not extend to
the momentum changes introduced by population resampling.

The combination of capabilities demonstrated here---three-dimensional moving-mesh
hydrodynamics, frequency-dependent transport with exact Lorentz transformations, \DDMC{}
acceleration of optically thick regions, and scaling to over ten thousand cores
\citep{Mizrachi_2026}---opens a range of astrophysical applications in which diffusion-based
radiation transport is inaccurate or of uncertain validity.  In tidal disruption events the
emission is produced by shocks in, and reprocessed through, optically thick and highly asymmetric
debris moving at a substantial fraction of the speed of light, so that the observed light depends
on the viewing angle \citep{Roth2016,Dai2018,SteinbergStone2024,Price2024}.  Shock breakout and
supernova light curves \citep{NakarSari2010,Sapir2011}, kilonovae \citep{Kasen2013,Metzger2019},
and super-Eddington accretion flows \citep{Ohsuga2005,Jiang2014} pose similar problems.  In all
of these settings the solver provides a reference that does not rely on a closure, and it can be
used both to compute the emergent spectra and light curves directly and to calibrate the cheaper
diffusion and moment methods that make larger simulations feasible.

\section*{Code Availability}
The \IMC{} and \DDMC{} solvers described here are part of \STORM{} \citep{Mizrachi_2026}, which
is publicly available as open-source software under the BSD 3-Clause licence.  Its first release
is permanently archived on Zenodo \citep{STORMZenodo}, and ongoing development, including the
solvers presented in this paper, is available at \url{https://github.com/maormizrachi/STORM}.
\STORM{} is distributed both as a standalone package and as an integrated module within the
\RICH{} three-dimensional moving-mesh hydrodynamics code \citep{Yalinewich2015}, which provides
the radiation--hydrodynamics coupling and is publicly available at
\url{https://gitlab.com/eladtan/RICH}.  Both codes build their meshes with the \MadVoro{} parallel
Voronoi tessellation builder \citep{Mizrachi2025MadVoro}, available under the BSD 3-Clause licence
at \url{https://github.com/maormizrachi/MadVoro}.  The codes may be freely used and modified; we
request that any published work that uses the radiation solvers cite this paper.

\begin{acknowledgments}
The benchmark simulations were run on the LIPS (Large-Scale Interdisciplinary Physical
Simulations) cluster of the Interdisciplinary Computational Physics Lab (ICPL) at the Racah
Institute of Physics, The Hebrew University of Jerusalem.\footnote{\url{https://icpl.huji.ac.il/}}
The authors used the large language models Claude (Fable~5.1 and Opus~5.5; Anthropic) and ChatGPT
(Sol~5.6 and Astra~6; OpenAI) to assist with generating the figures and with grammar and language
editing of the manuscript.  All output was reviewed and edited by the authors, who take full
responsibility for the content and writing of this paper.
\end{acknowledgments}

\bibliography{references}{}
\bibliographystyle{aasjournalv7}

\end{document}